\documentclass{lmcs}
\pdfoutput=1
\usepackage[utf8]{inputenc}

\usepackage{lastpage}
\lmcsdoi{22}{2}{34}
\lmcsheading{}{\pageref{LastPage}}{}{}%
{Apr.~15,~2025}{Jun.~29,~2026}{}


\usepackage{amsthm}
\usepackage{amsmath}
\usepackage{stmaryrd}

\usepackage{etoolbox}
\usepackage{xparse}
\usepackage{mleftright}

\usepackage{standalone}
\usepackage{wrapfig}
\usepackage{booktabs}
\usepackage{microtype}
\usepackage{multicol}
\usepackage{mathtools}

\usepackage{tikz}
\usetikzlibrary{bbox, automata, positioning, arrows}
\usepackage{tikz-cd}
\usepackage{figures/tikzit}

\usepackage{bussproofs}
\usepackage{mathpartir}
\usepackage{hyperref}
\usepackage[capitalise]{cleveref}

\newcommand{\mce}{\mathcal{E}}

\NewDocumentCommand\proj{mo}{%
  \pi_{#1}\IfValueT{#2}{\left(#2\right)}
}

\newcommand{\lub}{\sqcup}
\newcommand{\ljoin}{\lub}
\newcommand{\glb}{\sqcap}
\newcommand{\lmeet}{\glb}

\newcommand{\tuple}[2]{#1^{#2}}

\makeatletter
\newsavebox{\@brx}
\newcommand{\llangle}[1][]{\savebox{\@brx}{\(\m@th{#1\langle}\)}%
  \mathopen{\copy\@brx\kern-0.5\wd\@brx\usebox{\@brx}}}
\newcommand{\rrangle}[1][]{\savebox{\@brx}{\(\m@th{#1\rangle}\)}%
  \mathclose{\copy\@brx\kern-0.5\wd\@brx\usebox{\@brx}}}
\makeatother

\NewDocumentCommand\inl{o}{%
    \mathsf{inl}\IfValueT{#1}{(#1)}
}
\NewDocumentCommand\inr{o}{%
    \mathsf{inr}\IfValueT{#1}{(#1)}
}

\newcommand{\nat}{\mathbb{N}}
\newcommand{\natplus}{\mathbb{N}_{+}}

\newcommand{\freemon}[1]{{#1}^\star}
\newcommand{\concat}{\mathbin{+\hspace{-3pt}+}}

\newcommand{\listvar}[1]{\overline{#1}}

\newcommand{\listlistvar}[1]{\underline{\listvar{#1}}}

\newcommand{\morph}[3]{{{#1} \colon {#2 \to #3}}}
\newcommand{\seq}{\mathbin{\ \fatsemi\ }}
\NewDocumentCommand\id{o}{%
    \mathsf{id}\IfValueT{#1}{_{#1}}%
}

\DeclareMathOperator{\tr}{Tr}
\NewDocumentCommand\trace{moom}{%
    {\tr^{#1}\IfValueT{#2}{_{#2, #3}}\left(#4\right)}
}

\NewDocumentCommand\ccunit{o}{
    \eta\IfValueT{#1}{_{#1}}
}
\NewDocumentCommand\cccounit{o}{
    \varepsilon\IfValueT{#1}{_{#1}}
}

\NewDocumentCommand\ccopy{o}{
    {\Delta\IfValueT{#1}{_{#1}}}
}
\NewDocumentCommand\cdel{o}{
    {\scalebox{1.3}{\(\diamond\)}\IfValueT{#1}{_{#1}}}
}
\NewDocumentCommand\cmerge{o}{
    {\nabla\IfValueT{#1}{_{#1}}}
}
\NewDocumentCommand\cinit{o}{
    {\square\IfValueT{#1}{_{#1}}}
}

\NewDocumentCommand\fix{o}{
    \mathsf{fix}\IfValueT{#1}{\left(#1\right)}
}

\newcommand{\signature}{\Sigma}
\NewDocumentCommand\generators{o}{%
    \Sigma\IfValueT{#1}{_{#1}}
}

\NewDocumentCommand\equations{o}{%
    \mathcal{E}\IfValueT{#1}{_{#1}}
}

\NewDocumentCommand\dom{o}{%
    \mathsf{dom}\IfValueT{#1}{(#1)}
}
\NewDocumentCommand\domw{o}{%
    \overline{\mathsf{dom}}\IfValueT{#1}{(#1)}
}
\NewDocumentCommand\cod{o}{%
    \mathsf{cod}\IfValueT{#1}{(#1)}
}
\NewDocumentCommand\codw{o}{%
    \overline{\mathsf{cod}}\IfValueT{#1}{(#1)}
}
\NewDocumentCommand\eqaxioms{o}{%
    \IfValueTF{#1}{\stackrel{#1}{=}}{=}
}
\NewDocumentCommand\reduction{o}{%
    \IfValueTF{#1}{\stackrel{#1}{\rightsquigarrow}}{\rightsquigarrow}
}
\newcommand{\reductions}{\reduction[*]}

\NewDocumentCommand\spann{momom}{%
    {#1 \IfValueTF{#2}{\xleftarrow{#2}}{\leftarrow} #3 \IfValueTF{#4}{\xrightarrow{#4}}{\rightarrow} #5}
}
\NewDocumentCommand\cospan{momom}{%
    {#1 \IfValueTF{#2}{\xrightarrow{#2}}{\rightarrow} #3 \IfValueTF{#4}{\xleftarrow{#4}}{\leftarrow} #5}
}
\NewDocumentCommand\csp{om}{%
    \mathsf{Csp}\IfValueT{#1}{_{#1}}(#2)
}

\newcommand{\set}{\mathbf{Set}}

\newcommand{\pos}{\mathbf{Pos}}

\NewDocumentCommand\frob{o}{%
    \mathbf{Frob}\IfValueT{#1}{_{#1}}%
}

\NewDocumentCommand\ccomon{o}{%
    \mathbf{CComon}\IfValueT{#1}{_{#1}}%
}

\newcommand{\smc}[1]{\mathbf{S}_{#1}}

\newcommand{\stmc}[1]{\mathbf{T}_{#1}}

\NewDocumentCommand\idf{o}{%
    \mathsf{Id}\IfValueT{#1}{_{#1}}%
}

\newcommand{\monoidunitleqn}{\mathsf{MUL}}

\newcommand{\monoidunitreqn}{\mathsf{MUR}}

\newcommand{\monoidassoceqnletter}{\mathsf{MA}}

\newcommand{\monoidcommeqnletter}{\mathsf{MC}}

\newcommand{\comonoiduniteqnletter}{\mathsf{CU}}

\newcommand{\circuitsignature}{\Sigma}
\newcommand{\circuitsignaturevalues}{\mathcal{V}}
\newcommand{\values}{\mathbf{V}}

\newcommand{\circuitsignaturegates}{\mathcal{P}}

\newcommand{\circuitsignaturearity}{\dom}
\newcommand{\circuitsignaturecoarity}{\cod}

\newcommand{\disconnected}{\bullet}

\newcommand{\ccirc}[1]{\mathbf{CCirc}_{#1}}
\newcommand{\ccircsigma}{\ccirc{\Sigma}}
\newcommand{\ccircsigmag}{\ccirc{\Sigma}^{+}}

\newcommand{\scirc}[1]{\mathbf{SCirc}_{#1}}
\newcommand{\scircsigma}{\scirc{\Sigma}}
\newcommand{\scircsigmag}{\scirc{\Sigma}^{+}}
\newcommand{\scircsigmap}{\scirc{\Sigma}^\mathsf{P}}

\newcommand{\interpretation}{\mathcal{I}}
\NewDocumentCommand\valueinterpretation{o}{%
    \left\llbracket\IfValueTF{#1}{#1}{-}\right\rrbracket^{\mathbf{V}}
}
\NewDocumentCommand\gateinterpretation{o}{%
    \left\llbracket\IfValueTF{#1}{#1}{-}\right\rrbracket
}
\NewDocumentCommand\functostream{om}{%
    \left\ulcorner\IfValueTF{#1}{#1}{-}\right\urcorner_{#2}
}
\NewDocumentCommand\functostreami{o}{%
    \functostream[\IfValueTF{#1}{#1}{-}]{\interpretation}%
}
\NewDocumentCommand\circuittofunc{om}{%
    \left\llbracket\IfValueTF{#1}{#1}{-}\right\rrbracket^{\mathbf{C}}_{#2}
}
\NewDocumentCommand\circuittofuncg{om}{%
    \left\llbracket\IfValueTF{#1}{#1}{-}\right\rrbracket^{\mathbf{C}+}_{#2}
}
\NewDocumentCommand\circuittostream{om}{%
    \left\llbracket\IfValueTF{#1}{#1}{-}\right\rrbracket^{\mathbf{S}}_{#2}
}
\NewDocumentCommand\circuittostreamg{om}{%
    \left\llbracket\IfValueTF{#1}{#1}{-}\right\rrbracket^{\mathbf{S}+}_{#2}
}
\NewDocumentCommand\circuittostreami{o}{%
    \circuittostream[\IfValueTF{#1}{#1}{-}]{\interpretation}%
}
\NewDocumentCommand\circuittostreamig{o}{%
    \circuittostreamg[\IfValueTF{#1}{#1}{-}]{\interpretation}%
}
\NewDocumentCommand\circuittofunci{o}{%
    \circuittofunc[\IfValueTF{#1}{#1}{-}]{\interpretation}%
}
\NewDocumentCommand\circuittofuncig{o}{%
    \circuittofuncg[\IfValueTF{#1}{#1}{-}]{\interpretation}%
}
\NewDocumentCommand\circuittomealy{om}{%
    \left[\IfValueTF{#1}{#1}{-}\right]_{#2}%
}
\NewDocumentCommand\circuittomealyg{om}{%
    \left[\IfValueTF{#1}{#1}{-}\right]_{#2}^+%
}
\NewDocumentCommand\circuittomealyi{o}{%
    \circuittomealy[\IfValueTF{#1}{#1}{-}]{\interpretation}%
}
\NewDocumentCommand\circuittomealyig{o}{%
    \circuittomealyg[\IfValueTF{#1}{#1}{-}]{\interpretation}%
}

\newcommand{\scircq}[2]{\mathbf{SCirc}_{{#1} / {#2}}}
\newcommand{\scircgq}[2]{\mathbf{SCirc}^+_{{#1} / {#2}}}
\newcommand{\scircsigmai}{\scircq{\Sigma}{\approx_{\interpretation}}}
\newcommand{\scircsigmaig}{\scircgq{\Sigma}{\approx^+_{\interpretation}}}
\newcommand{\scircsigmae}{\scircq{\Sigma}{\mce_{\interpretation}}}
\newcommand{\scircsigmage}{\scircgq{\Sigma}{\mce^+_{\interpretation}}}
\newcommand{\scircsigmaobs}{\scircq{\Sigma}{\sim_{\interpretation}}}
\newcommand{\scircsigmagobs}{\scircgq{\Sigma}{\sim^+_{\interpretation}}}

\newcommand{\mealyequations}{\mathcal{M}}

\newcommand{\normalisingequations}{\mathcal{N}_\interpretation}
\newcommand{\encodingequation}{\mathsf{Enc}}
\newcommand{\encodingequations}{\mathcal{H}}
\newcommand{\restrictionequation}{\mathsf{Res}}

\newcommand{\instantfeedbackeqn}{\mathsf{IF}}

\newcommand{\gateeqn}{\mathsf{P}_{\interpretation}}
\newcommand{\forkeqn}{\mathsf{F}}

\newcommand{\joineqn}{\mathsf{J}}
\newcommand{\stubeqn}{\mathsf{E}}

\newcommand{\mealyeqn}{\mathsf{Mealy}}

\newcommand{\spliteqn}{\mathsf{Split}}
\newcommand{\combineeqn}{\mathsf{Comb}}

\newcommand{\streamingeqn}{\mathsf{Str}}

\newcommand{\bottomdelayeqn}{\mathsf{BD}}

\newcommand{\delayforkeqn}{\mathsf{DF}}

\DeclareMathOperator{\andgate}{AND}
\DeclareMathOperator{\orgate}{OR}
\DeclareMathOperator{\notgate}{NOT}

\DeclareMathOperator{\nandgate}{NAND}
\DeclareMathOperator{\norgate}{NOR}
\DeclareMathOperator{\xorgate}{XOR}

\newcommand{\booleans}{\mathbf{B}}

\newcommand{\belnapinterpretation}{\interpretation_{\belnap}}
\newcommand{\belnapfalse}{\mathsf{f}}
\newcommand{\belnaptrue}{\mathsf{t}}
\newcommand{\valuetuple}[1]{\tuple{\values}{#1}}

\newcommand{\valuetupleseq}[1]{(\valuetuple{#1})^\star}
\newcommand{\valuetuplestream}[1]{\stream{(\valuetuple{#1})}}
\newcommand{\belnap}{\mathsf{B}}
\newcommand{\belnapsignature}{\circuitsignature_{\belnap}}
\newcommand{\belnapvalues}{\values_{\belnap}}
\newcommand{\belnapgates}{\circuitsignaturegates_{\belnap}}
\newcommand{\belnaparity}{\circuitsignaturearity_{\belnap}}
\newcommand{\belnapcoarity}{\circuitsignaturecoarity_{\belnap}}
\NewDocumentCommand\belnapvalueinterpretation{o}{%
    \valueinterpretation\IfValueT{#1}{[#1]}_{\belnap}
}
\NewDocumentCommand\belnapgateinterpretation{o}{%
    \gateinterpretation\IfValueT{#1}{[#1]}_{\belnap}
}

\newcommand{\joinforkeqn}{\mathsf{JF}}

\newcommand{\forkuniteqn}{\mathsf{FU}}

\newcommand{\joinunitleqn}{\mathsf{JoinUnitL}}
\newcommand{\joinunitreqn}{\mathsf{JoinUnitR}}
\newcommand{\joinassoceqn}{\mathsf{JA}}
\newcommand{\joincommeqn}{\mathsf{JC}}

\newcommand{\equationdisplay}[3]{#1=#2\;(#3)}
\newcommand{\func}[1]{\mathbf{Func}_{#1}}
\newcommand{\funcg}[2]{\mathbf{Func}_{#1}^{#2}}
\newcommand{\funci}{\func{\interpretation}}
\newcommand{\funcig}{\funcg{\interpretation}{+}}

\NewDocumentCommand\mealyoutput{omm}{
    {#2}[{#3}]\IfValueT{#1}{^{#1}}
}
\NewDocumentCommand\mealytransition{omm}{
    \IfValueT{#1}{(}{{#2}_{#3}}\IfValueT{#1}{)^{#1}}
}

\NewDocumentCommand\mealyinitial{o}{
    \bar{s}\IfValueT{#1}{^{#1}}
}

\newcommand{\mealyfunctionoutput}[1]{#1_{1}}
\newcommand{\mealyfunctiontransition}[1]{#1_{0}}

\newcommand{\initialoutput}[2]{#1[#2]}
\newcommand{\streamderivative}[2]{#1_{#2}}

\NewDocumentCommand\streaminit{o}{\mathsf{hd}\IfValueT{#1}{\left(#1\right)}}
\NewDocumentCommand\streamderv{o}{\mathsf{tl}\IfValueT{#1}{\left(#1\right)}}
\newcommand{\streamcons}{\mathbin{::}}

\newcommand{\stream}[1]{{#1}^{\omega}}

\newcommand{\streamc}[1]{\mathbf{Stream}_{#1}}
\newcommand{\streamcg}[2]{\mathbf{Stream}_{#1}^{#2}}

\newcommand{\streami}{\streamc{\interpretation}}
\newcommand{\streamig}{\streamcg{\interpretation}{+}}

\newcommand{\mealyi}{\mathbf{Mealy}_{\interpretation}}
\newcommand{\mealyig}{\mathbf{Mealy}_{\interpretation}^+}
\newcommand{\trs}[2]{#1 \,|\, #2}

\newcommand{\mealyarrow}[2]{\xrightarrow{#1\,|\,#2}}

\NewDocumentCommand\pcsprop{o}{%
\mathbf{PCStream}_{\IfValueT{#1}{#1}}%
}
\newcommand{\uniquestream}{!}
\NewDocumentCommand\mealytostream{o}{%
    {\uniquestream\IfValueTF{#1}{#1}{(-)}}%
}
\NewDocumentCommand\mealytostreami{o}{%
    {\uniquestream_{\mathcal{I}}\IfValueTF{#1}{#1}{(-)}}%
}
\NewDocumentCommand\mealytostreamig{o}{%
{\uniquestream^+_{\mathcal{I}}\IfValueTF{#1}{#1}{(-)}}%
}
\NewDocumentCommand\mealytocircuit{omm}{%
    \left\lvert\left\lvert\IfValueTF{#1}{#1}{-}\right\rvert\right\rvert^{#2}_{#3}%
}
\NewDocumentCommand\mealytocircuitg{omm}{%
    \left\lvert\left\lvert\IfValueTF{#1}{#1}{-}\right\rvert\right\rvert^{#2+}_{#3}%
}
\NewDocumentCommand\mealytocircuiti{o}{%
    \mealytocircuit[\IfValueTF{#1}{#1}{-}]{\leq}{\interpretation}%
}
\NewDocumentCommand\mealytocircuitig{o}{%
    \mealytocircuitg[\IfValueTF{#1}{#1}{-}]{\leq}{\interpretation}%
}
\NewDocumentCommand\mealytofunc{o}{%
    \lvert\lvert\IfValueTF{#1}{#1}{-}\rvert\rvert%
}
\NewDocumentCommand\streamtomealy{om}{%
    \llangle\IfValueTF{#1}{#1}{-}\rrangle_{#2}%
}
\NewDocumentCommand\streamtomealyg{om}{%
    \llangle\IfValueTF{#1}{#1}{-}\rrangle_{#2}^+%
}
\NewDocumentCommand\streamtomealyi{o}{%
    \streamtomealy[\IfValueTF{#1}{#1}{-}]{\interpretation}%
}
\NewDocumentCommand\streamtomealyig{o}{%
    \streamtomealyg[\IfValueTF{#1}{#1}{-}]{\interpretation}%
}

\newcommand{\stateorder}{\preceq}

\NewDocumentCommand\termtohyp{om}{%
    \llbracket{\IfValueTF{#1}{#1}{-}}\rrbracket_{#2}
}
\NewDocumentCommand\termtohypsigma{o}{%
    \llbracket{\IfValueTF{#1}{#1}{-}}\rrbracket_{\Sigma}
}
\NewDocumentCommand\termtohypsigmac{o}{%
    \llbracket{\IfValueTF{#1}{#1}{-}}\rrbracket_{C,\Sigma}
}
\NewDocumentCommand\frobtohyp{om}{%
    \left[{\IfValueTF{#1}{#1}{-}}\right]_{#2}
}
\NewDocumentCommand\frobtohypsigma{o}{%
    \left[{\IfValueTF{#1}{#1}{-}}\right]_{\Sigma}
}
\NewDocumentCommand\frobtohypsigmac{o}{%
    \left[{\IfValueTF{#1}{#1}{-}}\right]_{C,\Sigma}
}
\NewDocumentCommand\tracedtosymandfrob{om}{
    \left\lfloor\IfValueTF{#1}{#1}{-}\right\rfloor^\mathbf{T}_{#2}
}
\NewDocumentCommand\tracedtosymandfrobsigma{o}{
    \left\lfloor\IfValueTF{#1}{#1}{-}\right\rfloor^\mathbf{T}_{\Sigma}
}
\NewDocumentCommand\tracedtosymandfrobsigmac{o}{
    \left\lfloor\IfValueTF{#1}{#1}{-}\right\rfloor^\mathbf{T}_{C, \Sigma}
}
\NewDocumentCommand\comonoidtofrob{o}{
    \left\lfloor\IfValueTF{#1}{#1}{-}\right\rfloor^\mathbf{C}
}
\NewDocumentCommand\comonoidtofrobc{o}{
    \left\lfloor\IfValueTF{#1}{#1}{-}\right\rfloor^\mathbf{C}_C
}
\NewDocumentCommand\tracedandcomonoidtofrob{om}{
    \left\lfloor\IfValueTF{#1}{#1}{-}\right\rfloor_{#2}
}
\NewDocumentCommand\tracedandcomonoidtofrobsigma{o}{
    \left\lfloor\IfValueTF{#1}{#1}{-}\right\rfloor_{\Sigma}
}
\NewDocumentCommand\tracedandcomonoidtofrobsigmac{o}{
    \left\lfloor\IfValueTF{#1}{#1}{-}\right\rfloor_{C, \Sigma}
}
\NewDocumentCommand\comonoidtohyp{om}{%
    \IfValueTF{#1}{\frobtohyp[#1]{#2}}{\frobtohyp{#2}}^\mathbf{C}
}
\NewDocumentCommand\comonoidtohypsigma{o}{%
    \IfValueTF{#1}{\frobtohypsigma[#1]}{\frobtohypsigma}^\mathbf{C}
}
\NewDocumentCommand\comonoidtohypsigmac{o}{%
    \IfValueTF{#1}{\frobtohypsigmac[#1]}{\frobtohypsigmac}^\mathbf{C}
}
\NewDocumentCommand\termandfrobtohyp{om}{%
    \llangle{\IfValueTF{#1}{#1}{-}}\rrangle_{#2}
}
\NewDocumentCommand\termandfrobtohypsigma{o}{%
    \llangle{\IfValueTF{#1}{#1}{-}}\rrangle_{\Sigma}
}
\NewDocumentCommand\termandfrobtohypsigmac{o}{%
    \llangle{\IfValueTF{#1}{#1}{-}}\rrangle_{C, \Sigma}
}
\NewDocumentCommand\foldinterfaces{o}{%
    \ulcorner{\IfValueTF{#1}{#1}{-}}\urcorner
}
\NewDocumentCommand\foldinterfacesc{o}{%
    \ulcorner{\IfValueTF{#1}{#1}{-}}\urcorner_C
}

\NewDocumentCommand\rewrite{o}{%
    \Rightarrow\IfValueT{#1}{_{#1}}
}
\NewDocumentCommand\rewrites{o}{%
    \Rightarrow^{\star}\IfValueT{#1}{_{#1}}
}
\NewDocumentCommand\grewrite{o}{%
    \rightsquigarrow\IfValueT{#1}{_{#1}}
}
\NewDocumentCommand\grewrites{o}{%
    \rightsquigarrow^{\star}\IfValueT{#1}{_{#1}}
}

\NewDocumentCommand\trgrewrite{o}{%
    \rightsquigarrow\IfValueT{#1}{_{#1}}
}

\newcommand{\ifdir}{(\Leftarrow)}
\newcommand{\onlyifdir}{(\Rightarrow)}
\input{figures/circuits.tikzdefs}

\tikzstyle{none}=[anchor=center]
\tikzstyle{vertex}=[anchor=center, fill=black, draw=black, shape=circle, minimum size=2.5mm, tikzit category=hypergraph, inner sep=0]
\tikzstyle{red outline vertex}=[anchor=center, fill=black, draw=palered, shape=circle, minimum size=2.5mm, tikzit category=hypergraph, inner sep=0, line width={\stringwidth}]
\tikzstyle{red vertex}=[anchor=center, fill=palered, draw=palered, shape=circle, minimum size=2.5mm, tikzit category=hypergraph, inner sep=0, line width={\stringwidth}]
\tikzstyle{edge}=[anchor=center, fill=neutral, draw=black, shape=rectangle, font={\boxsize}, tikzit category=hypergraph, minimum width=8mm, minimum height=8mm, rounded corners=3mm, very thick, anchor=center]
\tikzstyle{red outline edge}=[anchor=center, fill=neutral, draw=palered, shape=rectangle, font={\boxsize}, tikzit category=hypergraph, minimum width=8mm, minimum height=8mm, rounded corners=3mm, very thick, anchor=center]
\tikzstyle{edge subgraph}=[fill=neutral, draw=black, shape=rectangle, font={\boxsize}, tikzit category=hypergraph, minimum width=8mm, minimum height=8mm, rounded corners=3mm, very thick, dashed]
\tikzstyle{port}=[minimum size=2.5mm, fill=none, draw=none, shape=circle, tikzit draw={rgb,255: red,154; green,154; blue,154}, anchor=center, tikzit fill=neutral]
\tikzstyle{product}=[fill=neutral, draw=black, shape=circle, scale=0.66, tikzit category=string diagram]
\tikzstyle{type}=[fill=none, draw=none, shape=circle, font={\large}, tikzit fill={rgb,255: red,37; green,193; blue,141}, inner sep=0, anchor=center, tikzit category=string diagram]
\tikzstyle{tiny box white}=[{\corners}, font={\boxsize}, fill=neutral, draw=black, shape=rectangle, tikzit category=string diagram, minimum width={\tinywidth}, minimum height={\tinywidth}, anchor=center, line width={\stringwidth}, tikzit fill=neutral, inner sep={\innersep}]
\tikzstyle{tiny box comb}=[{\corners}, font={\boxsize}, fill=comb, draw=black, shape=rectangle, tikzit category=string diagram, minimum width={\tinywidth}, minimum height={\tinywidth}, anchor=center, line width={\stringwidth}, tikzit fill=comb, inner sep={\innersep}]
\tikzstyle{tiny box seq}=[{\corners}, font={\boxsize}, fill=seq, draw=black, shape=rectangle, tikzit category=string diagram, minimum width={\tinywidth}, minimum height={\tinywidth}, anchor=center, line width={\stringwidth}, inner sep={\innersep}]
\tikzstyle{tiny signal seq}=[{\corners}, font={\boxsize}, shape=signal, signal to=west, signal pointer angle=110, fill=seq, draw=black, tikzit category=string diagram, minimum width=6mm, minimum height=5mm, anchor=center, line width={\stringwidth}, inner sep={\innersep}]
\tikzstyle{small box white}=[{\corners}, font={\boxsize}, fill=neutral, draw=black, shape=rectangle, minimum height={\smallwidth}, minimum width={\tinywidth}, tikzit category=string diagram, anchor=center, line width={\stringwidth}, inner sep={\innersep}]
\tikzstyle{small square box white}=[{\corners}, font={\boxsize}, fill=neutral, draw=black, shape=rectangle, minimum height={\smallwidth}, minimum width={\smallwidth}, tikzit category=string diagram, anchor=center, line width={\stringwidth}, inner sep={\innersep}]
\tikzstyle{medium box}=[rounded corners, {\corners}, font={\boxsize}, fill=neutral, draw=black, shape=rectangle, tikzit category=string diagram, minimum height={\mediumwidth}, minimum width={\smallwidth}, anchor=center, line width={\stringwidth}, tikzit fill=neutral, inner sep={\innersep}]
\tikzstyle{medium box white}=[{\corners}, font={\boxsize}, fill=neutral, draw=black, shape=rectangle, tikzit category=string diagram, minimum height={\mediumwidth}, minimum width={\smallwidth}, anchor=center, line width={\stringwidth}, tikzit fill=comb, inner sep={\innersep}]
\tikzstyle{medium box comb}=[{\corners}, font={\boxsize}, fill=comb, draw=black, shape=rectangle, tikzit category=string diagram, minimum height={\mediumwidth}, minimum width={\smallwidth}, anchor=center, line width={\stringwidth}, tikzit fill=comb, inner sep={\innersep}]
\tikzstyle{medium square box white}=[rounded corners, {\corners}, font={\boxsize}, fill=neutral, draw=black, shape=rectangle, tikzit category=string diagram, minimum height={\mediumwidth}, minimum width={\mediumwidth}, line width={\stringwidth}, tikzit fill=neutral, inner sep={\innersep}]
\tikzstyle{medium square box comb}=[rounded corners, {\corners}, font={\boxsize}, fill=comb, draw=black, shape=rectangle, tikzit category=string diagram, minimum height={\mediumwidth}, minimum width={\mediumwidth}, line width={\stringwidth}, tikzit fill=comb, inner sep={\innersep}]
\tikzstyle{medium square box seq}=[rounded corners, {\corners}, draw=black, font={\boxsize}, fill=seq, shape=rectangle, tikzit category=string diagram, minimum height={\mediumwidth}, minimum width={\mediumwidth}, line width={\stringwidth}, tikzit fill=seq, inner sep={\innersep}]
\tikzstyle{medium square rounded box white}=[rounded corners, font={\boxsize}, fill=neutral, draw=black, shape=rectangle, tikzit category=string diagram, minimum height={\mediumwidth}, minimum width={\mediumwidth}, line width={\stringwidth}, tikzit fill=neutral, inner sep={\innersep}]
\tikzstyle{medium square rounded box comb}=[rounded corners, font={\boxsize}, fill=comb, draw=black, shape=rectangle, tikzit category=string diagram, minimum height={\mediumwidth}, minimum width={\mediumwidth}, line width={\stringwidth}, tikzit fill=comb, inner sep={\innersep}]
\tikzstyle{medium square rounded box seq}=[rounded corners, draw=black, font={\boxsize}, fill=seq, shape=rectangle, tikzit category=string diagram, minimum height={\mediumwidth}, minimum width={\mediumwidth}, line width={\stringwidth}, tikzit fill=seq, inner sep={\innersep}]
\tikzstyle{large box}=[{\corners}, font={\boxsize}, fill=neutral, draw=black, shape=rectangle, tikzit category=string diagram, minimum height=15mm, minimum width=10mm, anchor=center, line width={\stringwidth}, tikzit fill=neutral, inner sep={\innersep}]
\tikzstyle{large square box white}=[{\corners}, font={\boxsize}, fill=neutral, draw=black, shape=rectangle, tikzit category=string diagram, minimum width=15mm, minimum height=15mm, line width={\stringwidth}, tikzit fill=neutral, inner sep={\innersep}]
\tikzstyle{large square box comb}=[{\corners}, font={\boxsize}, fill=comb, draw=black, shape=rectangle, tikzit category=string diagram, minimum width=12mm, minimum height=12mm, line width={\stringwidth}, tikzit fill=comb, inner sep={\innersep}]
\tikzstyle{huge box}=[{\corners}, fill=neutral, draw=black, shape=rectangle, minimum height=40mm, minimum width=15mm, anchor=center, tikzit category=string diagram, line width={\stringwidth}, tikzit fill=neutral, inner sep={\innersep}]
\tikzstyle{output-node}=[fill=neutral, draw=black, shape=circle, anchor=west, inner sep=0.5]
\tikzstyle{delay}=[fill=neutral, draw=black, font={\boxsize}, shape=signal, tikzit category=string diagram, line width={\stringwidth}, minimum height={\tinywidth}, minimum width={\tinywidth}, inner sep=0.5mm, outer sep=0mm, minimum height=1em, minimum width=1em]
\tikzstyle{unit delay}=[fill=neutral, draw=unit, font={\boxsize}, shape=signal, tikzit category=string diagram, line width={\stringwidth}, minimum height={\tinywidth}, minimum width={\tinywidth}, inner sep=0.5mm, outer sep=0mm, minimum height=1em, minimum width=1em]
\tikzstyle{register}=[fill=seq, draw=black, font={\boxsize}, shape=signal, tikzit category=string diagram, line width={\stringwidth}, minimum width=1.5em, align=center, anchor=center, inner xsep=1mm, inner ysep=1mm, outer xsep=0mm]
\tikzstyle{waveform}=[fill=seq, draw=black, font={\boxsize}, shape=signal, signal from=west, signal to=east, tikzit category=string diagram, line width={\stringwidth}, minimum height=2em, minimum width=1.5em, align=center, anchor=center, inner xsep=1mm, inner ysep=-1mm, outer xsep=0mm]
\tikzstyle{bproduct}=[fill=black, draw=black, shape=circle, scale=0.5, tikzit category=string diagram]
\tikzstyle{red product}=[fill=palered, draw=palered, shape=circle, scale=0.5, tikzit category=string diagram]
\tikzstyle{wproduct}=[fill=neutral, draw=black, shape=circle, scale=0.5, line width=0.75, tikzit category=string diagram]
\tikzstyle{gproduct}=[fill=unit, draw=unit, shape=circle, scale=0.5, tikzit category=string diagram]
\tikzstyle{bport}=[minimum size=2.5mm, fill=none, draw=none, shape=circle, tikzit draw={rgb,255: red,154; green,154; blue,154}, anchor=center, tikzit fill=neutral, tikzit category=string diagram]
\tikzstyle{mux}=[fill=neutral, draw=black, shape=trapezium, tikzit category=circuits, rotate=-90, minimum height=1em, line width={\stringwidth}, scale=1.25]
\tikzstyle{fork}=[fill=black, draw=black, shape=circle, tikzit category=circuits, scale=0.25]
\tikzstyle{box}=[fill=neutral, draw=black, shape=rectangle, tikzit category=circuits]
\tikzstyle{dangling}=[fill=none, draw=none, shape=circle, anchor=east, scale=0.01, tikzit category=string diagram, tikzit fill={rgb,255: red,162; green,76; blue,77}]
\tikzstyle{label}=[fill=none, draw=none, shape=circle, align=center, inner sep=0, outer sep=0]
\tikzstyle{small label}=[scale=1.25, fill=none, draw=none, shape=circle, outer sep=0, inner sep=0, anchor=center]
\tikzstyle{wire label left}=[scale=1.25, fill=none, draw=none, shape=rectangle, outer sep=0, inner sep=0, anchor=east]
\tikzstyle{wire label right}=[scale=1.25, fill=none, draw=none, shape=rectangle, outer sep=0, inner sep=0, anchor=west]
\tikzstyle{wire label mid}=[scale=1.25, fill={\bgcolour}, shape=rectangle, outer sep=0, inner sep=0, minimum size=0]
\tikzstyle{tile}=[fill=neutral, draw=black, shape=rectangle, tikzit category=string diagram, {\corners}, minimum height=5mm, minimum width=5mm]
\tikzstyle{commuting label}=[fill=none, draw=none, shape=circle, scale=0.5]
\tikzstyle{and}=[fill=neutral, draw=black, shape=and gate, line width={\stringwidth}, and gate, scale=1.75, tikzit category=circuits]
\tikzstyle{or}=[fill=neutral, draw=black, or gate, scale=1.75, line width={\stringwidth}, tikzit category=circuits]
\tikzstyle{red or}=[fill=neutral, draw=palered, or gate, scale=1.75, line width={\stringwidth}, tikzit category=circuits]
\tikzstyle{not}=[fill=neutral, draw=black, not gate, scale=1.5, line width={\stringwidth}, tikzit category=circuits]
\tikzstyle{nor}=[fill=neutral, draw=black, nor gate, scale=1.75, line width={\stringwidth}, tikzit category=circuits]
\tikzstyle{nand}=[fill=neutral, draw=black, nand gate, scale=1.75, line width={\stringwidth}, tikzit category=circuits]
\tikzstyle{xor}=[fill=neutral, draw=black, xor gate, scale=1.75, line width={\stringwidth}, tikzit category=circuits]
\tikzstyle{xnor}=[fill=neutral, draw=black, xnor gate, scale=1.75, line width={\stringwidth}, tikzit category=circuits]
\tikzstyle{west}=[fill=none, draw=none, shape=circle, anchor=west]
\tikzstyle{short bundler}=[fill=neutral, draw=black, shape=rounded rectangle, minimum width=2.5em, rounded rectangle arc length=180, rotate=90, line width={\stringwidth}]
\tikzstyle{bundler}=[fill=neutral, draw=black, shape=rounded rectangle, minimum width=3em, rounded rectangle arc length=180, rotate=90, line width={\stringwidth}]
\tikzstyle{long bundler}=[fill=neutral, draw=black, shape=rounded rectangle, minimum width=5em, rounded rectangle arc length=180, rotate=90, line width={\stringwidth}]
\tikzstyle{box vertex}=[fill=white, draw=black, shape=circle, line width={\stringwidth}]

\tikzstyle{interface}=[-, fill={rgb,255: red,238; green,238; blue,255}, dashed, draw={rgb,255: red,170; green,170; blue,225}, ultra thick]
\tikzstyle{graph}=[-, fill={rgb,255: red,238; green,238; blue,238}, draw={rgb,255: red,191; green,191; blue,191}, dashed, ultra thick, anchor=center]
\tikzstyle{tentacle}=[-, very thick]
\tikzstyle{red tentacle}=[-, draw=palered, very thick]
\tikzstyle{wire}=[-, tikzit category=string diagram, line width={\stringwidth}]
\tikzstyle{red wire}=[-, draw=palered, tikzit category=string diagram, line width={\stringwidth}]
\tikzstyle{empty}=[-, densely dashed, {\corners}, dash pattern=on 0pt off 1.25pt on 1.25pt, line width={\stringwidth}]
\tikzstyle{transition}=[->]
\tikzstyle{output}=[->, decorate, decoration=zigzag]
\tikzstyle{gate}=[-, fill=neutral]
\tikzstyle{thicc}=[-, line width=1]
\tikzstyle{strikethrough}=[-, decoration={markings, mark=at position 0.5 with {
                \draw [blue, thick, - ] (-0.15,-0.15);
            }}, postaction=decorate]
\tikzstyle{traced}=[-, densely dashed, draw=gray]
\tikzstyle{arrow}=[<-, line width={\stringwidth}]
\tikzstyle{arrow up}=[->, line width={\stringwidth}]
\tikzstyle{dashed arrow}=[<-, dashed, line width={\stringwidth}]
\tikzstyle{unit wire}=[-, fill=none, draw=unit, line width={\stringwidth}]
\tikzstyle{interfacearrow}=[->, very thick, dashed]
\tikzstyle{tile none}=[-, draw=none, {\corners}, fill=none, line width={\stringwidth}]
\tikzstyle{tile white}=[-, {\corners}, fill=neutral, line width={\stringwidth}]
\tikzstyle{tile comb}=[-, {\corners}, fill=comb, line width={\stringwidth}]
\tikzstyle{tile seq}=[-, {\corners}, fill=seq, line width={\stringwidth}]
\tikzstyle{commute}=[->]
\tikzstyle{curved rectangle}=[-, rounded corners]
\tikzstyle{hasse}=[-]
\tikzstyle{wiredash}=[-, line width={\stringwidth}]
\tikzstyle{rewrite}=[-, dashed, line width={\stringwidth}]
\tikzstyle{juxtaposition}=[-, draw={rgb,255: red,128; green,128; blue,128}, densely dashed, line width={\stringwidth}]
\tikzstyle{functor box}=[-, thick, draw={rgb,255: red,83; green,83; blue,83}]
\tikzstyle{boundary box}=[-, draw=none, tikzit draw={rgb,255: red,255; green,0; blue,4}]
\tikzstyle{hom image}=[-, draw={rgb,255: red,191; green,191; blue,191}, fill={rgb,255: red,238; green,238; blue,238}, ultra thick, dashed]
\tikzstyle{tile boundary}=[-, draw={rgb,255: red,128; green,128; blue,128}, dashed]
\tikzstyle{lafont fork}=[-, fill=black]

\newcommand{\bgcolour}{white}

\begin{document}

\title[A Complete Theory of Digital Circuits]{%
    A Complete Theory of Digital Circuits: %
    Denotational, Operational, and Algebraic Semantics}%

\author[D.~R.~Ghica]{Dan R.\ Ghica\lmcsorcid{0000-0002-4003-8893}}[a]
\author[G.~Kaye]{George Kaye\lmcsorcid{0000-0002-0515-4055}}[a]
\author[D.~Sprunger]{David Sprunger\lmcsorcid{0000-0002-2518-4710}}[b]

\address{University of Birmingham}
\email{d.r.ghica@bham.ac.uk, georgejkaye@gmail.com}

\address{Indiana State University}
\email{david.sprunger@indstate.edu}

\begin{abstract}
    Digital circuits, despite having been studied for nearly a century and used
    at scale for about half that time, have until recently evaded a fully
    compositional theoretical understanding, in which arbitrary circuits may be
    freely composed together without consulting their internals.
    Recent work remedied this theoretical shortcoming by showing how digital
    circuits can be presented compositionally as morphisms in a freely generated
    symmetric traced category.
    However, this was done informally; in this paper we refine and expand the
    previous work in several ways, culminating in the presentation of three
    sound and complete semantics for digital circuits: denotational, operational
    and algebraic.
    For the denotational semantics, we establish a correspondence between
    stream functions with certain properties and circuits constructed
    syntactically.
    For the operational semantics, we present the reductions required to model
    how a circuit processes a value, including the addition of a new reduction
    for eliminating non-delay-guarded feedback; this leads to an adequate notion
    of observational equivalence for digital circuits.
    Finally, we define a new family of equations for translating circuits into
    bisimilar circuits of a `normal form', leading to a complete algebraic
    semantics for sequential circuits.
\end{abstract}

\maketitle

\section{Introduction}

Bothe was awarded the 1954 Nobel Prize in physics for creating the electronic
\(\andgate\) gate in 1924.
In the ensuing decades, exponential improvements in digital technology have led
to the development of the defining technologies of the modern world.
It may therefore seem improbable that there are theoretical gaps remaining in
our mathematical and logical understanding of digital circuits.

To be more precise, by `digital circuits' we primarily understand
\emph{electronic} circuits: deterministic circuits with clear notions of
input and output and which work on discrete signals.
These circuits are \emph{sequential} as outputs can be impacted by previous
inputs over time, and \emph{synchronous} because their state changes in time
with some global clock.
A classic example is that of logical gates and basic memory elements of known
and fixed delays, but there are alternatives such as CMOS transistors operating
in saturation mode.

What we do not attempt to handle are circuits operating on continuous signals
(such as amplifiers) or in continuous time (such as asynchronous circuits), nor
\emph{electrical} circuits of resistors and capacitors, which are quite
different~\cite{boisseau2022string}.
The difference between electrical circuits and electronic circuits boils down to
the difference between traced categories and compact closed categories: digital
circuits have a clear notion of \emph{causality} whereas electrical circuits are
\emph{relational} in nature.

Our goal is to devise a \emph{fully compositional} model of digital circuits.
By `fully compositional' we mean that a larger circuit can be built from smaller
circuits and interconnecting wires without paying heed to the internal structure
of these smaller circuits.
Of course, composition comes naturally to digital circuits and is widely used
informally~\cite{gordon1982model}.
Unfortunately one runs into an obstacle when trying to formalise this notion
mathematically: electrical connections can be created that inadvertently
connect the output of some elementary gate back to
its input such that no memory elements are encountered along the path.
Such a path, called a `combinational feedback loop' (or `cycle'), is not handled
by established mathematical theories of digital circuits, so conventional
digital design and engineering reject such circuits.
To enforce this restriction, we need to always look inside circuits as we
compose them, ensuring that no illegal combinational feedback loops are created,
or resort to some `safe' subset of circuits~\cite{christensen2021wire}.
This represents a failure of compositionality: what we want to do is to compose
\emph{any} circuits constructed from a fixed set of components.

On general principle, we have reason to expect that a compositional theory of
digital circuits may lead to more streamlined methods of analysis and
verification, which, in time, may also lead to new applications.
\emph{Combinational} circuits, which model functions, have an obvious
compositional syntax~\cite{lafont2003algebraic}, but \emph{sequential} circuits,
which contain delay and feedback, are more subtle.
The first forays towards a fully compositional syntactic and
categorical account of circuits have been made
recently~\cite{ghica2016categorical,ghica2017diagrammatic}, but they do not
paint a fully formal and coherent picture.
This paper develops the informal presentation into a mathematically
rigorous framework.

Our first contribution is to give, for the first time, a sound and complete
denotational semantics to digital circuits in the domain of causal and monotone
stream functions.
The completeness result depends on a novel albeit straightforward lifting of
Mealy machines~\cite{mealy1955method} to act on alphabets with a lattice
structure, utilising a handy coalgebraic perspective~\cite{rutten2006algebraic}.
Using Mealy machines to give a semantic interpretation to digital circuits is an
established methodology~\cite{kohavi2009switching}, and here they act as a
`bridge' between the syntactic and the semantic domain, showing how existing
circuit methodologies are compatible with our rigorous mathematical framework.

The second contribution of the paper is to generalise and systematise previous
efforts~\cite{ghica2017diagrammatic} to formulate a graph-rewriting-based
operational semantics for digital circuits.
The novelty is a new reduction rule for eliminating non-delay-guarded feedback
using a version of the Kleene fixpoint theorem, thus solving the problem of
productivity that previous operational semantics only solve partially.
The denotational and operational semantics together achieve the long-standing
goal of creating a semantic theory of digital circuits using the same
methodology as programming languages.

The methodological `glue' that binds together the two approaches is a new sound
and complete algebraic semantics, the third and final contribution of the paper.
This approach replaces the previous ad-hoc way of introducing equations for
digital circuits based on raw intuitions with a systematic approach guided by
the denotational semantics.
The key technical result of this method is deriving pseudo-normal forms of
digital circuits.

Although the motivation of this work is foundational, there are some early
hints of possible exciting applications, such as for partial evaluation and
blackboxing.
These are not yet industrial-strength applications, but the simplicity and
power of the framework must hold a certain degree of appeal and promise.

\section{Syntax}

\subsection{Circuit signatures}

We model circuits as morphisms in a symmetric monoidal (traced) category; a
morphism \(m \to n\) is a circuit with \(m\) inputs and \(n\) outputs.
Rather than restricting to any particular gate set, we parameterise a given
category of circuits over a \emph{circuit signature} containing details about
the available components.

\begin{defi}[Circuit signature, value, primitive symbol]
    A \emph{circuit signature} \(\circuitsignature\) is a tuple \((
    \values,
    \disconnected,
    \circuitsignaturegates,
    \circuitsignaturearity,
    \circuitsignaturecoarity
    )\) where \(\values\) is a finite set of \emph{values}, \(
    \disconnected \in \values
    \) is a \emph{disconnected} value, \(\circuitsignaturegates\) is a (usually
    finite) set of \emph{primitive symbols}, \(
    \morph{\circuitsignaturearity}{\circuitsignaturegates}{\nat}
    \) is an \emph{arity} function and \(
    \morph{\circuitsignaturecoarity}{\circuitsignaturegates}{\nat}
    \) is a \emph{coarity} function.
\end{defi}

A particularly important signature, and one which we will turn to for the
majority of examples in this thesis, is that of gate-level circuits.

\begin{exa}[Gate-level circuits]\label{ex:belnap-signature}
    The circuit signature for \emph{gate-level circuits} is \(
    \belnapsignature \coloneqq (
    \belnapvalues,
    \bot,
    \belnapgates,
    \belnaparity,
    \belnapcoarity
    )\), where \(
    \belnapvalues \coloneqq \{\bot, \belnapfalse, \belnaptrue, \top\}
    \), respectively representing \emph{no} signal, a \emph{false} signal, a
    \emph{true} signal and \emph{both} signals at once, \(
    \belnapgates \coloneqq \{\andgate,\orgate,\notgate\}
    \), \(
    \belnaparity \coloneqq
    \andgate \mapsto 2,
    \orgate \mapsto 2,
    \notgate \mapsto 1
    \) and \(
    \belnapcoarity \coloneqq - \mapsto 1,
    \)
\end{exa}

\begin{rem}
    Using four values may come as a surprise to those expecting the usual
    `true' and `false' logical values.
    This logical system is an old idea going back to
    Belnap~\cite{belnap1977useful} who remarked that this is `how a computer
    should think'.
    Rather than just thinking about how much \emph{truth content} a value
    carries, the four value system adds a notion of \emph{information content}:
    the \(\bot\) value is no information at all (a disconnected wire), whereas
    the \(\top\) value is `inconsistent' information, or both true and false
    information at once.
\end{rem}

\subsection{Combinational circuits}

A circuit signature freely generates symmetric monoidal categories (SMCs) of
digital circuits.
These are PROPS (categories of \emph{PRO}ducts and
\emph{P}ermutations)~\cite{maclane1965categorical}, SMCs with objects the
natural numbers and tensor product on objects as addition.

Instead of term strings, we employ the two dimensional syntax of
\emph{string diagrams}~\cite{joyal1991geometry,joyal1996traced,selinger2011survey}.
We draw an arbitrary generator \(\morph{\varphi}{m}{n}\) as a box \(
\adjustimage{valign=c,margin=0pt,page=1}{tikzfigures}
\) with \(m\) input wires and \(n\) output wires.
To avoid clutter and to enable reasoning about diagrams with arbitrary
inputs and outputs, wires may be collapsed into one wire and labelled
appropriately \(
\adjustimage{valign=c,margin=0pt,page=2}{tikzfigures}
\).
Composite morphisms, or `terms', are drawn as wider boxes with rounded corners
\(
\adjustimage{valign=c,margin=0pt,page=3}{tikzfigures}
\); composition as horizontal juxtaposition \(
\adjustimage{valign=c,margin=0pt,page=4}{tikzfigures}
\) and tensor product as vertical juxtaposition \(
\adjustimage{valign=c,margin=0pt,page=5}{tikzfigures}
\).
One of the advantages of this notation over standard term syntax is that
structural rules (identity, associativity, functoriality) are `absorbed' into
the diagrams, as illustrated in \cref{fig:smc-axioms}.

\begin{figure*}
    \centering
    \(
    \adjustimage{valign=c,margin=0pt,page=6}{tikzfigures}
    =
    \adjustimage{valign=c,margin=0pt,page=3}{tikzfigures}
    \quad
    \adjustimage{valign=c,margin=0pt,page=7}{tikzfigures}
    =
    \adjustimage{valign=c,margin=0pt,page=3}{tikzfigures}
    \)
    \\[1.5em]
    \(
    \adjustimage{valign=c,margin=0pt,page=8}{tikzfigures}
    =
    \adjustimage{valign=c,margin=0pt,page=9}{tikzfigures}
    \)
    \\[1.5em]
    \(
    \adjustimage{valign=c,margin=0pt,page=10}{tikzfigures}
    =
    \adjustimage{valign=c,margin=0pt,page=3}{tikzfigures}
    \quad
    \adjustimage{valign=c,margin=0pt,page=11}{tikzfigures}
    =
    \adjustimage{valign=c,margin=0pt,page=3}{tikzfigures}
    \)
    \\[1.5em]
    \(
    \adjustimage{valign=c,margin=0pt,page=12}{tikzfigures}
    =
    \adjustimage{valign=c,margin=0pt,page=13}{tikzfigures}
    \quad
    \adjustimage{valign=c,margin=0pt,page=14}{tikzfigures}
    =
    \adjustimage{valign=c,margin=0pt,page=15}{tikzfigures}
    \)
    \\[1.5em]
    \(
    \adjustimage{valign=c,margin=0pt,page=16}{tikzfigures}
    =
    \adjustimage{valign=c,margin=0pt,page=17}{tikzfigures}
    \)
    \\[1.5em]
    \(
    \adjustimage{valign=c,margin=0pt,page=18}{tikzfigures}
    =
    \adjustimage{valign=c,margin=0pt,page=19}{tikzfigures}
    \quad
    \adjustimage{valign=c,margin=0pt,page=20}{tikzfigures}
    =
    \adjustimage{valign=c,margin=0pt,page=21}{tikzfigures}
    \)
    \\[1.5em]
    \(
    \adjustimage{valign=c,margin=0pt,page=22}{tikzfigures}
    =
    \adjustimage{valign=c,margin=0pt,page=23}{tikzfigures}
    \quad
    \adjustimage{valign=c,margin=0pt,page=24}{tikzfigures}
    =
    \adjustimage{valign=c,margin=0pt,page=25}{tikzfigures}
    \quad
    \adjustimage{valign=c,margin=0pt,page=26}{tikzfigures}
    =
    \adjustimage{valign=c,margin=0pt,page=27}{tikzfigures}
    \)
    \vspace{1em}
    \caption{
        The equations of PROPS in string diagram notation
    }
    \label{fig:smc-axioms}
\end{figure*}

We will first define a category of \emph{combinational circuits} with no state;
these circuits compute \emph{functions} of their inputs.

\begin{defi}[Combinational generators]
    For a circuit signature \(
    \circuitsignature = (
    \circuitsignaturevalues,
    \bullet,
    \circuitsignaturegates,
    \circuitsignaturearity,
    \circuitsignaturecoarity
    )
    \), let the set \(\generators[\ccirc{}]\) of
    \emph{combinational circuit generators} be defined as the set containing \(
    \adjustimage{valign=c,margin=0pt,page=28}{tikzfigures}
    \) for each \(g \in \circuitsignaturegates\),
    \(\adjustimage{valign=c,margin=0pt,page=29}{tikzfigures}\),
    \(\adjustimage{valign=c,margin=0pt,page=30}{tikzfigures}\),
    \(\adjustimage{valign=c,margin=0pt,page=31}{tikzfigures}\), and
    \(\adjustimage{valign=c,margin=0pt,page=32}{tikzfigures}\).
    We write \(\ccircsigma\) for the freely generated PROP
    \(\smc{\generators[\ccirc{}]}\).
\end{defi}

Each primitive symbol \(p \in \circuitsignaturegates\) has a corresponding
generator in \(\ccircsigma\).
The remaining generators are \emph{structural} generators
for manipulating
wires: these are present regardless of the signature.
In order, they are for \emph{introducing} wires, \emph{forking}
wires, \emph{joining} wires and \emph{eliminating} wires.

\begin{exa}
    The gate generators of \(\ccirc{\belnapsignature}\) are \(
    \adjustimage{valign=c,margin=0pt,page=33}{tikzfigures}
    \), \(
    \adjustimage{valign=c,margin=0pt,page=34}{tikzfigures}
    \), and \(
    \adjustimage{valign=c,margin=0pt,page=35}{tikzfigures}
    \).
\end{exa}

When drawing circuits, the coloured backgrounds of generators will often be
omitted in the interests of clarity.
Since the category is freely generated, morphisms are defined by
juxtaposing the generators in a given signature sequentially or in parallel with
each other, the identity and the symmetry.
Arbitrary combinational circuit morphisms defined in this way are drawn as light
blue boxes \adjustimage{valign=c,margin=0pt,page=36}{tikzfigures}.

\begin{nota}\label{not:arbitrary-width-structure}
    The structural generators are defined on single bits, but it is
    straightforward to define versions for arbitrary bit wires.
    For \(0\)-bit wires they are drawn as empty space or as
    `faded' wires, and `thicker' constructs have their wires annotated with
    their widths.
    \begin{gather*}
        \adjustimage{valign=c,margin=0pt,page=37}{tikzfigures}
        \quad
        \adjustimage{valign=c,margin=0pt,page=38}{tikzfigures}
        \quad
        \adjustimage{valign=c,margin=0pt,page=39}{tikzfigures}
        \quad
        \adjustimage{valign=c,margin=0pt,page=40}{tikzfigures}
    \end{gather*}
    These composite constructs are defined inductively over the width of the
    wires.

    \begin{center}
        \begin{minipage}{0.48\textwidth}
            \centering
            \(\adjustimage{valign=c,margin=0pt,page=41}{tikzfigures}
            \coloneqq
            \adjustimage{valign=c,margin=0pt,page=42}{tikzfigures}
            \)

            \vspace{1em}

            \(
            \adjustimage{valign=c,margin=0pt,page=43}{tikzfigures}
            \coloneqq
            \adjustimage{valign=c,margin=0pt,page=44}{tikzfigures}
            \)

            \vspace{1em}

            \(
            \adjustimage{valign=c,margin=0pt,page=45}{tikzfigures}
            \coloneqq
            \adjustimage{valign=c,margin=0pt,page=42}{tikzfigures}
            \)\quad\(
            \adjustimage{valign=c,margin=0pt,page=46}{tikzfigures}
            \coloneqq
            \adjustimage{valign=c,margin=0pt,page=47}{tikzfigures}
            \)
        \end{minipage}
        \quad
        \begin{minipage}{0.48\textwidth}
            \centering
            \(\adjustimage{valign=c,margin=0pt,page=48}{tikzfigures}
            \coloneqq
            \adjustimage{valign=c,margin=0pt,page=42}{tikzfigures}
            \)

            \vspace{1em}

            \(
            \adjustimage{valign=c,margin=0pt,page=49}{tikzfigures}
            \coloneqq
            \adjustimage{valign=c,margin=0pt,page=50}{tikzfigures}
            \)

            \vspace{1em}

            \(\adjustimage{valign=c,margin=0pt,page=51}{tikzfigures}
            \coloneqq
            \adjustimage{valign=c,margin=0pt,page=42}{tikzfigures}
            \)\quad\(
            \adjustimage{valign=c,margin=0pt,page=52}{tikzfigures}
            \coloneqq
            \adjustimage{valign=c,margin=0pt,page=53}{tikzfigures}
            \)
        \end{minipage}
    \end{center}
\end{nota}

\begin{exa}[More logic gates]
    The \(\andgate\), \(\orgate\), and \(\notgate\) gates are not the only logic
    gates used in circuit design.
    A \(\nandgate\) gate \(
    \adjustimage{valign=c,margin=0pt,page=54}{tikzfigures}
    \) acts as the inverse of an \(\andgate\) gate: it
    outputs true if none of the inputs are true.
    Similarly, a \(\norgate\) gate \(
    \adjustimage{valign=c,margin=0pt,page=55}{tikzfigures}
    \) is the inverse of an \(\orgate\) gate.
    These two gates can be constructed in terms of the primitive gates in
    \(\belnapsignature\):

    \[
        \adjustimage{valign=c,margin=0pt,page=54}{tikzfigures}
        \coloneqq
        \adjustimage{valign=c,margin=0pt,page=56}{tikzfigures}
        \qquad
        \adjustimage{valign=c,margin=0pt,page=55}{tikzfigures}
        \coloneqq
        \adjustimage{valign=c,margin=0pt,page=57}{tikzfigures}
    \]

    Another type of gate is the \(\xorgate\) gate \(
    \adjustimage{valign=c,margin=0pt,page=58}{tikzfigures}
    \), which outputs true if and only if exactly one of the inputs is
    true.
    In \(\ccirc{\belnapsignature}\) this is constructed as
    \[
        \adjustimage{valign=c,margin=0pt,page=58}{tikzfigures}
        \coloneqq
        \adjustimage{valign=c,margin=0pt,page=59}{tikzfigures}
        =
        \adjustimage{valign=c,margin=0pt,page=60}{tikzfigures}.
    \]
\end{exa}

\begin{exa}[Half adder]\label{ex:half-adder}
    The \(\xorgate\) gate is used in a classic combinational circuit known as a
    \emph{half adder}, the basic building block of circuit arithmetic.
    A half adder takes two inputs and computes their \emph{sum} modulo
    \(2\) and the resulting \emph{carry}.
    That is to say, \(0+0\) has sum \(0\) and carry \(0\), \(1+0\) and \(0+1\)
    have sum \(1\) and carry \(0\), and \(1+1\) has sum \(0\) and carry \(1\).

    The sum is computed using an \(\xorgate\) gate and the carry by an
    \(\andgate\) gate.
    The design of a half adder along with its construction in
    \(\ccirc{\belnapsignature}\) is shown below.
    \[
        \adjustimage{valign=c,margin=0pt,page=61}{tikzfigures}
        \qquad
        \adjustimage{valign=c,margin=0pt,page=62}{tikzfigures}
        =
        \adjustimage{valign=c,margin=0pt,page=63}{tikzfigures}
    \]
\end{exa}

\subsection{Sequential circuits}

Combinational circuits compute functions of their inputs, but have no internal
state.
This is all very well for doing simple calculations, but for all but the most
simple of circuits we need to be able to have \emph{memory}.
As we have mentioned earlier, such circuits are called
\emph{sequential circuits}.

Circuits gain state with \emph{delay} and \emph{feedback}.
The latter means we need to move into a symmetric \emph{traced} monoidal
category.
In string diagram notation, the trace is represented as a loop from outputs to
inputs; much like with the symmetric monoidal setting, this gives the equations
of STMCs intuitive graphical interpretations, illustrated in
\autoref{fig:stmc-axioms}.

\begin{figure*}
    \def\arraystretch{2}
    \begin{tabular}{lccc}
        \textbf{Tightening}
         &
        \(
        \adjustimage{valign=c,margin=0pt,page=64}{tikzfigures}
        \)
         &
        \(
        =
        \)
         &
        \(
        \adjustimage{valign=c,margin=0pt,page=65}{tikzfigures}
        \)
        \\
        \textbf{Sliding}
         &
        \(
        \adjustimage{valign=c,margin=0pt,page=66}{tikzfigures}
        \)
         &
        \(=\)
         &
        \(
        \adjustimage{valign=c,margin=0pt,page=67}{tikzfigures}
        \)
        \\
        \textbf{Vanishing}
         &
        \(
        \adjustimage{valign=c,margin=0pt,page=68}{tikzfigures}
        \)
         &
        \(
        =
        \)
         &
        \(
        \adjustimage{valign=c,margin=0pt,page=69}{tikzfigures}
        \)
        \\
        \textbf{Superposing}
         &
        \(
        \adjustimage{valign=c,margin=0pt,page=70}{tikzfigures}
        \)
         &
        \(
        =
        \)
         &
        \(
        \adjustimage{valign=c,margin=0pt,page=71}{tikzfigures}
        \)
        \\
        \textbf{Yanking}
         &
        \(
        \adjustimage{valign=c,margin=0pt,page=72}{tikzfigures}
        \)
         &
        \(
        =
        \)
         &
        \(
        \adjustimage{valign=c,margin=0pt,page=73}{tikzfigures}
        \)
    \end{tabular}
    \def\arraystretch{1}
    \caption{The trace equations in string diagram notation}
    \label{fig:stmc-axioms}
\end{figure*}

\begin{defi}[Sequential circuits]
    For a circuit signature \(\circuitsignature\) with value set \(\values\),
    let \(\scircsigma\) be the STMC freely generated over the generators of
    \(\ccircsigma\) along with new generators \(
    \adjustimage{valign=c,margin=0pt,page=74}{tikzfigures}
    \) for each \(v \in \values \setminus \bullet\), and a generator \(
    \adjustimage{valign=c,margin=0pt,page=75}{tikzfigures}
    \).
\end{defi}

The first set of generators are \emph{instantaneous values} for each value in
\(\values \setminus \bullet\).
Value generators are intended to be interpreted as an \emph{initial state}:
in the first cycle of execution they will emit their value, and produce the
disconnected \(\bullet\) value after that.
This is why there is no \(\bullet\) value generator; instead it is a
\emph{combinational} generator \(
\adjustimage{valign=c,margin=0pt,page=29}{tikzfigures}
\) intended to always produce the \(\bullet\) value.

\begin{nota}
    Although \(
    \adjustimage{valign=c,margin=0pt,page=29}{tikzfigures}
    \) is itself not a sequential value, when we refer to an arbitrary value
    \(
    \adjustimage{valign=c,margin=0pt,page=74}{tikzfigures}
    \), \(v\) can be any value in \(\values\) including \(\bullet\).
    For a word of values \(\listvar{v} \in \valuetuple{n}\) (again possibly
    including \(\bullet\)), we may draw multiple value generators collapsed into
    one as \(
    \adjustimage{valign=c,margin=0pt,page=76}{tikzfigures}
    \), defined inductively over \(\listvar{v}\) as
    \begin{gather*}
        \adjustimage{valign=c,margin=0pt,page=77}{tikzfigures}
        \coloneqq
        \adjustimage{valign=c,margin=0pt,page=42}{tikzfigures}
        \qquad
        \adjustimage{valign=c,margin=0pt,page=78}{tikzfigures}
        \coloneqq
        \adjustimage{valign=c,margin=0pt,page=79}{tikzfigures}
    \end{gather*}
\end{nota}

\begin{exa}
    The `values' of \(\scirc{\belnapsignature}\) are \(
    \adjustimage{valign=c,margin=0pt,page=29}{tikzfigures}
    \), \(
    \adjustimage{valign=c,margin=0pt,page=80}{tikzfigures}
    \), \(
    \adjustimage{valign=c,margin=0pt,page=81}{tikzfigures}
    \), \(
    \adjustimage{valign=c,margin=0pt,page=82}{tikzfigures}
    \); the first is a combinational generator and the latter three are
    sequential.
\end{exa}

In the first cycle of execution the delay is intended to produce the \(\bullet\)
value, but in future cycles it outputs the signal it received in the previous
cycle.

\begin{rem}
    While the mathematical interpretation of a delay is straightforward, the
    physical aspect of a digital circuit it models is less clear.
    An obvious interpretation could be that the delay models a D flipflop in
    a clocked circuit, so the delay is one clock cycle.
    A more subtle interpretation is the `minimum observable duration'; in this
    case the delay models inertial delay on wires up to some fixed precision.
\end{rem}

\begin{nota}
    We define delay components for arbitrary-bit wires
    \(
    \adjustimage{valign=c,margin=0pt,page=83}{tikzfigures}
    \) as follows:
    \begin{gather*}
        \adjustimage{valign=c,margin=0pt,page=84}{tikzfigures}
        \coloneqq
        \adjustimage{valign=c,margin=0pt,page=42}{tikzfigures}
        \qquad
        \adjustimage{valign=c,margin=0pt,page=85}{tikzfigures}
        \coloneqq
        \adjustimage{valign=c,margin=0pt,page=86}{tikzfigures}
    \end{gather*}
\end{nota}

Often one may also want to think about delays with some explicit `initial
value', like a sort of register.
This is so common that we introduce special notation for it.

\begin{nota}[Register]\label{not:register}
    For a word \(\listvar{v} \in \valuetuple{m}\), let \(
    \adjustimage{valign=c,margin=0pt,page=87}{tikzfigures}
    \coloneqq
    \adjustimage{valign=c,margin=0pt,page=88}{tikzfigures}
    \).
\end{nota}

To distinguish them from combinational circuits, arbitrary sequential circuit
morphisms are drawn as green boxes \(
\adjustimage{valign=c,margin=0pt,page=89}{tikzfigures}
\).

\begin{exa}[SR latch]\label{ex:sr-latch}
    A sequential circuit one might come across early on in an electronics
    textbook is the \emph{SR NOR latch}, one of the simplest registers.
    A possible design in usual circuit diagram notation is shown below on the
    left, with an interpretation in \(\scirc{\belnapsignature}\) using string
    diagram notation to the right.
    \begin{gather*}
        \adjustimage{valign=c,margin=0pt,page=90}{tikzfigures}
        \qquad
        \adjustimage{valign=c,margin=0pt,page=91}{tikzfigures}
    \end{gather*}

    SR NOR latches are used to hold state.
    They have two inputs: Reset (\(\mathsf{R}\)) and Set (\(\mathsf{S}\)), and
    two outputs \(\mathsf{Q}\) and \(\overline{\mathsf{Q}}\) which are always
    negations of each other.
    When Set receives a true signal, the \(\mathsf{Q}\) output is forced true,
    and will remain as such even if the Set input stops being pulsed true.
    It is only when the Reset input is pulsed true that the \(\mathsf{Q}\)
    output will return to false.
    (It is illegal for both Set and Reset to be pulsed high simultaneously; this
    issue is fixed in more complicated latches).

    SR latches work because of delays in how gates and wires transmit signals;
    one of the feedback loops between the two \(\norgate\) gates will `win'.
    We can model this in \(\scircsigma\) by using a different number of delay
    generators between the top and the bottom of the latch.
    We have opted for just the one because otherwise the upcoming examples
    become excessively complicated, but any number would do, so long as the top
    and bottom differ.
\end{exa}

\subsection{Generalised circuit signatures}

In a circuit signature, gates are assigned a number of input and output wires.
This serves us well when we want to model lower level circuits in which we
really are dealing with single-bit wires.
However, when designing circuits it is often advantageous to work at a higher
level of abstraction with `thicker' wires carrying words of information.
For example, the values in the circuits could be used to represent binary
numbers.

This can still be modelled in \(\scircsigma\) `as is' by using lots of parallel
wires to connect to the various primitives, but this can get messy with wires
all over the place.
Instead, we will introduce a generalisation of circuit signatures in which these
thicker buses of wires are treated as first-class objects.

\begin{defi}[Generalised circuit signature]
    A \emph{generalised circuit signature} \(\circuitsignature\) is a tuple \((
    \values,
    \disconnected,
    \circuitsignaturegates,
    \circuitsignaturearity,
    \circuitsignaturecoarity
    )\) where \(\values\) is a finite set of values, \(
    \disconnected \in \values
    \) is a \emph{disconnected} value, \(\circuitsignaturegates\) is a (usually
    finite) set of \emph{gate symbols}, \(
    \morph{\circuitsignaturearity}{\circuitsignaturegates}{\natplus^\star}
    \) is an \emph{arity} function and \(
    \morph{\circuitsignaturecoarity}{\circuitsignaturegates}{\natplus^\star}
    \) is a \emph{coarity} function.
\end{defi}

In a generalised circuit signature, primitives are typed with input and output
\emph{words} rather than just natural numbers.

\begin{exa}
    The generalised circuit signature for \emph{simple arithmetic circuits} is
    \(
    \belnapsignature^+ \coloneqq \left(
    \belnapvalues,
    \bot,
    \belnapgates^+,
    \belnaparity^+,
    \belnapcoarity^+
    \right)
    \), where \begin{gather*}
        \belnapgates
        \coloneqq \{
        \andgate_{k,n},
        \orgate_{k,n},
        \notgate_{n},
        \mathsf{add}_n
        \,|\,
        n \in \natplus
        \}
        \\[0.5em]
        \belnaparity^+(\andgate_{k,n}) \coloneqq [n \,|\, i < k]
        \quad
        \belnaparity^+(\orgate_{k,n}) \coloneqq [n \,|\, i < k]
        \\[0.5em]
        \belnaparity^+(\notgate_{n}) \coloneqq [n]
        \quad
        \belnaparity^+(\mathsf{add}_n) \coloneqq [n,n]
        \\[0.5em]
        \belnapcoarity^+
        \coloneqq
        \andgate_{k,n} \mapsto [n],
        \orgate_{k,n} \mapsto [n],
        \notgate_n \mapsto [n],
        \mathsf{add}_n \mapsto [n]
    \end{gather*}
    The gates \(\andgate_{k,n}\) and \(\orgate_{k,n}\) are gates that operate
    on \(k\) input wires of width \(n\); similarly the \(\notgate_n\) gate
    operates on input wires of width \(n\).
    The \(\mathsf{add}_n\) component represents an adder that takes as input
    two \(n\)-bit wires and outputs their \(n\)-bit sum.
\end{exa}

When viewed graphically, a regular circuit signature generates a PROP in which
the string diagrams have wires of a single width, or `colour'.
A generalised circuit signature generates a PROP in which wires can be of any
width \(n \in \natplus\); such a category is known as a
\emph{\(\natplus\)-coloured PROP}.

\begin{defi}
    For a generalised circuit signature \(\circuitsignature\), let the set
    \(\generators[\ccirc{}^+]\) of
    \emph{generalised combinational circuit generators} be defined as the set
    containing
    \begin{gather*}
        \adjustimage{valign=c,margin=0pt,page=28}{tikzfigures}
        \,
        \text{for each}\ g \in \circuitsignaturegates
        \\[1em]
        \adjustimage{valign=c,margin=0pt,page=37}{tikzfigures}
        \quad
        \adjustimage{valign=c,margin=0pt,page=38}{tikzfigures}
        \quad
        \adjustimage{valign=c,margin=0pt,page=39}{tikzfigures}
        \quad
        \adjustimage{valign=c,margin=0pt,page=40}{tikzfigures}
        \quad
        \adjustimage{valign=c,margin=0pt,page=92}{tikzfigures}
        \quad
        \adjustimage{valign=c,margin=0pt,page=93}{tikzfigures}
        \quad
        \text{for each}\ n \in \natplus
    \end{gather*}
    We write \(\ccircsigmag\) for the freely generated \(\natplus\)-coloured
    PROP \(\smc{\generators[\ccirc{}^+]}\).
\end{defi}

Most of the generators  in \(\ccircsigmag\) are fairly straightforward
generalisations of the primitives in \(\ccircsigma\) to act on each
colour (width) of wires.
The only new generators are the \emph{bundlers} at the end of the bottom row;
their intended meaning is that they can be used to \emph{split} and
\emph{combine} bundles of wire buses into bundles with different widths.
These constructs were first proposed by Wilson et al.~\cite{wilson2023string}
as a notation for \emph{non-strict monoidal categories}.
We take inspiration from their observation that a similar idea could also be
applied to strict monoidal categories.

\begin{exa}[ALU]
    The computation of a CPU is performed by an \emph{arithmetic logic unit},
    or ALU for short.
    An ALU takes some input wires of a fixed width and performs an operation
    on them given some control signal.
    While ALUs can often perform many different operations, we will look at an
    example operating on four-bit wires that performs a bitwise \(\andgate\)
    operation when the control is false, and an addition when the control is
    true.
    This ALU will also produce an output indicating if the sum is zero, and
    the sign of the sum; these auxiliary outputs only produce useful output when
    the addition operation is selected.

    \begin{center}
        \adjustimage{valign=c,margin=0pt,page=94}{tikzfigures}
    \end{center}

    To apply the single-bit control to the four-bit \(\andgate\) gates, the
    top bundler and forks are used to create a wire containing only the
    original bit.

    The sum is zero if all of the bits are false.
    To determine this, the \(\orgate_{1,4}\) gate folds the four-bit sum into
    a one-bit value which is true if at least one of the bits is true.
    The \(\notgate_{1,1}\) inverts the output to produce true if there are no
    true bits.

    In two's complement representation, the most significant bit indicates if
    the sum is negative.
    To model this, the lower bundler splits the four-bit sum into its
    constituent bits, discarding the least significant three.
\end{exa}

Sequential circuits are generalised in the same way.

\begin{defi}
    For a generalised circuit signature \(\circuitsignature\), let the set
    \(\generators[\scirc{}^+]\) of
    \emph{generalised sequential circuit generators} be the set of
    generalised combinational circuit generators along with
    \(
    \adjustimage{valign=c,margin=0pt,page=76}{tikzfigures}
    \) for each \(n \in \natplus\) and \(\listvar{v} \in \valuetuple{n}\), and
    \(
    \adjustimage{valign=c,margin=0pt,page=83}{tikzfigures}
    \) for each \(n \in \natplus\).
    We write \(\scircsigmag\) for the freely generated PROP
    \(\stmc{\generators[\scirc{}^+]}\).
\end{defi}

Most of the upcoming results will be shown for the single-width case, as the
proofs are more elegant.
However, most of the results also generalise to the coloured case, and this will
be remarked upon throughout.

\section{Denotational semantics}

Circuits in \(\scircsigma\) are purely \emph{syntax}; the only equational
properties they satisfy are the `structural equations' of traced PROPs: moving
boxes around while preserving connectivity.
In this section we will present a fully compositional denotational
semantics for sequential circuits based on \emph{causal}, \emph{monotone} and
\emph{finitely specified} stream functions.
This denotational semantics is defined using a functor from \(\scircsigma\) to a
PROP of stream functions with the desired properties; a reverse functor is then
defined which maps a stream function \(f\) to a circuit in \(\scircsigma\) with
\(f\) as its denotation, showing that this denotational semantics is
\emph{sound and complete}.

We will interpret digital circuits as a certain class of
\emph{stream functions}, functions that operate on infinite sequences of values.
This represents how the output of a circuit may not just operate on the current
input, but all of the previous ones as well.

\begin{rem}
    In \cite{mendler2012constructive}, the semantics of digital circuits with
    delays and cycles are presented using \emph{timed ternary simulation}, an
    algorithm to compute how a sequence of circuit outputs stabilises over time
    given the inputs and value of the current state.
    This differs from our approach as we assign each circuit a concrete stream
    function describing its behaviour, rather determining its behaviour by
    solving a system of equations in terms of its gates.
\end{rem}

Recall that a function \(f\) between two posets is \emph{monotone}
if \(x \leq y \Rightarrow f(x) \leq f(y)\), and a \emph{lattice} is a poset
in which each pair of elements has a least upper bound \(\vee\) (a \emph{join})
and a greatest lower bound \(\wedge\) (a \emph{meet}); subsequently
every \emph{finite} lattice has an \emph{greatest element} \(\bot\) and a
\emph{least element} \(\top\).
We write \(v^n\) for the \(n\)-tuple containing only \(v\), and call a function
\(\morph{f}{\valuetuple{m}}{\valuetuple{n}}\) \emph{\(\bot\)-preserving} if
\(f(\bot^m) = \bot^n\).

\subsection{Interpreting circuit components}\label{sec:interpreting-components}

Before assigning stream functions to a given circuit in \(\scircsigma\), we will
first decide how to interpret the individual \emph{components} of a given
circuit signature.
First we consider the interpretation of the \emph{values} that flow through the
wires in our circuits.
In the denotational semantics the set of values will need to have a bit more
structure, as it must be ordered by how much \emph{information} each value
carries.
In our context of digital circuits the least and greatest elements respectively
represent signals with a complete \emph{lack} of information and \emph{every}
piece of information at once; for this reason, we model the values as elements
of a \emph{lattice}.

\begin{defi}[Interpretation]
    For a signature \(
    \signature = (
    \values, \bullet, \circuitsignaturegates, \circuitsignaturearity,
    \circuitsignaturecoarity
    )\), an \emph{interpretation} of
    \(\signature\) is a tuple \((\sqsubseteq, \gateinterpretation)\) where
    \((\values, \sqsubseteq)\) is a lattice with \(\bullet\) as the least
    element, and \(\gateinterpretation\) maps each
    \(p \in \circuitsignaturegates\) to a \(\bot\)-preserving monotone function
    \(
    \valuetuple{\circuitsignaturearity(p)}
    \to
    \valuetuple{\circuitsignaturecoarity(p)}
    \).
\end{defi}

\begin{exa}\label{ex:belnap-interpretation}
    Recall the gate-level signature \(
    \belnapsignature = (
    \belnapvalues, \bot, \belnapgates, \belnaparity, \belnapcoarity
    )
    \) from \autoref{ex:belnap-signature}.
    We assign a partial order \(\leq_\mathsf{B}\) to values in
    \(\belnapvalues\) as showing on the left of
    \autoref{fig:belnap-interpretation}.
    The gate interpretation \(\belnapgateinterpretation\) has action \(
    \andgate \mapsto \land, \orgate \mapsto \lor, \notgate \mapsto \neg
    \) where \(\land\), \(\lor\) and \(\neg\) are defined by the truth tables
    on the right of \autoref{fig:belnap-interpretation}~\cite{belnap1977useful}.

    \begin{figure*}
        \centering
        \begin{tikzcd}
            & \top & \\
            \belnapfalse \arrow[dash]{ur} & & \belnaptrue \arrow[dash]{ul} \\
            & \bot \arrow[dash]{ul} \arrow[dash]{ur} &
        \end{tikzcd}
        \qquad\qquad
        \begin{tabular}{|c|cccc|}
            \hline
            \(\land\)        & \(\bot\)         & \(\belnapfalse\) & \(\belnaptrue\)  & \(\top\)         \\
            \hline
            \(\bot\)         & \(\bot\)         & \(\belnapfalse\) & \(\bot\)         & \(\belnapfalse\) \\
            \(\belnapfalse\) & \(\belnapfalse\) & \(\belnapfalse\) & \(\belnapfalse\) & \(\belnapfalse\) \\
            \(\belnaptrue\)  & \(\bot\)         & \(\belnapfalse\) & \(\belnaptrue\)  & \(\top\)         \\
            \(\top\)         & \(\belnapfalse\) & \(\belnapfalse\) & \(\top\)         & \(\top\)         \\
            \hline
        \end{tabular}
        \quad
        \begin{tabular}{|c|cccc|}
            \hline
            \(\lor\)         & \(\bot\)        & \(\belnapfalse\) & \(\belnaptrue\) & \(\top\)        \\
            \hline
            \(\bot\)         & \(\bot\)        & \(\bot\)         & \(\belnaptrue\) & \(\belnaptrue\) \\
            \(\belnapfalse\) & \(\bot\)        & \(\belnapfalse\) & \(\belnaptrue\) & \(\top\)        \\
            \(\belnaptrue\)  & \(\belnaptrue\) & \(\belnaptrue\)  & \(\belnaptrue\) & \(\belnaptrue\) \\
            \(\top\)         & \(\belnaptrue\) & \(\top\)         & \(\belnaptrue\) & \(\top\)        \\
            \hline
        \end{tabular}
        \quad
        \begin{tabular}{|c|c|}
            \hline
            \(\neg\)         &                  \\
            \hline
            \(\bot\)         & \(\bot\)         \\
            \(\belnaptrue\)  & \(\belnapfalse\) \\
            \(\belnapfalse\) & \(\belnaptrue\)  \\
            \(\top\)         & \(\top\)         \\
            \hline
        \end{tabular}
        \caption{
            The partial order \(\leq_\mathsf{B}\), and interpretations
            of primitives in \(\belnapgates\).
        }
        \label{fig:belnap-interpretation}
    \end{figure*}

    The gate-level interpretation is then \(
    (\leq_\mathsf{B}, \belnapgateinterpretation)
    \).
    As it can be unintuitive to those familiar with two-value logic, tools have
    been developed for experimenting with the gate-level
    interpretation~\cite{kaye2026belnap}.
\end{exa}

\begin{rem}
    When discussing interpretations of circuit signatures, we prefer to use
    \(\sqsubseteq\) rather than \(\leq\), as this means that the corresponding
    lattice operations \(\ljoin\) and \(\lmeet\) do not clash with the usual
    notation for the logical operations for \(\orgate\) and \(\andgate\), as
    used in the interpretations on primitives in
    \autoref{fig:belnap-interpretation}.

    This clash of notation is actually not a coincidence; if one consults the
    lattice diagram in \autoref{fig:belnap-interpretation} but rotated by 90
    degrees anticlockwise, the \(\land\) operation is the meet and the
    \(\lor\) operation is the join.
\end{rem}

\subsection{Denotational semantics of combinational circuits}

The semantic domain for \emph{combinational} circuits is straightforward: each
circuit maps to a monotone function.

\begin{defi}
    For a circuit signature \(\circuitsignature\) and interpretation
    \(\interpretation\) of \(\circuitsignature\) where \(\values\) is the
    underlying values lattice, let
    \(\funci\) be the PROP in which the morphisms \(m \to n\) are the
    \(\bot\)-preserving monotone functions
    \(\valuetuple{m} \to \valuetuple{n}\).
\end{defi}

To map between PROPs we must use a \emph{PROP morphism}, a strict symmetric
monoidal functor between PROPs.

\begin{defi}
    For a circuit signature \(\circuitsignature\) and interpretation
    \(\interpretation = \left(\sqsubseteq, \gateinterpretation\right)\) of
    \(\circuitsignature\), let \(\morph{\circuittofunci}{\ccircsigma}{\funci}\)
    be the PROP morphism with action%
    \vspace{-\abovedisplayskip}
    \begin{center}
        \begin{minipage}{0.32\textwidth}
            \centering
            \begin{align*}
                \circuittofunci[
                    \adjustimage{valign=c,margin=0pt,page=30}{tikzfigures}
                ]
                 & \coloneqq
                (v) \mapsto (v, v)
                \\
                \circuittofunci[
                    \adjustimage{valign=c,margin=0pt,page=31}{tikzfigures}
                ]
                 & \coloneqq
                (v, w) \mapsto v \sqcup w
            \end{align*}
        \end{minipage}
        \quad
        \begin{minipage}{0.25\textwidth}
            \centering
            \begin{align*}
                \circuittofunci[
                    \adjustimage{valign=c,margin=0pt,page=32}{tikzfigures}
                ]
                 & \coloneqq
                (v) \mapsto ()
                \\
                \circuittofunci[
                    \adjustimage{valign=c,margin=0pt,page=29}{tikzfigures}
                ]
                 & \coloneqq
                () \mapsto \bot
            \end{align*}
        \end{minipage}
        \quad
        \begin{minipage}{0.25\textwidth}
            \centering
            \vspace{1.5em}
            \(\circuittofunci[
                \adjustimage{valign=c,margin=0pt,page=95}{tikzfigures}
            ]
            \coloneqq
            \gateinterpretation[p]
            \)
        \end{minipage}
    \end{center}
\end{defi}

\begin{rem}
    One might wonder why the fork and the join have different semantics, as they
    would be physically realised by the same wiring.
    This is because digital circuits have a notion of \emph{static causality}:
    outputs can only connect to inputs.
    This is why the semantics of combinational circuits is \emph{functions} and
    not \emph{relations}.

    In real life one could force together two digital devices, but this might
    lead to undefined behaviour in the digital realm.
    This is reflected in the semantics by the use of the join; for example, in
    the gate-level interpretation if one tries to join together \(\belnaptrue\) and
    \(\belnapfalse\) then the overspecified \(\top\) value is produced.
\end{rem}

\subsection{Denotational semantics of sequential circuits}

In a combinational circuit, the output only depends on the inputs at the current
cycle, but for sequential circuits inputs can affect outputs many cycles after
they occur.

We therefore have to reason with \emph{infinite sequences} of inputs rather than
individual values; these are known as \emph{streams}.

\begin{nota}
    Given a set \(A\), we denote the set of streams (infinite sequences) of
    \(A\) by \(\stream{A}\).
    As a stream can equivalently be viewed as a function \(\nat \to A\), we
    write \(\sigma(i) \in A\) for the \(i\)-th element of stream
    \(\sigma \in \stream{A}\).
    Individual streams are written as \(
    \sigma \in \stream{A}
    \coloneqq
    \sigma(0) \streamcons \sigma(1) \streamcons
    \sigma(2) \streamcons \cdots
    \).
\end{nota}

Streams can be viewed a bit like lists, in that they have a head component and
an (infinite) tail component.

\begin{defi}[Operations on streams]\label{def:stream-operations}
    Given a stream \(\sigma \in \stream{A}\), its \emph{initial value}
    \(\streaminit(\sigma) \in A\) is defined as \(\sigma \mapsto \sigma(0)\)
    and its \emph{stream derivative} \(\streamderv(\sigma) \in \stream{A}\) is
    defined as \(\sigma \mapsto (i \mapsto \sigma(i+1))\).
\end{defi}

\begin{nota}
    A stream \(\sigma\) with initial value \(a\) and stream derivative
    \(\tau\) is written \(\sigma \coloneqq a \streamcons \tau\).
\end{nota}

Streams will serve as the inputs and outputs to circuits, so the denotations of
sequential circuits will be \emph{stream functions}, which consume and produce
streams.
When we discussed the denotations of combinational circuits, we noted that only
the \(\bot\)-preserving monotone functions were valid denotations; when
considering stream function denotations of sequential circuits, there are a few
more properties we need to consider.

\begin{defi}[Causal stream function~\cite{rutten2006algebraic}]
    A stream function \(\morph{f}{\stream{A}}{\stream{B}}\) is \emph{causal} if,
    for all \(i \in \nat\) and \(\sigma,\tau \in \stream{A}\) we have that
    \(\sigma(j) = \tau(j)\) for all \(j \leq i\), then
    \(f(\sigma)(i) = f(\tau)(i)\).
\end{defi}

Causality is a form of continuity; a causal stream function is a stream function
in which the \(i\)-th element of its output stream only depends on elements
\(0\) through \(i\) inclusive of the input stream; it cannot look into the
future.
A neat consequence of causality is that it enables us to lift the stream
operations of initial value and stream derivative to stream \emph{functions}.

\begin{defi}[Initial output~\cite{rutten2006algebraic}]
    For a causal stream function \(\morph{f}{\stream{A}}{\stream{B}}\) and
    \(a \in A\), the \emph{initial output of \(f\) on input \(a\)} is an element
    \(\initialoutput{f}{a} \in B\) defined as
    \(\initialoutput{f}{a} \coloneqq \streaminit(f(a \streamcons \sigma))\) for
    an arbitrary \(\sigma \in \stream{A}\).
\end{defi}

Since \(f\) is causal, the stream \(\sigma\) in the definition of initial
output truly is arbitrary; the \(\streaminit\) function only depends on the
first element of the stream.

\begin{defi}[Functional stream derivative~\cite{rutten2006algebraic}]
    For a stream function \(\morph{f}{\stream{A}}{\stream{B}}\) and
    \(a \in A\), the
    \emph{functional stream derivative of \(f\) on input \(a\)} is a stream
    function \(\streamderivative{f}{a}\) defined as \(
    \streamderivative{f}{a}
    \coloneqq
    \sigma \mapsto \streamderv(f(a \streamcons \sigma))
    \).
\end{defi}

The functional stream derivative \(\streamderivative{f}{a}\) is a new stream
function which acts as \(f\) would `had it seen \(a\) first'.

\begin{rem}
    One intuitive way to view stream functions is to think of them as the states
    of some Mealy machine; the initial output is the output given some input,
    and the functional stream derivative is the transition to a new state.
    As with most things in mathematics, this is no coincidence; there is a
    homomorphism from any Mealy machine to a stream function.
    We will exploit this fact in the next section.
\end{rem}

This leads us to the next property of denotations of sequential circuits.
Although they may have infinitely many inputs and outputs, circuits themselves
are built from a finite number of components.
This means they cannot specify infinite \emph{behaviour}.

\begin{nota}
    Given a finite word \(w \in \freemon{A}\), we abuse notation
    and write \(\streamderivative{f}{\listvar{a}}\) for the repeated
    application of the functional stream derivative for the elements of
    \(\listvar{a}\), i.e.\ \(
    \streamderivative{f}{\varepsilon} \coloneqq f
    \) and \(
    \streamderivative{f}{wa} \coloneqq
    \streamderivative{(\streamderivative{f}{w})}{a}
    \)
    if \(w \in \freemon{A}\) and \(a \in A\).
\end{nota}

\begin{defi}
    Given a stream function \(\morph{f}{\stream{A}}{\stream{B}}\), we say it is
    \emph{finitely specified} if the set \(\{
    \streamderivative{f}{w} \,|\, w \in \freemon{A}
    \}\) is finite.
\end{defi}

As the components of our circuits are monotone and \(\bot\)-preserving, a
denotational semantics for circuits must also be monotone and
\(\bot\)-preserving.
This means we need to lift the order on values to an order on streams.

\begin{nota}
    For a poset \((A, \leq_A)\) and streams \(\sigma,\tau \in \stream{A}\), we
    say that \(\sigma \leq_{\stream{A}} \tau\) if \(\sigma(i) \leq_A \tau(i)\)
    for all \(i \in \nat\).
\end{nota}

For these properties to be suitable as a denotational semantics for
sequential circuits, we must show that stream functions with these
properties form a category we can map into from \(\scircsigma\).
We will first show that these categories form a symmetric monoidal category, so
we need a suitable candidate for composition and tensor.
There are fairly obvious choices here: for the former we use regular function
composition and for the latter we use the Cartesian product.

\begin{lem}\label{lem:causality-preserved}
    Causality is preserved by composition and Cartesian product.
\end{lem}
\begin{proof}
    If the \(i\)-th element of two stream functions \(f\) and \(g\) only depends
    on the first \(i+1\) elements of the input, then so will their composition
    and product.
\end{proof}

\begin{lem}\label{lem:finitely-specified-preserved}
    Finite specification is preserved by composition and Cartesian
    product.
\end{lem}
\begin{proof}
    For both the composition and product of two stream functions \(f\) and
    \(g\), the largest the set of stream derivatives could be is the product of
    stream derivatives of \(f\) and \(g\), so this will also be finite.
\end{proof}

\begin{lem}\label{lem:monotonicity-preserved}
    \(\bot\)-preserving monotonicity is preserved by composition and Cartesian
    product.
\end{lem}
\begin{proof}
    The composition and product of any monotone function is monotone, and must
    preserve the \(\bot\) element.
\end{proof}

As the categorical operations preserve the desired properties, these stream
functions form a PROP.

\begin{prop}
    There is a PROP \(\streami\) in which the morphisms \(m \to n\) are the
    causal, finitely specified and \(\bot\)-preserving monotone stream
    functions \(\valuetuplestream{m} \to \valuetuplestream{n}\).
\end{prop}
\begin{proof}
    Identity is the identity function, the symmetry swaps streams, composition
    is composition of functions, and tensor product on
    \(\morph{f}{\valuetuplestream{m}}{\valuetuplestream{n}}\) and
    \(\morph{g}{\valuetuplestream{p}}{\valuetuplestream{q}}\) is the Cartesian
    product of functions composed with the components of the isomorphism
    \(\valuetuplestream{m} \times \valuetuplestream{n}
    \cong \valuetuplestream{m+n}\).

    As these constructs satisfy the categorical axioms, and as function
    composition and Cartesian product preserve causality
    (\autoref{lem:causality-preserved}),
    finite specification (\autoref{lem:finitely-specified-preserved}),
    and monotonicity (\autoref{lem:monotonicity-preserved}), this data defines a
    symmetric monoidal category.
\end{proof}

Modelling sequential circuits as stream functions deals with temporal
aspects, but what about feedback?
As the assignment of denotations needs to be compositional, we need
to map the trace on \(\scircsigma\) to a trace on \(\streami\).
A usual candidate for the trace when considering partially ordered settings is
the \emph{least fixed point}.
There is an important result concerning fixed points that will be crucial to
the remainder of this section.

\begin{thm}[Kleene fixed-point theorem~\cite{tarski1955latticetheoretical}]
    Let \((A, \leq)\) be a poset in which each of its directed subsets has a
    join, and let \(\morph{f}{L}{L}\) be a Scott-continuous function.
    Then \(f\) has a least fixed point, defined as \(
    \bot \vee f(\bot) \vee f(f(\bot)) \vee \cdots
    \).
\end{thm}

So far we have not explicitly enforced Scott-continuity on stream functions; it
turns out that it is implied by causality and monotonicity.

\begin{prop}\label{prop:monotone-causal-scott}
    Let \((A, \leq_A)\) and \((B, \leq_B)\) be finite lattices, and let
    \((\stream{A}, \leq_{\stream{A}})\) and \((\stream{B}, \leq_{\stream{B}})\)
    be the induced lattices on streams.
    Then a causal and monotone function \(\stream{A} \to \stream{B}\) must also
    be Scott-continuous.
\end{prop}
\begin{proof}
    Consider a directed subset \(C \subseteq \stream{A}\); we need to show that
    for an arbitrary causal and monotone \(f\) we
    have that \(f\left(\bigvee C\right) = \bigvee f[C]\).

    Let \(\morph{\proj{n}}{\stream{B}}{B^n}\) be defined as the
    function that takes the first \(n\) elements of an input stream i.e.\
    \(
    \proj{n}\left(\sigma\right)
    \coloneqq \left(
    \sigma(0),
    \sigma(1),
    \ldots,
    \sigma(n-1)\right)\).
    For some \(d \in \stream{B}\) and \(D \subseteq \stream{B}\), we have that
    \(d = \bigvee D\) if \(
    \proj{n}\left(d\right)
    =
    \bigvee\left(\proj{n}[D]\right)
    \) for all \(n \in \nat\).
    Therefore to show our desired result we substitute \(D \coloneqq f[C]\) and
    \(d \coloneqq f\left(\bigvee C\right)\), and show that \(
    \proj{n}\left(f\left(\bigvee C\right)\right)
    =
    \bigvee\left(\proj{n}\left(f[C]\right)\right)
    \).

    Next, for a function \(\morph{f}{\stream{A}}{\stream{B}}\), let
    \(\morph{f_n}{A^n}{B^n}\) be defined as \(
    f_n(\listvar{a})
    \coloneqq
    \pi_n\left(f(\listvar{a} \streamcons \bot^\omega)\right)
    \).
    Note that when \(f\) is causal, we have for \(
    \sigma \in \stream{A}
    \) and \(
    \listvar{a} \in A^n
    \) that \(
    \proj{n}\left(f(\listvar{a} \streamcons \bot^\omega)\right)
    =
    \proj{n}\left(f(\listvar{a} \streamcons \sigma)\right)
    \), so the following line of reasoning holds for an arbitrary stream
    \(\tau \in \stream{A}\):
    \[
        \proj{n}\left(f(\sigma)\right)
        =
        \proj{n}\left(f(\proj{n}(\sigma) \streamcons \tau)\right)
        =
        \proj{n}\left(f(\proj{n}(\sigma) \streamcons \bot^\omega)\right)
        =
        f_n\left(\proj{n}(\sigma)\right)
    \]
    i.e.\ \(\proj{n} \circ f = f^n \circ \proj{n}\).
    As each of these functions \(f_n\) is a monotone function between finite
    lattices, they preserve directed joins.
    Putting everything together, we have that
    \[
        \proj{n}\left(f\left(\bigvee C\right)\right)
        =
        f_n\left(\proj{n}\left(\bigvee C\right)\right)
        =
        f_n\left(\bigvee \left(\proj{n}[C]\right)\right)
        =
        \bigvee \left(\proj{n}\left(f[C]\right)\right)
    \]
    and subsequently that \(f\) is Scott-continuous.
\end{proof}

This means we can use the Kleene fixed point theorem as a tool to show that the
least fixed point is a trace on \(\streami\).

\begin{nota}[Concatenation]
    For a set \(A\), \(\listvar{a} \in A^m\), and \(\listvar{b} \in A^n\), we
    write \(\listvar{a} \concat \listvar{b}\) for the \emph{concatenation}
    of \(\listvar{a}\) and \(\listvar{b}\): the tuple of length \(m+n\)
    containing the elements of \(\listvar{a}\) followed by the elements of
    \(\listvar{b}\).
\end{nota}

\begin{nota}[Projection]
    For a set \(A\), let \(\listvar{a} \in A^m\) and \(\listvar{b} \in A^n\).
    Then for their concatenation \(
    \listvar{c} \coloneqq \listvar{a} \concat \listvar{b} \in A^{m+n}\),
    we write \(\proj{0}(\listvar{c}) = \listvar{a}\) and
    \(\proj{1}(\listvar{c}) = \listvar{b}\) for the \emph{projections}.
\end{nota}

\begin{lem}\label{lem:lfp-stream-function}
    Given a morphism \(
    \morph{f}{\valuetuplestream{x+m}}{\valuetuplestream{x+n}}
    \in \streami
    \), and stream \(\sigma \in \valuetuplestream{m}\), the function \(
    \tau \mapsto \proj{0}\mleft(f(\tau,\sigma)\mright)
    \) has a least fixed point.
\end{lem}
\begin{proof}
    The function \(\tau \mapsto \proj{0}\mleft(f(\tau,\sigma)\mright)\) is
    causal and monotone because \(f\) and the projection function are
    causal and monotone, so it is Scott-continuous by
    \autoref{prop:monotone-causal-scott}.
    By the Kleene fixed point theorem, this function has a least fixed point,
    namely \(
    \proj{0}\mleft(f(\bot^\omega, \sigma)\mright) \ljoin
    \proj{0}\mleft(f(\proj{0}\mleft(f(\bot^\omega, \sigma)\mright), \sigma)\mright) \ljoin
    \cdots
    \).
\end{proof}

We show that this notion of least fixed point is a trace on \(\streami\).
The first step is to show that taking the least fixed point of a stream function
in \(\streami\) produces another causal, finitely specified,
\(\bot\)-preserving, and monotone stream function.

\begin{defi}\label{def:state-order}
    Let \(A\) and \(B\) be posets and let
    \(\morph{f, g}{\stream{A}}{\stream{B}}\) be stream functions.
    We say \(f \stateorder g\) if \(f(\sigma) \leq_{\stream{B}} g(\sigma)\)
    for all \(\sigma \in \stream{A}\).
\end{defi}

\begin{thm}\label{thm:trace-well-defined}
    For a function \(
    \morph{f}{\valuetuplestream{x+m}}{\valuetuplestream{x+n}} \in \streami
    \)
    where $\values$ is a lattice satisfying the ascending chain condition,
    let \(\mu_f(\sigma)\) be the least fixed point of the function \(
    \tau \mapsto \proj{0}\mleft(f(\tau \concat \sigma)\mright)
    \).
    Then, the stream function \(
    \sigma \mapsto \proj{1}\mleft(f(\mu_f(\sigma) \concat \sigma)\mright)
    \) is in \(\streami\).
\end{thm}
\begin{proof}
    We prove that
    \(\sigma \mapsto \proj{1}\mleft(f(\mu_f(\sigma) \concat \sigma)\mright)\)
    is causal, finitely specified, and \(\bot\)-preserving monotone.
    The principal difficulty is establishing that the map
    \(\sigma \mapsto \mu_f(\sigma)\) is in \(\streami\);
    the desired result follows immediately after this since \(f\) and
    \(\proj{1}\) are \(\streami\) morphisms.

    Define \(\mu_f^i: \valuetuplestream{m} \to \valuetuplestream{x}\)
    for all \(i \in \nat\)
    by \(\mu_f^0(\sigma) = (\bot^x)^\omega\) and
    \(\mu_f^{i+1}(\sigma) = \proj{0}(f(\mu_f^i(\sigma) \concat \sigma))\) for \(i \geq 0\).
    Then by the Kleene fixed point theorem,
    \(\mu_f(\sigma) = \bigsqcup_{i \in \nat} \mu_f^i(\sigma)\).

    It is a straightforward induction to establish that \(\mu_f^i\)
    is in \(\streami\) for all \(i \in \nat\).
    Further, join preserves causality, \(\bot\)-preservation, and monotonicity,
    so \(\mu_f\) is causal and \(\bot\)-preserving monotone.
    It remains to show that \(\mu_f\) is finitely specified;
    we thus investigate its derivatives.

    Next, we claim that for all \(i \in \nat\) and all
    \(w \in \freemon{(\valuetuple{m})}\) that
    \begin{align}
        \streamderivative{(\mu_f^{i+1})}{w}(\sigma) = \proj{0}(
        \streamderivative{f}{\mu_f^i[w]\concat w}(
        \streamderivative{(\mu_f^i)}{w}(\sigma) \concat \sigma)). \label{eq:twdf-claim}
    \end{align}
    We prove this by induction on \(w\). When \(w = \varepsilon\), this
    claim is precisely the definition of \(\mu_f^{i+1}\). When \(w = ua\)
    where \(u \in \freemon{(\valuetuple{m})}\) and \(a \in \valuetuple{m}\),
    this claim is
    \begin{align*}
        \streamderivative{(\mu_f^{i+1})}{w}(\sigma)                                       & = \streamderivative{(\mu_f^{i+1})}{ua}(\sigma) = \streamderv{(\streamderivative{(\mu_f^{i+1})}{u}(a\streamcons\sigma))} \\
                                                                                          & = \streamderv(\proj{0}(
        \streamderivative{f}{\mu_f^i[u]\concat u}(
        \streamderivative{(\mu_f^i)}{u}(a\streamcons\sigma) \concat a\streamcons\sigma))) & \text{(IH)}                                                                                                             \\
                                                                                          & = \proj{0}(\streamderv(\streamderivative{f}{\mu_f^i[u]\concat u}(
        (\streamderivative{(\mu_f^i)}{u}[a]\concat a)\streamcons (\streamderivative{(\mu_f^i)}{ua}(\sigma) \concat \sigma)
        )))                                                                               & \text{($\streaminit, \streamderv$ decomp.)}                                                                             \\
                                                                                          & = \proj{0}(\streamderivative{
        (\streamderivative{f}{\mu_f^i[u]\concat u})}{
        \streamderivative{(\mu_f^i)}{u}[a]\concat a} (
        \streamderivative{(\mu_f^i)}{ua}(\sigma) \concat \sigma))
                                                                                          & \text{(def'n $\streamderivative{f}{-}$)}
        \\
                                                                                          & = \proj{0}(
        (\streamderivative{f}{\mu_f^i[ua]\concat ua})(
        \streamderivative{(\mu_f^i)}{ua}(\sigma) \concat \sigma))                         & (*)
    \end{align*}
    as desired. Note the starred step is an application of $g[u] \streamcons g_u[a] = g[ua]$ when $g = \mu_f^i$.

    Note that $\mu_f^0[w] = (\bot^x)^{|w|}$, where $|w|$ is
    the length of $w$ as a word in $\freemon{(\valuetuple{m})}$. Thus,
    $\mu_f^0[w] = (\bot^x)^{|w|} \leq \mu_f^1[w]$. By induction,
    $\mu_f^i[w] = \pi_0[f[\mu_f^{i-1}[w], w]]
        \leq \pi_0[f[\mu_f^i[w], w]] = \mu_f^{i+1}[w]$
    since $\pi_0$ and $f$ are monotone.
    Since all ascending chains in $\values$ stabilize, this chain of fixed length
    words in $\freemon{(\valuetuple{m})}$ must also stabilize.
    Note also if $\mu_f^i[w] = \mu_f^{i+1}[w]$ for some $i \in \nat$, then
    $\mu_f^i[w] = \mu_f^j[w]$ for all $j \geq i$.

    It is straightforward to check that
    $f_{\mu_f^i[w]\concat w} \stateorder f_{\mu_f^{i+1}[w]\concat w}$
    since $\mu_f^i[w] \leq \mu_f^{i+1}[w]$.

    From our claim (\ref{eq:twdf-claim}) we know
    \[
        (\mu_f)_w(\sigma) = \bigsqcup_{i \in \nat} (\mu_f^i)_w(\sigma)
        = \left(\bigsqcup_{i = 1}^\infty \proj{0}(
            \streamderivative{f}{\mu_f^{i}[w]\concat w}(
            \streamderivative{(\mu_f^i)}{w}(\sigma),\sigma))\right) \sqcup \mu_f^0(\sigma),
    \]
    and additionally for all $w \in \freemon{(\valuetuple{m})}$ we know the sequence
    $\{\streamderivative{f}{\mu_f^{i}[w]\concat w}\}_{i \in \nat}$ is an
    ascending infinite chain in
    $\stateorder$ with the property that
    if $\streamderivative{f}{\mu_f^{n}[w]\concat w} = \streamderivative{f}{\mu_f^{n+1}[w]\concat w}$, then the chain stabilizes at $n$.
    Hence this ascending infinite chain is actually a finite strictly ascending
    chain followed by infinitely many instances of the last element of the finite chain.

    If $w, w' \in \freemon{(\valuetuple{m})}$ are two words such that
    $\streamderivative{f}{\mu_f^{i}[w]\concat w} = \streamderivative{f}{\mu_f^{i}[w']\concat w'}$ for all $i \in \nat$, then certainly $(\mu_f)_w = (\mu_f)_{w'}$
    Hence, the number of distinct derivatives of $\mu_f$ is bounded above by
    the number of strictly ascending finite chains in the set
    $\{f_z \,|\, z \in \freemon{(\valuetuple{x+m})}\}$.
    However, since $f$ is finitely specified, this is a finite set
    and therefore there are only finitely many strictly ascending finite chains
    in this set.
\end{proof}

Even if \(\streami\) is closed under least fixed point, this does not mean that
it is a valid trace.
To verify this we must establish that the categorical axioms of the trace hold.

\begin{thm}
    A trace on \(\streami\) is given for a function \(
    \morph{f}{\valuetuplestream{x+m}}{\valuetuplestream{x+n}}
    \) by the stream function \(
    \sigma \mapsto \proj{1}(f(\mu_f(\sigma), \sigma))
    \), where \(\mu_f(\sigma)\) is the least fixed point of the function \(
    \tau \mapsto \proj{0}\mleft(f(\tau,\sigma)\mright)
    \) for fixed \(\sigma\).
\end{thm}
\begin{proof}
    By \autoref{thm:trace-well-defined}, \(\streami\) is closed under taking the
    least fixed point, so we just need to show that the axioms of STMCs hold.
    Most of these follow in a straightforward way; the only interesting one is
    the sliding axiom.
    We need to show that for stream functions \(
    \morph{f}{\valuetuplestream{x+m}}{\valuetuplestream{y+n}}
    \) and \(
    \morph{g}{\valuetuplestream{y}}{\valuetuplestream{x}}
    \), \[
        \trace{y}{(\tau, \sigma) \mapsto f(g(\tau), \sigma)}
        =
        \trace{x}{
            (\tau, \sigma)
            \mapsto
            g\left(
            \proj{0}\mleft(f(\tau, \sigma)\mright)
            \right)
            \concat
            \proj{1}\mleft(f(\tau, \sigma)\mright)
        }.
    \]

    To this end, let
    \(
    l_\sigma \coloneqq \tau \mapsto \proj{0}(f(g(\tau), \sigma))
    \)
    and let
    \(
    r_\sigma \coloneqq \tau \mapsto g(\proj{0}(f(\tau, \sigma)))
    \).
    As per the usual notation, this means that \(\mu_{l_\sigma}(\sigma)\) and
    \(\mu_{r_\sigma}(\sigma)\) are the least fixed points of \(l_\sigma\) and
    \(r_\sigma\) respectively.
    By applying the definition of the proposed trace operation, we reduce the
    original required equation to
    \(
    \proj{1}(f(g(\mu_{l_\sigma}(\sigma)), \sigma))
    =
    \proj{1}\mleft(f(\mu_{r_\sigma}(\sigma), \sigma)\mright)
    \).
    Therefore the required equation holds if
    \(
    g(\mu_{l_\sigma}(\sigma))
    =
    \mu_r(\sigma)
    \).

    Now let \(
    h_\sigma \coloneqq \tau
    \mapsto
    \proj{0}\left(f\left(\tau, \sigma\right)\right)
    \); note that
    \(g \circ h_\sigma = r_\sigma\)
    and
    \(h_\sigma \circ g = l_\sigma\).
    Let \(\upsilon_l\) be a fixed point of \(l_\sigma\); we show that
    \(g\) maps this to a fixed point of \(r_\sigma\):
    \[
        r_\sigma\left(g\left(\upsilon_l\right)\right)
        =
        g\mleft(\proj{0}\mleft(f\mleft(g\mleft(\upsilon_l\mright), \sigma\mright)\mright)\mright)
        =
        g\mleft(h_\sigma\mleft(g\mleft(\upsilon_k\mright)\mright)\mright)
        =
        g\mleft(l_\sigma\mleft(\upsilon_l\mright)\mright)
        =
        g\mleft(\upsilon_l\mright)
    \]
    Similarly, let \(\upsilon_r\) be a fixed point of \(r_\sigma\);
    we show that \(h_\sigma\) maps fixed points of \(r_\sigma\) to fixed
    points of \(l_\sigma\):
    \[
        l_\sigma\mleft(h_\sigma\mleft(\upsilon_r\mright)\mright)
        =
        \proj{0}\mleft(f\mleft(g\mleft(h_\sigma\mleft(\upsilon_r\mright)\mright), \sigma\mright)\mright)
        =
        \proj{0}\mleft(f\mleft(r_\sigma\mleft(\upsilon_r\mright), \sigma\mright)\mright)
        =
        \proj{0}\mleft(f\mleft(\upsilon_r, \sigma\mright)\mright)
        =
        h_\sigma\mleft(\upsilon_r\mright)
    \]

    In particular, we have that \(g\left(\mu_l(\sigma)\right)\) is a fixed point
    of \(r_\sigma\), so \(g(\mu_l(\sigma)) \geq \mu_r(\sigma)\).
    Conversely, \(h_\sigma(\mu_r(\sigma))\) is a fixed point of \(l_\sigma\),
    so \(h_\sigma(\mu_r(\sigma)) \geq \mu_l(\sigma)\).
    Subsequently we have that
    \(\mu_r(\sigma) = g(h_\sigma(\mu_r(\sigma))) \geq g(\mu_l(\sigma))\).
    As we have that
    \(g(\mu_l(\sigma)) \leq \mu_r(\sigma)\)
    and
    \(\mu_r(\sigma) \leq g(\mu_l(\sigma))\),
    it follows that
    \(g(\mu_l(\sigma)) = \mu_r(\sigma)\) as originally required.
\end{proof}

We now have two traced PROPs: a \emph{syntactic} PROP of sequential circuit
terms \(\scircsigma\) and a \emph{semantic} PROP of causal, finitely
specified, monotone stream functions \(\streami\).
Before mapping from circuits to streams we will examine an intermediate
structure with close links to both circuits and stream functions; that of
\emph{Mealy machines}.

\subsection{Monotone Mealy machines}\label{sec:mealy}

While it would be straightforward to map from circuits to stream functions by
mapping from each generator of \(\scircsigma\) to stream functions in
\(\streami\), it is not immediately obvious how to define a reverse mapping.
Even though every stream function in \(\streami\) has finitely many stream
derivatives, how does one encapsulate this behaviour into a circuit?
To answer this question, we will factor our proof through
\emph{Mealy machines}.

Mealy machines are used in circuit design to specify the behaviour of a circuit
without having to use concrete components.
They also have a very useful \emph{coalgebraic} viewpoint which we will wield in
order to build a bridge from circuits into stream functions.
In particular, there is a unique homomorphism from a Mealy machine to a causal,
finitely specified stream function.
Our strategy is to assemble a special class of Mealy machines which we dub
\emph{monotone Mealy machines} into another traced PROP.

\begin{defi}[Mealy machine~\cite{mealy1955method}]\label{def:mealy}
    Let \(A\) and \(B\) be finite sets.
    A (finite) \((A,B)\)-\emph{Mealy machine} is a tuple \((S, f)\) where
    \(S\) is a finite set called the \emph{state space},
    \(\morph{f}{S}{(S \times B)^A}\) is the \emph{Mealy function}.
\end{defi}

An \((A,B)\)-Mealy machine is parameterised over a set \(A\) of \emph{inputs}
and a set \(B\) of \emph{outputs}, and is comprised of a set \(S\) of
\emph{states} and a function transforming a pair \((s, a)\) of a current state
and an input into a pair \(\langle{t,b}\rangle\) of a transition state and an
output.

\begin{nota}
    We will use the shorthand \(
    \mealyfunctiontransition{f} \coloneqq (s, a) \mapsto \proj{0}(f(s)(a))
    \) and \(
    \mealyfunctionoutput{f} \coloneqq (s, a) \mapsto \proj{1}(f(s)(a))
    \) for the transition and output component of the Mealy function respectively.
\end{nota}

\begin{exa}\label{ex:mealy}
    Let the set of Booleans be defined as
    \(\booleans \coloneqq \{\mathsf{f},\mathsf{t}\}\).
    We define a \((\booleans,\booleans)\)-Mealy machine \((S, f)\) as follows:
    \begin{center}
        \begin{minipage}{0.33\textwidth}
            \[S \coloneqq \{s_0, s_1\}\]
        \end{minipage}
        \begin{minipage}{0.66\textwidth}
            \begin{gather*}
                f(s_0, \mathsf{f}) \mapsto \langle{s_0, \mathsf{f}}\rangle
                \qquad
                f(s_0, \mathsf{t}) \mapsto \langle{s_1, \mathsf{t}}\rangle
                \\
                f(s_1, \mathsf{f}) \mapsto \langle{s_1, \mathsf{t}}\rangle
                \qquad
                f(s_1, \mathsf{t}) \mapsto \langle{s_0, \mathsf{f}}\rangle
            \end{gather*}
        \end{minipage}
    \end{center}
    This is a Mealy machine with two states; at state \(s_0\) the output is the
    input, and at state \(s_1\) the output is the negation.
    If the input is true then the state switches.
    To illustrate Mealy machines we draw states as circles; an arrow between
    states labelled \(v|w\) represents a transition on input \(v\) producing
    output \(w\).
    \begin{center}
        \begin{tikzpicture}[
        ->,
        >=stealth,
        node distance=0.25cm,
        every node/.style={font=\scriptsize},
        every state/.style={font=\normalsize, thick, fill=gray!10, minimum size=2pt, ellipse, minimum size=2pt},
        initial text=$ $,
        initial distance=0.5cm,
        every initial by arrow/.style={*->},
        x=20pt,
        y=20pt
    ]
    \node[state] (s0) at (0,0) {\(s_0\)};
    \node[state] (s1) at (3,0) {\(s_1\)};

    \draw (s0) edge[bend left, above] node{\(\trs{\mathsf{t}}{\mathsf{t}}\)} (s1);
    \draw (s0) edge[out=135, in=225, loop, distance=1cm] node[left]{
            \(\trs{\mathsf{f}}{\mathsf{f}}\)
        } (s0);

    \draw (s1) edge[bend left, below] node{\(\trs{\mathsf{t}}{\mathsf{f}}\)} (s0);
    \draw (s1) edge[out=315, in=45, loop, distance=1cm] node[right]{
            \(\trs{\mathsf{f}}{\mathsf{t}}\)
        } (s1);
\end{tikzpicture}
    \end{center}
\end{exa}

\subsection{The coalgebraic perspective}

It has been shown that there are close links between Mealy machines and certain
types of stream
functions~\cite{rutten2006algebraic,bonsangue2008coalgebraic}, which we will
exploit to show the correspondence between \(\scircsigma\) and \(\streami\).
These results use an alternative definition of Mealy machines as types of
\emph{coalgebra}.

\begin{defi}
    For sets \(A\) and \(B\), an \emph{\((A,B)\)-Mealy coalgebra} is a coalgebra
    of the functor \(\morph{Y}{\set}{\set}\), defined as
    \(S \mapsto (S \times B)^A\).
\end{defi}

\begin{exaC}[\cite{bonsangue2008coalgebraic}]
    Given sets \(A\) and \(B\), let \(\Gamma\) be the set of causal stream
    functions \(\stream{A} \to \stream{B}\), and let
    \(\morph{\nu}{\Gamma}{(\Gamma \times B)^A}\) be the function defined as \(
    f \mapsto a \mapsto \langle{
        \streamderivative{f}{a},\initialoutput{f}{a}
    }\rangle\).
    Then \((\Gamma,\nu)\) is a \((A,B)\)-Mealy coalgebra.
\end{exaC}

\begin{rem}
    Note that the interpretation of a Mealy function as a coalgebra differs from
    how we have presented Mealy functions so far as it is curried
    (i.e.\ \(S \to (A \to (S \times B)\) as opposed to
    \(S \times A \to S \times B\).
    In general we prefer to use the latter as it corresponds closer to how we
    will later interpret Mealy functions string diagrammatically as boxes
    with state, input and output wires.
\end{rem}

The above example lays the groundwork to establish connections between circuits,
stream functions and Mealy machines.
If we inspect it a little closer, we find that stream functions are even more
special than just being `an' \((A,B)\)-Mealy coalgebra.

\begin{defi}[Mealy homomorphism]\label{def:mealy-homomorphism}
    For sets \(A\) and \(B\), a \emph{Mealy homomorphism} between two
    \((A,B)\)-Mealy coalgebra \((S,f)\) and \((T,g)\) is a function
    \(\morph{h}{S}{T}\) preserving transitions and
    outputs, i.e.\ if \(f(s,a) = (r,b)\), then \(g(h(s),a) = (h(r),b)\).
\end{defi}

The \emph{final} \((A,B)\)-Mealy coalgebra is a \((A,B)\)-Mealy coalgebra to
which every other \((A,B)\)-Mealy coalgebra has a unique homomorphism.

\begin{propC}[{\cite[Proposition 2.2]{rutten2006algebraic}}]
    \label{prop:final-coalgebra}
    For every \((A,B)\)-Mealy coalgebra \((S,f)\), there exists a
    unique \((A,B)\)-Mealy homomorphism \(\morph{{!}}{(S,f)}{(\Gamma,\nu)}\).
\end{propC}
\begin{proof}
    An \((A,B)\)-Mealy homomorphism \(\morph{g}{(S, f)}{(\Gamma, \nu)}\) is a
    function \(S \to \Gamma\), so for a state \(s_0 \in S\), \(g(s)\) will be a
    stream function \(\stream{A} \to \stream{B}\).
    Let \(\sigma \in \stream{A}\) be an input stream; there is a (unique) series
    of transitions \[
        s_0
        \mealyarrow{\sigma(0)}{b_0}
        s_1
        \mealyarrow{\sigma(1)}{b_1}
        s_2
        \mealyarrow{\sigma(2)}{b_2}
        s_3
        \mealyarrow{\sigma(3)}{b_3}
        \cdots
    \]
    Then \(!(s)\) is defined for input \(\sigma\) and
    index \(i \in \nat\) as \(!(s)(\sigma)(i) \coloneqq b_i\).
\end{proof}

For a Mealy coalgebra \((S, f)\) and a start state \(s_0\),
\(!(s_0)(\sigma)\) maps to the stream of outputs that \((S, f)\) would produce
by applying \(f\) to each element of \(\sigma\), starting from \(s_0\).

\subsection{Mealy machines on posets}

To use Mealy machines as a bridge between \(\scircsigma\) and \(\streami\) they
must be assembled into another traced PROP.
Not all Mealy machines defined so far correspond to circuits in
\(\scircsigma\); we must refine our notion of Mealy machine in order to find
those that do: those that map to stream functions in \(\streami\).

\begin{lem}
    For a Mealy machine \((S, f)\) and state \(s_0 \in S\), \(!(s_0)\)
    is finitely specified.
\end{lem}
\begin{proof}
    \(S\) is finite, and \(\mealytostream\) must preserve transitions.
\end{proof}

As we must also impose a monotonicity condition, we move from Mealy machines
over sets and functions to Mealy machines over posets and monotone functions.

\begin{defi}[Monotone Mealy machine]
    Let \(A\), \(B\) be lattices; an \((A,B)\)-Mealy machine
    \((S, f)\) is called a \emph{monotone} Mealy machine if \(S\) is a
    lattice and \(f\) is \(\bot\)-preserving monotone with respect to the
    ordering on \(A\), \(B\), and \(S\).
\end{defi}

To map from Mealy machines to circuits we need to assemble the former into
another PROP, in which the morphisms \(m \to n\) are
\((\valuetuple{m},\valuetuple{n})\)-Mealy machines; we must also take into
account the `initial state' of circuits in \(\scircsigma\).

\begin{defi}[Initialised Mealy machine]
    An \emph{initialised} Mealy machine is a tuple \((S, f, s_0)\), where
    \((S, f)\) is a Mealy machine, and \(s_0 \in S\) is an \emph{initial state}.
\end{defi}

\begin{exa}\label{ex:mealy-init}
    We can initialise the \((\booleans,\booleans)\)-Mealy machine
    \((\{s_0,s_1\},f)\) from \autoref{ex:mealy} in two ways; here we will choose to
    initialise it as \((S,f,s_0)\).
    In the diagrams, we label the initial state with an arrow.
    \begin{center}
        \begin{tikzpicture}[
        ->,
        >=stealth,
        node distance=0.25cm,
        every node/.style={font=\scriptsize},
        every state/.style={font=\normalsize, thick, fill=gray!10, minimum size=2pt, ellipse, minimum size=2pt},
        initial text=$ $,
        initial distance=0.5cm,
        every initial by arrow/.style={*->},
        x=20pt,
        y=20pt
    ]
    \node[state, initial above] (s0) at (0,0) {\(s_0\)};
    \node[state] (s1) at (3,0) {\(s_1\)};

    \draw (s0) edge[bend left, above] node{\(\trs{\mathsf{t}}{\mathsf{t}}\)} (s1);
    \draw (s0) edge[out=135, in=225, loop, distance=1cm] node[left]{
            \(\trs{\mathsf{f}}{\mathsf{f}}\)
        } (s0);

    \draw (s1) edge[bend left, below] node{\(\trs{\mathsf{t}}{\mathsf{f}}\)} (s0);
    \draw (s1) edge[out=315, in=45, loop, distance=1cm] node[right]{
            \(\trs{\mathsf{f}}{\mathsf{t}}\)
        } (s1);
\end{tikzpicture}
    \end{center}
\end{exa}

All that remains to define is the composition of Mealy machines, which is
standard.

\begin{defi}[Cascade product of Mealy machines~\cite{ginzburg2014algebraic}]
    Given an initialised \((A,B)\)-Mealy machine \((S,f,s_0)\) and an
    initialised \((B,C)\)-Mealy machine \((T,g,t_0)\), their
    \emph{cascade product} is an initialised \((A,C)\)-Mealy machine defined as
    \[
        \left(S \times T, ((s, t), a) \mapsto \left\langle
        \left(
        \mealyfunctiontransition{f}(s,a),
        \mealyfunctiontransition{g}(t, \mealyfunctionoutput{f}(s, a))
        \right),
        \mealyfunctionoutput{g}(t, \mealyfunctionoutput{f}(s, a))
        \right\rangle,
        (s_0, t_0)\right).
    \]
\end{defi}

The cascade product of two Mealy machines effectively executes the first on the
inputs, then executes the second on the outputs of the first.

\begin{exa}\label{ex:mealy-cascade}
    Recall the initialised \((\booleans,\booleans)\)-Mealy machine
    \((S, f, s_0)\) from \autoref{ex:mealy-init}; we will now compose this with
    \((\{t_0,t_1\},g,t_0)\) where \(g\) is defined as follows:
    \begin{center}
        \begin{tikzpicture}[
        ->,
        >=stealth,
        node distance=0.25cm,
        every node/.style={font=\scriptsize},
        every state/.style={font=\normalsize, thick, fill=gray!10, minimum size=2pt, ellipse},
        initial text=$ $,
        initial distance=0.5cm,
        every initial by arrow/.style={*->},
        x=20pt,
        y=20pt
    ]
    \node[state, initial above] (s0) at (0,0) {\(t_0\)};
    \node[state] (s1) at (3,0) {\(t_1\)};

    \draw (s0) edge[above] node{\(\trs{\mathsf{t}}{\mathsf{t}}\)} (s1);
    \draw (s0) edge[out=135, in=225, loop, distance=1cm] node[left]{
            \(\trs{\mathsf{f}}{\mathsf{f}}\)
        } (s0);

    \draw (s1) edge[out=315, in=45, loop, distance=1cm] node[right]{
            \(\trs{\mathsf{f},\mathsf{t}}{\mathsf{t}}\)
        } (s1);
\end{tikzpicture}
    \end{center}
    The cascade product \((R, h, r_0)\) of these two machines is defined as
    follows:
    \begin{center}
        \begin{tikzpicture}[
        bezier bounding box=true,
        ->,
        >=stealth,
        node distance=0.25cm,
        every node/.style={font=\scriptsize},
        every state/.style={font=\normalsize, thick, fill=gray!10, minimum size=2pt, ellipse},
        initial text=$ $,
        initial distance=0.5cm,
        every initial by arrow/.style={*->},
        x=20pt,
        y=20pt
    ]
    \node[state, initial above] (s0) at (0,0) {\((s_0,t_0)\)};
    \node[state] (s1) at (4,0) {\((s_1,t_0)\)};
    \node[state] (s2) at (0,-4) {\((s_0,t_1)\)};
    \node[state] (s3) at (4,-4) {\((s_1,t_1)\)};

    \draw (s0) edge[left] node{\(\trs{\mathsf{t}}{\mathsf{t}}\)} (s3);
    \draw (s0) edge[out=135, in=225, loop, distance=2cm] node[left]{
            \(\trs{\mathsf{f}}{\mathsf{f}}\)
        } (s0);

    \draw (s1) edge[above] node{\(\trs{\mathsf{t}}{\mathsf{f}}\)} (s0);
    \draw (s1) edge[right] node{\(\trs{\mathsf{f}}{\mathsf{t}}\)} (s3);

    \draw (s2) edge[bend left, above] node{\(\trs{\mathsf{t}}{\mathsf{t}}\)} (s3);
    \draw (s2) edge[out=135, in=225, loop, distance=2cm] node[left]{
            \(\trs{\mathsf{f}}{\mathsf{t}}\)
        } (s2);

    \draw (s3) edge[bend left, below] node{\(\trs{\mathsf{f}}{\mathsf{t}}\)} (s2);
    \draw (s3) edge[out=315, in=45, loop, distance=2cm] node[right]{
            \(\trs{\mathsf{t}}{\mathsf{t}}\)
        } (s3);
\end{tikzpicture}
    \end{center}
\end{exa}

Tensor product is far more straightforward.

\begin{defi}[Direct product of Mealy machines]
    Given an initialised \((A,B)\)-Mealy machine \((S,f,s_0)\) and an
    initialised \((C,D)\)-Mealy machine \((T,g,t_0)\), their
    \emph{direct product} is an initialised \((A \times C,B \times D)\)-Mealy
    machine defined as \[
        (S \times T, \left((s, t), (a, c)\right) \mapsto \left\langle
        \left(
        \mealyfunctiontransition{f}\mleft(s, a\mright),
        \mealyfunctiontransition{g}\mleft(t, c\mright)
        \right),
        \left(
        \mealyfunctionoutput{f}\mleft(s, a\mright),
        \mealyfunctionoutput{g}\mleft(t, c\mright)
        \right)\right\rangle,
        (s_0, t_0)
        ).
    \]
\end{defi}

\begin{exa}\label{ex:mealy-direct}
    The direct product of the two initialised \((\booleans,\booleans)\)-Mealy
    machines introduced in \autoref{ex:mealy-init} and \autoref{ex:mealy-cascade} is
    a \((\booleans^2,\booleans^2)\)-Mealy machine \((Q,k,q_0)\) defined as
    follows:
    \begin{center}
        \begin{tikzpicture}[
        bezier bounding box=true,
        ->,
        >=stealth,
        node distance=0.25cm,
        every node/.style={font=\scriptsize},
        every state/.style={font=\normalsize, thick, fill=gray!10, minimum size=2pt, ellipse},
        initial text=$ $,
        initial distance=0.5cm,
        every initial by arrow/.style={*->},
        x=20pt,
        y=20pt
    ]
    \node[state, initial above] (s0) at (0,0) {\((s_0,t_0)\)};
    \node[state] (s1) at (5,0) {\((s_1,t_0)\)};
    \node[state] (s2) at (0,-5) {\((s_0,t_1)\)};
    \node[state] (s3) at (5,-5) {\((s_1,t_1)\)};

    \draw (s0) edge[out=135, in=225, loop, distance=2cm] node[left]{
            \(\trs{\mathsf{ff}}{\mathsf{ff}}\)
        } (s0);
    \draw (s0) edge[left] node{\(\trs{\mathsf{ft}}{\mathsf{ft}}\)} (s2);
    \draw (s0) edge[bend left, above] node{\(\trs{\mathsf{tf}}{\mathsf{tf}}\)} (s1);
    \draw (s0) edge[left, near start] node{\(\trs{\mathsf{tt}}{\mathsf{tt}}\)} (s3);

    \draw (s1) edge[out=315, in=45, loop, distance=2cm] node[right]{
            \(\trs{\mathsf{ff}}{\mathsf{tf}}\)
        } (s1);
    \draw (s1) edge[bend left, above] node{\(\trs{\mathsf{tf}}{\mathsf{ff}}\)} (s0);
    \draw (s1) edge[right] node{\(\trs{\mathsf{ft}}{\mathsf{tt}}\)} (s3);
    \draw (s1) edge[right, near start] node{\(\trs{\mathsf{tt}}{\mathsf{ft}}\)} (s2);

    \draw (s2) edge[out=135, in=225, loop, distance=2cm] node[left]{
            \(\trs{\mathsf{ff},\mathsf{ft}}{\mathsf{ft}}\)
        } (s2);
    \draw (s2) edge[bend left, below] node{\(\trs{\mathsf{tf},\mathsf{tt}}{\mathsf{tt}}\)} (s3);

    \draw (s3) edge[out=315, in=45, loop, distance=2cm] node[right]{
            \(\trs{\mathsf{ff},\mathsf{ft}}{\mathsf{tt}}\)
        } (s3);
    \draw (s3) edge[bend left, below] node{\(\trs{\mathsf{tf},\mathsf{tt}}{\mathsf{ft}}\)} (s2);
\end{tikzpicture}
    \end{center}
\end{exa}

With cascade product as composition and direct product as tensor, initialised
monotone Mealy machines form a PROP.

\begin{defi}
    Let \(\mealyi\) be the PROP in which the morphisms
    \(m \to n\) are the initialised monotone
    \((\valuetuple{m}, \valuetuple{n})\)-Mealy machines \((S, f, s_0)\) in which
    \(S\) is a lattice.
    Composition is by cascade product and tensor on morphisms is by
    direct product.
    The identity and the symmetry are the single-state machines that output the
    input and swap the inputs respectively.
\end{defi}

Once again, we must show that this category has a trace.
This can be computed in much the same way as it was for stream functions.

\begin{defi}
    Let \((S, f)\) be a monotone \(
    (\valuetuple{x+m}, \valuetuple{x+n})
    \)-Mealy machine.
    For a state \(s \in S\) and input \(\listvar{a} \in \valuetuple{m}\), let
    \(\mu_{s}(\listvar{a})\) be the least fixed point of \(
    \listvar{r}
    \mapsto
    \proj{0}\mleft(
    \mealyfunctionoutput{f}\mleft(s, \listvar{r} \concat \listvar{a}\mright)\mright)
    \).
    The \emph{least fixed point} of an initialised Mealy machine \((S, f, s_0)\)
    is a \((\valuetuple{m}, \valuetuple{n})\)-Mealy machine \[\left(
        S, (s, \listvar{a})
        \mapsto
        \langle{
            \mealyfunctiontransition{f}\mleft(s, \mu_{s}\mleft(\listvar{a}\mright) \concat \listvar{a}\mright),
            \proj{1}\mleft(\mealyfunctionoutput{f}\mleft(s, \mu_{s}\mleft(\listvar{a}\mright) \concat \listvar{a}\mright)\mright)
        }\rangle,
        s_0
        \right).
    \]
\end{defi}

\begin{exa}\label{ex:trace-mealy}
    Consider the monotone \((\valuetuple{2},\valuetuple{2})\)-Mealy machine with
    state set \(\belnapvalues\), initial state \(\bot\), and Mealy function \[
        f
        \coloneqq
        (s, (a, b))
        \mapsto
        \langle{s \land b, (a \land b, b \lor \neg a)}\rangle.
    \]
    To compute the least fixed point of this machine we must first compute the
    least fixed point of the function \(a \mapsto a \land b\), which
    is clearly just \(
    \bot
    \ljoin (\bot \land b)
    \ljoin ((\bot \land b) \land b)
    \ljoin \cdots
    =
    (\bot \land b)
    \ljoin (\bot \land b)
    \ljoin \cdots
    = \bot \land b
    \).
    Therefore the least fixed point of the original machine is a
    \((\values, \values)\)-Mealy machine with Mealy function
    \begin{align*}
        (s, b)
         & \mapsto
        \langle{
            \mealyfunctiontransition{f}\mleft(s, \mleft(\mu_{s}\mleft(b\mright), b\mright)\mright),
            \proj{1}\mealyfunctionoutput{f}\mleft(s, \mleft(\mu_{s}\mleft(b\mright), b\mright)\mright)
        }\rangle   \\
        (s, b)
         & \mapsto
        \langle{
            s \land b,
            b \lor \neg \mleft(\bot \land b\mright)
        }\rangle   \\
        (s, b)
         & \mapsto
        \langle{
            s \land b,
            b \lor \bot \lor \neg b
        }\rangle   \\
        (s, b)
         & \mapsto
        \langle{
            s \land b,
            \belnaptrue
        }\rangle
    \end{align*}
\end{exa}

We must show that the least fixed point on Mealy machines is the trace
on \(\mealyi\).

\begin{prop}
    The least fixed point is a trace on \(\mealyi\).
\end{prop}
\begin{proof}
    Let \((S, f)\) be a monotone
    \((\valuetuple{x+m}, \valuetuple{x+n})\)-Mealy machine; this means that
    \(S\) is a lattice.
    The Mealy function \(
    \morph{f}{
        S \times \valuetuple{x+m}
    }{
        S \times \valuetuple{x+n}
    }
    \) is monotone with regards to the orders on \(S\) and
    \(\valuetuple{x+m}\) and \(S \times \valuetuple{x+n}\) is a finite lattice, so
    \(f\) has a least fixed point.
    The function \(
    f^\prime \coloneqq (s, \listvar{a})
    \mapsto
    \proj{1}\mleft(f(s, \mu_{s}(\listvar{a}) \concat \listvar{a})\mright)
    \) is a composition of \(\bot\)-preserving monotone functions, so it is
    itself \(\bot\)-preserving monotone; this makes \((S, f^\prime)\) a monotone
    \((\valuetuple{m}, \valuetuple{n})\)-Mealy machine.
    This construction is a trace for the same reason as the trace of
    \(\streami\) is.
\end{proof}

With monotone Mealy machines in a PROP, we can now represent the unique
homomorphism from a Mealy machine to a set of state functions as a PROP
morphism.

\begin{prop}\label{prop:mealy-to-stream}
    There is a traced PROP morphism
    \(\morph{\mealytostreami}{\mealyi}{\streami}\) sending a monotone Mealy
    machine \(\morph{\left(S, f, s_0\right)}{m}{n}\) to \(!(s_0)\), where \(!\)
    is the unique homomorphism \((S,f) \to (\Gamma,\nu)\).
\end{prop}
\begin{proof}
    Since every stream function also has a Mealy coalgebra structure and Mealy
    homomorphisms preserve transitions and outputs,
    composition of Mealy machines also coincides with composition of stream
    functions.
\end{proof}

\subsection{Streams to Mealy machines}

So far we have seen how a causal, finitely specified, and \(\bot\)-preserving
monotone stream function can be retrieved from a monotone Mealy machine.
It is also possible to retrieve a Mealy machine for a given stream function in
\(\streami\) by repeatedly taking stream derivatives; since we know there are
finitely many we will be able to compute a finite set of states in a Mealy
machine.

\begin{defi}
    For a causal, finitely specified, stream function \(
    \morph{f}{\stream{M}}{\stream{N}}
    \), let \(S\) be the least set of
    causal stream functions including \(f\) and closed under stream derivatives:
    i.e.\ for all \(h \in S\) and \(a \in M\), \(h_a \in S\).
    Then let \(
    \streamtomealy[f]{}
    \) be defined as the initialised Mealy machine \((S, g, f)\), where \(
    g(h, a) = \langle\mealytransition{h}{a}, \mealyoutput{h}{a}\rangle
    \).
\end{defi}

For a finitely specified stream function \(f\) it is a straightforward procedure
to find all the states of the corresponding Mealy machine
\(\streamtomealy[f]{}\).

\begin{exa}
    Let \(\morph{f}{\belnapvalues}{\belnapvalues}\) be a stream function defined
    as \(f(\sigma)(0) = \sigma(0)\) and
    \(f(\sigma)(k+1) = f(\sigma)(k) \land \sigma(k+1)\).
    We will derive a Mealy machine in \(\mealyi\) from this stream function.
    The complete set of states is \(\{
    f, \streamderivative{f}{\bot}, \streamderivative{f}{\belnapfalse},
    \streamderivative{f}{\top}
    \}\):
    \begin{itemize}
        \item \(\streamderivative{f}{\belnaptrue} = f\);
        \item \(\streamderivative{(\streamderivative{f}{\bot})}{\bot}
              =
              \streamderivative{(\streamderivative{f}{\bot})}{\belnaptrue}
              =
              \streamderivative{f}{\bot}
              \);
        \item \(\streamderivative{(\streamderivative{f}{\top})}{\top}
              =
              \streamderivative{(\streamderivative{f}{\top})}{\belnaptrue}
              =
              \streamderivative{f}{\top}
              \); and
        \item \(
              \streamderivative{(\streamderivative{f}{\bot})}{\belnapfalse}
              =
              \streamderivative{(\streamderivative{f}{\bot})}{\top}
              =
              \streamderivative{(\streamderivative{f}{\top})}{\bot}
              =
              \streamderivative{(\streamderivative{f}{\top})}{\belnapfalse}
              =
              \streamderivative{f}{\belnapfalse}
              \).
    \end{itemize}
    The Mealy function is defined for each state as the initial value and
    stream derivative of the original stream function.
    The initial state of the Mealy machine is \(f\).
\end{exa}

This procedure does not just find an arbitrary Mealy machine with the same
behaviour as the stream function.
In fact, it produces the \emph{minimal} such Mealy machine.

\begin{corC}[{\cite[Corollary 2.3]{rutten2006algebraic}}]\label{cor:minimal-mealy}
    For a stream function \(\morph{f}{\stream{M}}{\stream{N}}\),
    \(\mealytostream[f]{}\) has the smallest state space of all Mealy machines
    with the property \(
    \mealytostream[\streamtomealy[f]{}]{} = f
    \).
\end{corC}
\begin{proof}
    Since \(S\) is generated from the function \(f\) and is the \emph{smallest}
    possible set, there are no unreachable states in \(S\).
    Since the output and transition of states in
    \(\streamtomealyi[f]\) are the initial output and stream derivative, two
    states can only have the same `behaviour' if they are derived from the same
    original stream function.
\end{proof}

We will encode this data into a PROP morphism from \(\streami\) to \(\mealyi\);
in order to do this we must verify that the produced Mealy machine is monotone.

\begin{lem}\label{lem:head-tail-monotone}
    The functions \(\sigma \mapsto \streaminit(\sigma)\) and
    \(\sigma \mapsto \streamderv(\sigma)\) are monotone.
\end{lem}
\begin{proof}
    Let \(\sigma \coloneqq a \streamcons \sigma^\prime\) and
    \(\tau \coloneqq b \streamcons \tau^\prime\) be streams in \(\stream{A}\)
    such that \(\sigma \leq_{\stream{A}} \tau\); subsequently \(a \leq b\) and
    \(\sigma^\prime \leq_{\stream{A}} \tau^\prime\).
    So \(
    \streaminit(\sigma) \coloneqq
    \streaminit(a \streamcons \sigma^\prime) =
    a \leq_{A}
    b =
    \streaminit(b \streamcons \tau^\prime) \coloneqq
    \streaminit(\tau)
    \) and \(
    \streamderv(\sigma) \coloneqq
    \streamderv(a \streamcons \sigma^\prime) =
    \sigma^\prime \leq_{\stream{A}}
    \tau^\prime =
    \streamderv(b \streamcons \tau^\prime) =
    \streamderv(\tau)
    \).
\end{proof}

\begin{lem}\label{lem:initial-derivative-monotone}
    For posets \(A\) and \(B\) and a monotone causal stream function
    \(\morph{f}{\stream{A}}{\stream{B}}\), the functions
    \(a \mapsto \initialoutput{f}{a}\) and \(a \mapsto \streamderivative{f}{a}\)
    are monotone.
\end{lem}
\begin{proof}
    Let \(a, b \in A\) such that \(a \leq_A b\); then by monotonicity
    \(f(a \streamcons \sigma) \leq_{\stream{B}} f(b \streamcons \sigma)\) for
    all \(\sigma \in \stream{A}\).
    By \autoref{lem:head-tail-monotone}, \(\streaminit \circ f\) and
    \(\streamderv \circ f\) are monotone.
    First we show that the initial output is monotone: \(
    \initialoutput{f}{a} \coloneqq
    \streaminit(f(a \streamcons \sigma)) \leq_{A}
    \streaminit(f(b \streamcons \sigma)) =
    \initialoutput{f}{b}
    \).
    For the stream derivative, \(
    \streamderivative{f}{a}(\sigma) \coloneqq
    \streamderv(f(a \streamcons \sigma)) \leq_{\stream{B}}
    \streamderv(f(b \streamcons \sigma)) \coloneqq
    \streamderivative{f}{b}(\sigma)
    \).
\end{proof}

As the procedure mapping stream functions to Mealy machines preserves
monotonicity, we can use it to define a PROP morphism between \(\streami\) and
\(\mealyi\).

\begin{lem}\label{lem:stream-to-mealy-is-monotone}
    Given \(f \in \streami\), \(\streamtomealy[f]{}\) is
    a monotone Mealy machine.
\end{lem}
\begin{proof}
    Each state in the derived Mealy machine is a monotone stream function, so
    this is a poset ordered by \(\stateorder\) as defined in
    \autoref{def:state-order}. and
    The Mealy function is the pairing of the initial output and stream
    derivative; by \autoref{lem:initial-derivative-monotone} these are monotone
    so the Mealy function must also be monotone.
\end{proof}

\begin{defi}
    Let \(\morph{\streamtomealyi}{\streami}{\mealyi}\) be the traced PROP
    morphism with action \(f \mapsto \streamtomealy[f]{}\).
\end{defi}

This means we can map between monotone Mealy machines and causal, finitely
specified, monotone stream functions in either direction.
Mealy machines are perhaps more intuitive to work with, but stream functions
are the `purest' specification of the behaviour in that they have the smallest
possible state set.
Ideally we would be able to work in whichever setting is most beneficial at a
given time, so we need to show that the translations are
\emph{behaviour-preserving}.

\begin{prop}\label{prop:mealy-stream-id}
    \(\mealytostreami \circ \streamtomealyi = \id[\streami]\).
\end{prop}
\begin{proof}
    Stream functions are equal if they have the same initial output and
    stream derivative.
    \(\streamtomealyi\) preserves outputs and derivatives by definition, and
    \(\mealytostreami\) preserves transitions and outputs because it is a Mealy
    homomorphism.
\end{proof}

The reverse does not hold as many Mealy machines may map to the same stream
function, but the result of
\(\mealytostreami \circ \streamtomealyi \circ \mealytostreami\) is of course
equal to \(\mealytostreami\).

\subsection{From circuits to Mealy machines}\label{sec:synthesis}

The close links between \(\streami\) and \(\mealyi\) are nice to have but are
unsurprising; the main contribution of this chapter is to define maps
\(\scircsigma \to \mealyi\) and \(\mealyi \to \scircsigma\).
This allows us to use monotone Mealy machines as a stepping stone in the
correspondence between sequential circuits and stream functions.
By exploiting the coalgebraic properties shared between Mealy machines and
stream functions, this can be used to show that for every stream function
\(f \in \streami\) we can recover a circuit \(
\adjustimage{valign=c,margin=0pt,page=96}{tikzfigures}
\in \scircsigma
\) such that the interpretation of \(
\adjustimage{valign=c,margin=0pt,page=96}{tikzfigures}\) as a stream
function is \(f\).

\subsection{Circuits to monotone Mealy machines}

Circuits have a natural interpretation as Mealy machines, so the action
of a PROP morphism from \(\scircsigma\) to \(\mealyi\) is intuitive.

\begin{defi}
    Let \(\morph{\circuittomealyi}{\scircsigma}{\mealyi}\) be the traced PROP
    morphism defined on generators as
    \begin{align*}
        \circuittomealy[
            \adjustimage{valign=c,margin=0pt,page=97}{tikzfigures}
        ]{\interpretation}
         & \coloneqq (
        \{s\},
         &             & \left(s, \listvar{v}\right) \mapsto
        \left\langle{s, \gateinterpretation[g](\listvar{v})}\right\rangle,
         &             & s
        )
        \\
        \circuittomealy[
            \adjustimage{valign=c,margin=0pt,page=30}{tikzfigures}
        ]{\interpretation}
         & \coloneqq (
        \{s\},
         &             & (s, v) \mapsto \left\langle{s, (v, v)}\right\rangle,
         &             & s
        )
        \\
        \circuittomealy[
            \adjustimage{valign=c,margin=0pt,page=31}{tikzfigures}
        ]{\interpretation}
         & \coloneqq (
        \{s\},
         &             & (s, (v, w)) \mapsto
        \left\langle{s, (v \ljoin w)}\right\rangle,
         &             & s
        )
        \\
        \circuittomealy[
            \adjustimage{valign=c,margin=0pt,page=32}{tikzfigures}
        ]{\interpretation}
         & \coloneqq (
        \{s\},
         &             & (s, v) \mapsto
        \left\langle{s, ()}\right\rangle,
         &             & s
        )
        \\
        \circuittomealy[
            \adjustimage{valign=c,margin=0pt,page=74}{tikzfigures}
        ]{\interpretation}
         & \coloneqq
        (
        \{s_v, s_\bot\},
         &             & \{
        s_v \mapsto \left\langle{s_\bot,v}\right\rangle,
        s_\bot \mapsto \left\langle{s_\bot,\bot}\right\rangle
        \},
         &             & s_v
        )
        \\
        \circuittomealy[
            \adjustimage{valign=c,margin=0pt,page=75}{tikzfigures}
        ]{\interpretation}
         & \coloneqq
        (
        \{ s_v \,|\, v \in \values\},
         &             & (s_v, a) \mapsto \left\langle{v,s_a}\right\rangle,
         &             & s_\bot
        )
    \end{align*}
\end{defi}

\begin{exa}
    The action of \(\circuittomealy{\belnapinterpretation}\) on values and
    delays is illustrated in
    \autoref{fig:belnap-machines}.
\end{exa}

\begin{figure}
    \centering
    \begin{tikzpicture}[
        ->,
        >=stealth,
        node distance=0.25cm,
        every node/.style={font=\scriptsize},
        every state/.style={font=\normalsize, thick, fill=gray!10, minimum size=2pt, ellipse},
        initial text=$ $,
        initial distance=0.5cm,
        every initial by arrow/.style={*->},
        x=20pt,
        y=20pt
    ]
    \node[state, initial] (v) at (0,0) {\(s_v\)};
    \node[state] (bot) at (5,0) {\(s_\bot\)};

    \draw (v) edge[above] node{\( () \,|\, v\)} (bot);
    \draw (bot) edge[loop right, out=40, in=320, distance=1.5cm] node{\( () \,|\, \bot\)} (bot);
\end{tikzpicture}

    \vspace{1em}

    \begin{tikzpicture}[
        ->,
        >=stealth,
        node distance=0.25cm,
        every node/.style={font=\scriptsize},
        every state/.style={font=\normalsize, thick, fill=gray!10, minimum size=2pt, ellipse},
        initial text=$ $,
        initial distance=0.5cm,
        every initial by arrow/.style={*->},
        x=20pt,
        y=15pt
    ]
    \node[state] (bot) at (0,0) {\(s_\bot\)};
    \node[state] (f) at (0,-5) {\(s_\mathsf{f}\)};
    \node[state] (t) at (-5,-5) {\(s_\mathsf{t}\)};
    \node[state] (top) at (5,-5) {\(s_\top\)};

    \node[shape=circle,fill=black,inner sep=1.75pt,outer sep=0pt] (x) at (-1,2) {};

    \draw (bot) edge[bend right, pos=0.25, above] node{$ \mathsf{t} \,|\, \bot$} (t)
    (bot) edge[bend right, pos=0.25, left] node {$ \mathsf{f} \,|\, \bot$} (f)
    (bot) edge[bend left, pos=0.25, above] node{$ \top \,|\, \bot$} (top)
    (bot) edge[loop above] node{$ \bot \,|\, \bot$} (bot)
    (t) edge[loop left] node{$ \mathsf{t} \,|\, \mathsf{t}$} (t)
    (t) edge[bend left, pos=0.2, above] node{$ \mathsf{f} \,|\, \mathsf{t}$} (f)
    (t) edge[bend left=80, pos=0.25, above left] node{$ \bot \,|\, \mathsf{t}$} (bot)
    (t) edge[bend right=50, pos=0.2, right] node{$ \top \,|\, \mathsf{t}$} (top)
    (f) edge[loop below] node{$ \mathsf{f} \,|\, \mathsf{f}$} (f)
    (f) edge[bend left, pos=0.4, above] node{$ \top \,|\, \mathsf{f}$} (top)
    (f) edge[bend left, pos=0.2, above] node{$ \mathsf{t} \,|\, \mathsf{f}$} (t)
    (f) edge[bend right, pos=0.4, left] node{$ \bot \,|\, \mathsf{f}$} (bot)
    (top) edge[loop right] node{$ \top \,|\, \top$} (top)
    (top) edge[bend left=80, pos=0.2, below right] node{$ \mathsf{f} \,|\, \bot$} (t)
    (top) edge[bend left, pos=0.2, above] node{$ \mathsf{f} \,|\, \top$} (f)
    (top) edge[bend right=80, pos=0.25, above right] node{$ \bot \,|\, \top$} (bot)
    (x) edge (bot)
    ;
\end{tikzpicture}
    \caption{
        Mealy machines for gate-level values and delays
    }
    \label{fig:belnap-machines}
\end{figure}

\begin{exa}\label{ex:mealy-translation}
    Applying \(\circuittomealy{\belnapinterpretation}\) to the SR NOR latch from
    \autoref{ex:sr-latch} produces the monotone Mealy machine in
    \autoref{ex:trace-mealy}, which is illustrated in \autoref{fig:latch-machine}.
\end{exa}

\begin{figure}
    \centering
    \begin{tikzpicture}[
        bezier bounding box=true,
        ->,
        >=stealth,
        node distance=0.25cm,
        every node/.style={font=\scriptsize},
        every state/.style={font=\Large, thick, fill=gray!10, minimum size=2pt, ellipse},
        initial text=$ $,
        initial distance=1.5cm,
        every initial by arrow/.style={*->},
        every edge/.append style={},
        yscale=-1,
        x=20pt,
        y=17pt
    ]
    \node[state, initial below] (bot) at (0,0) {\(\bot\)};
    \node[state] (true) at (-5,5) {\(\mathsf{t}\)};
    \node[state] (top) at (5,5) {\(\top\)};
    \node[state] (false) at (0,10) {\(\mathsf{f}\)};
    \draw (bot) edge[loop, out=300, in=240, distance=2.1cm] node[above]{
            \(\trs{\bot\mathsf{t},\bot\top}{\bot\mathsf{f}}\)
        } (bot);
    \draw (bot) edge[loop, out=315, in=225, distance=3.6cm] node[above] {
            \(\trs{\bot\bot,\bot\mathsf{t},\mathsf{f}\bot,\mathsf{f}\mathsf{f}}{\bot\bot}\)
        } (bot);
    \draw (bot) edge node[right]{
            \(\trs{\mathsf{f}\mathsf{t},\mathsf{f}\top}{\bot\mathsf{f}}\)
        } (true);
    \draw (bot) edge[bend right] node[right] {
            \(\trs{\top\mathsf{t},\top\top}{\mathsf{f}\mathsf{f}}\)
        } (top);
    \draw (bot) edge[distance=9cm, out=0, in=0] node[very near start, above]{
            \(\trs{\mathsf{t}\mathsf{t},\mathsf{t}\top}{\bot\mathsf{f}}\)
        } (false);
    \draw (bot) edge[distance=10.5cm, out=340, in=20] node[near start, above,yshift=0.5cm]{
            \(\trs{\mathsf{t}\bot,\mathsf{f}\mathsf{f},\top\bot,\top\mathsf{f}}{\bot\bot}\)
        } (false);
    \draw (true) edge[loop left, distance=1cm, out=135, in=225] node[left]{
            \(\trs{\mathsf{f}v}{\mathsf{t}\mathsf{f}}\)
        } (true);
    \draw (true) edge[bend right] node[left]{
            \(\trs{\bot v}{\mathsf{t}\mathsf{f}}\)
        } (bot);
    \draw (true) edge[bend left] node[left]{
            \(\trs{\mathsf{t}v}{\mathsf{t}\mathsf{f}}\)
        } (false);
    \draw (true) edge node[above]{
            \(\trs{\top v}{\mathsf{t}\mathsf{f}}\)
        } (top);

    \draw (top) edge[loop right, distance=1cm, out=45, in=315] node[near start,yshift=-0.25cm,xshift=-0.25cm]{
            \(\trs{\top\bot,\top\mathsf{t}}{\top\mathsf{f}}\)
        } (top);
    \draw (top) edge[loop right, distance=5cm, out=55, in=305] node[near start, below, xshift=-0.5cm]{
            \(\trs{\mathsf{f}\mathsf{f},\mathsf{f}\top,\top\mathsf{f},\top\top}{\top\top}\)
        } (top);
    \draw (top) edge[bend left] node[below] {
            \(\trs{\mathsf{f}\top,\mathsf{t}\mathsf{t}}{\top\mathsf{t}}\)
        } (true);
    \draw (top) edge[left] node{
            \(\trs{\bot\bot,\bot\mathsf{t}}{\top\mathsf{f}}\)
        } (bot);
    \draw (top) edge[out=165,in=270,distance=2.5cm] node[left,near start,yshift=0.05cm]{
            \(\trs{\bot\mathsf{f},\bot\top,\mathsf{t}\mathsf{f},\mathsf{f}\top}{\top\top}\)
        } (false);
    \draw (top) edge[bend right] node[right]{
            \(\trs{\mathsf{t}\bot,\mathsf{t}\mathsf{t}}{\top\mathsf{f}}\)
        } (false);

    \draw (false) edge[loop, distance=1.25cm, out=80, in=100] node[below, yshift=0.1cm]{
            \(\trs{\mathsf{f}\mathsf{f}}{\mathsf{t}\mathsf{t}}\)
        } (false);
    \draw (false) edge[loop, distance=2.5cm, out=65, in=115] node[below, yshift=0.1cm]{
            \(\trs{\bot\mathsf{f},\mathsf{f}\mathsf{f},\mathsf{t}\mathsf{f},\top\mathsf{t}}{\mathsf{f}\mathsf{t}}\)
        } (false);
    \draw (false) edge[loop, distance=5cm] node[below]{
            \(\trs{\bot\top,\mathsf{f}\top,\mathsf{t}\top,\top\top}{\mathsf{f}\top}\)
        } (false);
    \draw (false) edge node[left]{
            \(\trs{\mathsf{f}\mathsf{t}}{\mathsf{f}\mathsf{f}}\)
        } (true);
    \draw (false) edge[out=160, in=200, distance=8cm] node[very near start, left, xshift=-0.5cm]{
            \(\trs{\bot\bot,\mathsf{f}\bot}{\mathsf{f}\bot}\)
        } (bot);
    \draw (false) edge[distance=6.5cm, out=180, in=180] node[near start, below, xshift=-0.33cm]{
            \(\trs{\bot\top}{\mathsf{f}\mathsf{f}}\)
        } (bot);
    \draw (false) edge[out=320, in=110] node[left]{
            \(\trs{\top\mathsf{t}}{\mathsf{f}\mathsf{f}}\)
        } (top);
    \draw (false) edge[bend left, out=350, in=190] node[left]{
            \(\trs{\top\top}{\mathsf{f}\top}\)
        } (top);
\end{tikzpicture}
    \caption{
        Mealy machine for the SR NOR latch
    }
    \label{fig:latch-machine}
\end{figure}

Mealy machines are a reasonable semantics for sequential circuits, but the
image of \(\circuittomealyi\) does not always lead to minimal Mealy machines,
and there are many Mealy machines that may correspond to the same behaviour.
The `purest' semantics of a sequential circuit is a stream function in
\(\streami\).

\begin{defi}
    Let \(\morph{\circuittostreami}{\scircsigma}{\streami}\) be defined as
    \(\mealytostreami \circ \circuittomealyi\).
\end{defi}

We have now finally established the \emph{denotation} of a sequential circuit \(
\adjustimage{valign=c,margin=0pt,page=89}{tikzfigures}
\): it is the stream function \(
\morph{\circuittostreami[\adjustimage{valign=c,margin=0pt,page=98}{tikzfigures}]}{\valuetuplestream{m}}{\valuetuplestream{n}}
\).
The existence of the PROP morphism \(\circuittostreami\) confirms that causal,
finitely specified and \(\bot\)-preserving monotone stream functions are a
\emph{sound} denotational semantics for sequential circuits: every circuit in
\(\scircsigma\) has a corresponding stream function in \(\streami\).

It is useful to verify that this denotational semantics of sequential circuits
agrees with the denotational semantics we defined earlier for
\emph{combinational} circuits in \autoref{sec:interpreting-components}.

\begin{lem}\label{lem:sequential-combinational-semantics}
    Let \(\adjustimage{valign=c,margin=0pt,page=36}{tikzfigures}\) be
    a combinational circuit; for \(\sigma \in \valuetuplestream{m}\) and
    \(i \in \nat\), \(
    \circuittostreami[\adjustimage{valign=c,margin=0pt,page=99}{tikzfigures}](\sigma)(i)
    =
    \circuittofunci[\adjustimage{valign=c,margin=0pt,page=99}{tikzfigures}](\sigma(i))
    \).
\end{lem}
\begin{proof}
    Since \(\adjustimage{valign=c,margin=0pt,page=36}{tikzfigures}\) is
    combinational, \(
    \circuittomealyi[
        \adjustimage{valign=c,margin=0pt,page=36}{tikzfigures}
    ]
    \) is a Mealy machine with a single state \(s\), i.e.\ there is a function
    \(\morph{g}{\valuetuple{m}}{\valuetuple{n}}\) such that  \(
    \circuittomealyi[
        \adjustimage{valign=c,margin=0pt,page=36}{tikzfigures}
    ] = (
    \{s\},
    (s, \listvar{v}) \mapsto \left\langle{s, g(\listvar{v})}\right\rangle,
    s
    )\).
    By definition of \(\mealytostreami\), we have that \(\mealytocircuiti[
        \circuittomealyi[
            \adjustimage{valign=c,margin=0pt,page=36}{tikzfigures}
        ]
    ](\sigma)(i) = g(\sigma(i))\).
    To complete the proof, we need to show that \(
    g(\sigma)(i) =
    \circuittofunci[\adjustimage{valign=c,margin=0pt,page=99}{tikzfigures}](\sigma(i))
    \); this holds because \(\circuittomealyi\) and \(\circuittofunci\) freely
    build functions using the same constructs.
\end{proof}

Using this idea, it will be convenient to have a mapping from functions to
these constant stream functions.

\begin{defi}
    Let \(\morph{\functostreami}{\funci}{\streami}\) be defined as the PROP
    morphism with action \(
    \functostreami[f] \coloneqq \sigma \mapsto i \mapsto f(\sigma)(i)
    \)
\end{defi}

\subsection{Monotone Mealy machines to circuits}

We now need a way to retrieve a circuit morphism in \(\scircsigma\) from a
stream function \(f \in \streami\).
To prevent us from picking an arbitrary circuit, the denotation of the circuit
must also be \(f\).

We already know by \autoref{cor:minimal-mealy} that given a stream function
\(f\) we can retrieve a monotone Mealy machine \(\streamtomealyi[f]\).
All that remains is to translate this into a circuit morphism.
For regular Mealy machines, there is a standard procedure in circuit
design~\cite{kohavi2009switching} in which each state of a Mealy machine is
\emph{encoded} as a power of values, and the Mealy function is interpreted as
a circuit using combinational logic.

\begin{exa}\label{ex:boolean-to-circuit}
    Consider the following Mealy machine operating on Boolean values.
    \begin{center}
        \begin{tikzpicture}[
        ->,
        >=stealth,
        node distance=0.25cm,
        every node/.style={font=\scriptsize},
        every state/.style={font=\normalsize, thick, fill=gray!10, minimum size=2pt, ellipse},
        initial text=$ $,
        initial distance=0.5cm,
        every initial by arrow/.style={*->},
        x=20pt,
        y=20pt
    ]
    \node[state, initial below] (s0) at (0,0) {\(s_0\)};
    \node[state] (s1) at (3,0) {\(s_1\)};

    \draw (s0) edge[bend left, above] node{\(\trs{\mathsf{f}}{\mathsf{t}}\)} (s1);
    \draw (s0) edge[out=135, in=225, loop, distance=1cm] node[left]{
            \(\trs{\mathsf{t}}{\mathsf{f}}\)
        } (s0);

    \draw (s1) edge[bend left, below] node{\(\trs{\mathsf{t}}{\mathsf{t}}\)} (s0);
    \draw (s1) edge[out=315, in=45, loop, distance=1cm] node[right]{
            \(\trs{\mathsf{f}}{\mathsf{f}}\)
        } (s1);
\end{tikzpicture}
    \end{center}
    \vspace{-\belowdisplayskip}
    To convert this machine to a circuit, we encode each state as a boolean
    value:
    in this case \(s_0 \mapsto \belnapfalse, s_1 \mapsto \belnaptrue\).
    We can now construct a truth table to show how a state and an input map to
    a transition and an output:
    \begin{center}
        \begin{tabular}{cc|cc}
            \(\belnapfalse\) & \(\belnapfalse\) & \(\belnaptrue\)  & \(\belnaptrue\)  \\
            \(\belnapfalse\) & \(\belnaptrue\)  & \(\belnapfalse\) & \(\belnapfalse\) \\
            \(\belnaptrue\)  & \(\belnapfalse\) & \(\belnaptrue\)  & \(\belnapfalse\) \\
            \(\belnaptrue\)  & \(\belnaptrue\)  & \(\belnapfalse\) & \(\belnaptrue\)  \\
        \end{tabular}
    \end{center}
    It is possible to describe these truth tables as logical expressions by
    creating a disjunction in which each clause corresponds to a row in the
    truth table that produces \(\belnaptrue\), and each clause is a conjunction
    of each input, negated if the corresponding column of the row in the table
    is \(\belnapfalse\).

    Applying this strategy to the table above, the expression for the next state
    is \(
    (v_0, v_1)
    \mapsto
    (\neg v_0 \land \neg v_1) \lor (v_0 \land \neg v_1)
    \) (which can further be simplified to \((v_0, v_1) \mapsto \neg v_1\))
    and the expression for the output is \(
    (v_0, v_1)
    \mapsto
    (\neg v_0 \land \neg v_1) \lor (v_0 \land v_1)
    \).
    These expressions can clearly be constructed as combinational circuits using
    \(\andgate\), \(\orgate\) and \(\notgate\) gates; the entire circuit
    corresponding to the Mealy machine is constructed by combining the
    combinational logic with registers to hold the state.
    \[\adjustimage{valign=c,margin=0pt,page=100}{tikzfigures}\]
    Using this strategy even complicated Mealy machines can be translated into
    circuits in a systematic way.
\end{exa}

We will use a variation of this procedure to map from \(\mealyi\) to
\(\scircsigma\).
However, when considering \emph{monotone} Mealy machines, this procedure must
additionally respect monotonicity as the combinational logic is constructed
using monotone components.
This means that an arbitrary encoding cannot be used; we will now show how to
select something suitable.

\begin{defi}[Encoding]\label{def:encoding}
    Let \(S\) be a set equipped with a partial order \(\stateorder\) and a total
    order \(\leq\) such that \(S\) can be represented as
    \(s_0 \leq s_1 \leq \dots s_{k-1}\).
    The \emph{\(\leq\)-encoding} for this assignment is a function
    \(\morph{\gamma_\leq}{S}{\valuetuple{k}}\) defined as
    \(\gamma_\leq(s)(i) \coloneqq \top\) if \(s_i \stateorder s\) and
    \(\gamma_\leq(s)(i) \coloneqq \bot\) otherwise.
\end{defi}

\begin{exa}
    Recall the monotone Mealy machine from \autoref{ex:mealy-translation}, which
    has state set \(
    \belnapvalues \coloneqq \{\bot,\belnapfalse,\belnaptrue,\top\}
    \).
    We choose the total order on \(\belnapvalues\) as
    \(\bot \leq \belnapfalse \leq \belnaptrue \leq \top\); subsequently, the
    \(\leq\)-encoding is defined as \(
    \bot \mapsto \top\bot\bot\bot, \belnapfalse \mapsto \top\top\bot\bot,
    \belnaptrue \mapsto \top\bot\top\bot, \top \mapsto \top\top\top\top
    \).
\end{exa}

It is essential that a \(\leq\)-encoding respects the original ordering of the
states.

\begin{lem}
    For an ordered state space \((S,\stateorder)\) and a \(\leq\)-encoding
    \(\gamma_\leq\), \(s \stateorder s^\prime\) if and only if
    \(\gamma_\leq(s) \sqsubseteq \gamma_\leq(s^\prime)\).
\end{lem}
\begin{proof}
    First the \(\onlyifdir\) direction.
    Let \(s_i, s_j \in S\) such that \(s_i \stateorder s_j\); we need to show
    that for every \(l < |S|\),
    \(\gamma_\leq(s_i)(l) \sqsubseteq \gamma_\leq(s_j)(l)\).
    The only way this can be violated is if \(\gamma_\leq(s_i)(l) = \top\) and
    \(\gamma_\leq(s_j)(l) = \bot\); this can only occur if
    \(s_l \stateorder s_i\) and \(s_l \not\stateorder s_j\).
    But since \(s_i \stateorder s_j\), this is a contradiction due to
    transitivity so \(\gamma_\leq(s_l) \sqsubseteq \gamma_\leq(s_j)\) also
    holds.

    Now the \(\ifdir\) direction.
    Assume that \(\gamma_\leq(s_i) \sqsubseteq \gamma_\leq(s_j)\); we need to
    show that \(s_i \stateorder s_j\); i.e.\ that \(\gamma_\leq(s_j)(i) = \top\)
    If \(\gamma_\leq(s_i) \sqsubseteq \gamma_\leq(s_j)\), then for each
    \(l < k\) then \(\gamma_\leq(s_i)(l) \sqsubseteq \gamma_\leq(s_j)(l)\);
    in particular \(\gamma_\leq(s_i)(i) \sqsubseteq \gamma_\leq(s_j)(i)\)
    By definition of \(\gamma_\leq\), \(\gamma_\leq(s_i)(i) = \top\), so if
    \(\gamma_\leq(s_i) \sqsubseteq \gamma_\leq(s_j)\) then
    \(\gamma_\leq(s_j)(i)\) is also \(\top\).
\end{proof}

Using this encoding, we will construct a circuit morphism that,
when interpreted as a function, implements the output and transition function
of the Mealy machine.
There is no reason for such a morphism to exist for an arbitrary interpretation:
why should we expect some collection of gates to be able to model every
function?
The useful interpretations are those that \emph{can} model every function.

\begin{defi}[Functional completeness]\label{def:functional-completeness}
    A \emph{complete interpretation} is a tuple
    \((\interpretation,\mealytofunc)\) in which \(\interpretation\) is an
    interpretation of a signature \(\signature\) and \(
    \morph{\mealytofunc}{\funci}{\scircsigma}
    \) is a map that sends functions \(
    \morph{f}{\valuetuple{m}}{\valuetuple{n}}
    \), to circuits of the form \(
    \adjustimage{valign=c,margin=0pt,page=101}{tikzfigures}
    \) for some word \(\listvar{v} \in \freemon{\values}\) such that
    \(\circuittostreami[\mealytofunc[f]](\sigma)(i) = f(\sigma(i))\).
\end{defi}

For a given complete interpretation \((\interpretation,\mealytofunc)\), we refer
to a circuit \(\mealytofunc[f]\) as the \emph{normalised circuit for \(f\)}.\

\begin{rem}
    Even though \(\mealytofunc\) maps combinational functions, its codomain is
    the category of \emph{sequential} circuits \(\scircsigma\).
    This is because instantaneous values may be required to create the
    normalised circuit.
    Despite the use of sequential components, the loop enforces that the state
    is \emph{constant}: it will always produce the word \(\listvar{v}\), so the
    the circuit still has combinational behaviour.

    Sometimes this is the only way to ensure every function can be modelled.
    For example, consider the Boolean function \(\booleans \to \booleans\) that
    always produces \(\belnapfalse\).
    Using the strategy from \autoref{ex:boolean-to-circuit}, no lines of the truth
    table are true, so the expression can only be defined using the unit of the
    disjunction, the constant false.

    Note also that this sequential component is by no means mandatory: the
    functional completeness map may actually map only to combinational circuits,
    in which case the width of the sequential component would be \(0\).
\end{rem}

\begin{exa}
    The gate-level interpretation from \autoref{ex:belnap-interpretation} is
    functionally complete, due to a variation of the standard functional
    completeness method for Boolean values.
    In the interests of space, we omit the proof and point the interested reader
    to \cite[Sec.\ 4.5]{kaye2024foundations}.
    A tool has been developed to generate gate-level expressions from functions
    and truth tables~\cite{kaye2026belnap}.
\end{exa}

With the knowledge that any monotone function has a corresponding circuit
in \(\scircsigma\), we set about encoding an arbitrary Mealy function
\(S \times \valuetuple{m} \to S \times \valuetuple{n}\) into a function
\(\valuetuple{k} \times \valuetuple{n} \to \valuetuple{k} \times \valuetuple{n}\).
One point to note here is that there may be more values in \(\valuetuple{k}\)
than there are states in \(S\), so we may need to `fill in the gaps' in a way
that is compatible with monotonicity.

\begin{defi}[Monotone completion]\label{def:monotone-completion}
    Let \(A\) be a finite poset and let \(B\) be a finite lattice such that
    \(A \subseteq B\).
    Then for another finite lattice \(C\) and a monotone function
    \(\morph{f}{A}{C}\), let the \emph{monotone \(B\)-completion} of \(f\) be
    the function \(\morph{f_\mathsf{m}}{B}{C}\) recursively defined as \[
        f_\mathsf{m}(v) = \begin{cases}
            f(v)
             &
            \text{if}\ v \in A
            \\
            \bot_C
             &
            \text{if}\ v = \bot_B, \bot \not\in A
            \\
            \bigvee \{ f_\mathsf{m}(w) \,|\, w \leq_B v, w \neq v \}
             &
            \text{otherwise}
        \end{cases}
    \]
\end{defi}

\begin{rem}
    Phrased more categorically, the monotone \(B\)-completion of
    \(\morph{f}{A}{C}\) can be viewed as the
    left Kan extension along the inclusion \(A \hookrightarrow B\).
\end{rem}

\begin{exa}
    For \(n \in \nat\), let \(\nat_{n}\) be the subset of the natural numbers
    containing the numbers \(0,1,\dots,n-1\) with the usual order.
    Let \(\morph{f}{\{2,4\}}{\nat}\) be defined as \(2 \mapsto 6\) and
    \(4 \mapsto 7\).
    The monotone \(\nat_5\)-completion of \(f\) is a function
    \(\morph{f_\mathsf{m}}{\nat_5}{\nat}\), defined as follows:
    \(f_\mathsf{m}(0) = 0\) as \(0\) is the least element of \(\nat_5\);
    \(f_\mathsf{m}(1) = 0\) as \(0 \leq 1\) and \(g_\mathsf{m}(1) = 0\);
    \(f_\mathsf{m}(2) = 6\) as \(2 \in \{2, 4\}\) and \(g(2) = 6\);
    \(f_\mathsf{m}(3) = 6\) because \(
    f_\mathsf{m}(3) =
    f_\mathsf{m}(0) \vee f_\mathsf{m}(1) \vee f_\mathsf{m}(2)
    = 0 \vee 0 \vee 6 = 6
    \); and \(f_\mathsf{m}(4) = 7\) because \(f(4) = 7\).
\end{exa}

A Mealy function can now be encoded over powers of \(\values\) by using the
monotone completion of some encoding function.

\begin{defi}[Monotone Mealy encoding]\label{def:mealy-encoding}
    For a monotone Mealy machine \((S, f, s_0)\) with \(k\) states and an
    encoding \(\morph{\gamma_\leq}{S}{\valuetuple{k}}\), let
    \(\morph{\gamma^p_\leq}{\gamma_\leq[S] \times \valuetuple{m}}{\valuetuple{k} \times \valuetuple{n}}\)
    be defined as the function \(
    (\gamma_\leq(s), \listvar{x}) \mapsto
    (\gamma_\leq(\mealyfunctiontransition{f}(s, \listvar{x})),
    \mealyfunctionoutput{f}(s, \listvar{x}))
    \).
    The \emph{monotone Mealy encoding} of \((S, f, s_0)\) is a function
    \(
    \morph{
        \gamma_\leq(f)
    }{
        \valuetuple{k} \times \valuetuple{m}
    }{
        \valuetuple{k} \times \valuetuple{n}
    }
    \) defined as the \((\valuetuple{k} \times \valuetuple{m})\)-completion of
    \(\gamma_\leq^p\).
\end{defi}

To obtain the syntactic circuit for a monotone Mealy function encoded
in this way, it needs to be a morphism in \(\funci\).
It is monotone by definition, but we need to make sure it is also
\(\bot\)-preserving.

\begin{lem}
    A monotone Mealy encoding is in \(\funci\).
\end{lem}
\begin{proof}
    A Mealy encoding is monotone as it is a monotone completion.
    There cannot be a state \(\bot^k\) since at least one bit must
    be \(\top\); this means the monotone completion will send the input
    \((\bot^k, \bot^m)\) to \((\bot^k, \bot^n)\): it is
    \(\bot\)-preserving.
\end{proof}

The foundations are now set for establishing the image of a PROP morphism from
Mealy machines to circuit terms.
There is one more thing to consider: \autoref{def:encoding} depends on some
arbitrary total ordering on the states in a given monotone Mealy machine.
While this may not seem much of an issue, when
defining a PROP morphism this must be \emph{fixed}, otherwise the mapping of
Mealy machines to circuits might be nondeterministic.

\begin{defi}[Chosen state order]
    Let \((S, f, s_0)\) be a monotone Mealy machine with input space
    \(\valuetuple{m}\), and let \(\leq\) be a total order on \(\values\);
    \(\leq\) can be extended to a total order \(\leq_\star\) on
    \(\freemon{(\values^m)}\) using the lexicographic order.
    Let \(S^\prime\) be the set of accessible states of \(S\).
    For each state \(s \in S^\prime\), let
    \(t_{s,\leq} \in \freemon{(\values^m)}\) be the minimal element of the
    subset of words that transition from \(s_0\) to \(s\), ordered by \(\leq\).
    Then the \emph{chosen state order} \(\leq_{S^\prime}\) is a total order on
    \(S^\prime\) defined as \(s \leq_{S^\prime} s^\prime\) if
    \(t_{s,\leq} \leq_\star t_{s^\prime,\leq}\).
\end{defi}

The PROP morphism from monotone Mealy machines to circuits can then be
parameterised by some ordering on the set of values \(\values\), ensuring that
there is a canonical term in \(\scircsigma\) for each monotone Mealy machine.

\begin{defi}\label{def:mealy-to-circuit}
    For a functionally complete interpretation \(\interpretation\) and total
    order \(\leq\) on \(\values\), let \(
    \morph{
        \mealytocircuiti
    }{
        \mealyi
    }{
        \scircsigma
    }
    \) be the traced PROP morphism with action defined for a monotone Mealy
    machine \((S,f,s)\) as producing \(
    \adjustimage{valign=c,margin=0pt,page=102}{tikzfigures}
    \).
\end{defi}

Before proceeding to the result that this PROP morphism is behaviour-preserving,
we must show a lemma linking the behaviour of circuits in the image of
\(\mealytocircuiti\) to initial outputs and stream derivatives.

\begin{prop}
    \label{prop:mealy-form-image}
    For a combinational circuit \(
    \adjustimage{valign=c,margin=0pt,page=103}{tikzfigures}
    \), let \(\mathsf{mf}(f)\) be the map with action \(
    (\listvar{s}) \mapsto
    \circuittostreami[
        \adjustimage{valign=c,margin=0pt,page=104}{tikzfigures}
    ]
    \) and let \(
    g
    :=
    \circuittofunci[
        \adjustimage{valign=c,margin=0pt,page=105}{tikzfigures}
    ]
    \).
    Then, \(
    \mealyoutput{\mathsf{mf}(f)(\listvar{s})}{\listvar{a}}
    =
    \proj{1}(g(\listvar{s}, \listvar{a}))
    \) and \(
    \mealytransition{\mathsf{mf}(f)(\listvar{s})}{\listvar{a}}
    =
    \mathsf{mf}(f)(\proj{0}(g(\listvar{s}, \listvar{a})))
    \).
\end{prop}
\begin{proof}
    The machine \(\circuittomealyi[
        \adjustimage{valign=c,margin=0pt,page=104}{tikzfigures}
    ]\) is computed as the fixed point of the machine \(
    \left(
    \valuetuple{x},
    (\listvar{r}, (\listvar{a}, \listvar{b}))
    \mapsto
    \left\langle
    \listvar{a}, g(\listvar{r}, \listvar{b})
    \right\rangle,
    \listvar{s}
    \right)
    \),
    which is
    \(
    \left(
    \valuetuple{x},
    \left(\listvar{r}, \listvar{b}\right)
    \mapsto
    \left\langle
    \proj{0}\mleft(g(\listvar{r}, \listvar{b})\mright),
    \proj{1}\mleft(g(\listvar{r}, \listvar{b})\mright)
    \right\rangle, \listvar{s}
    \right)
    \).
    The output and derivative of \(
    \mealytocircuiti[
        \circuittomealyi[
            \adjustimage{valign=c,margin=0pt,page=104}{tikzfigures}
        ]
    ]
    \) are the output and transition of the Mealy machine, so the original
    statement holds by \autoref{lem:sequential-combinational-semantics}.
\end{proof}

The goal of this section is to show that the translation from Mealy machines to
circuits and back again using \(\circuittomealyi \circ \mealytocircuiti\) is
\emph{behaviour-preserving}: while the mapping may not be the identity in
\(\mealyi\), the stream functions obtained using \(\streamtomealyi\) should
be equal.
This is an important new result, as it means that rather than showing
results about the denotational semantics of circuits in \(\scircsigma\) by
interpreting them in \(\streami\), we can view morphisms of the former as
Mealy machines instead.

\begin{thm}\label{thm:mealy-to-circuit}
    \(
    \mealytostream = \circuittostreami \circ \mealytocircuiti
    \).
\end{thm}
\begin{proof}
    Let \((S ,f)\) be a monotone Mealy machine and let \(s \in S\) be an
    arbitrary state.
    By \autoref{prop:mealy-to-stream}, the initial output of
    \(\mealytostreami[(S, f, s)]\) is
    \(\listvar{a} \mapsto \mealyfunctionoutput{f}\mleft(s, \listvar{a}\mright)\)
    and the stream derivative of \(\mealytostreami[(S, f, s)]\) is \(
    \listvar{a}
    \mapsto
    \mealytostreami[(S, f, \mealyfunctiontransition{f}\mleft(s, \listvar{a}\mright))]
    \).

    Now we consider the composite
    \(\circuittostreami[\mealytocircuiti[(S, f, s)]]\).
    By \autoref{def:mealy-to-circuit} we have that \(
    \mealytocircuiti[(S, f, s_0)]
    =
    \adjustimage{valign=c,margin=0pt,page=102}{tikzfigures}
    \); by applying \(\mealytocircuiti\) and \(\mealytofunc\),
    there exists a combinational circuit \(
    \adjustimage{valign=c,margin=0pt,page=106}{tikzfigures}
    \) such that \[
        \mealytocircuiti[(S, f, s_0)]
        =
        \adjustimage{valign=c,margin=0pt,page=102}{tikzfigures}
        =
        \adjustimage{valign=c,margin=0pt,page=107}{tikzfigures}.
    \]
    Let \(
    g^\prime
    =
    \circuittofunci[\adjustimage{valign=c,margin=0pt,page=106}{tikzfigures}]
    \); note that for all \(\listvar{r} \in \valuetuple{x}\) and
    \(\listvar{a} \in \valuetuple{m}\),
    \(g^\prime(\listvar{v}, \listvar{r}, \listvar{a})
    =
    \gamma_\leq(f)(\listvar{r},\listvar{a})
    \).

    We can now use \autoref{prop:mealy-form-image} to compute the initial output
    and stream derivative of \(\circuittostreami[\mealytocircuiti[(S, f, s)]]\).
    To show that \(\mealytostreami = \circuittostreami \circ \mealytocircuiti\),
    we need to show that these `agree' with those of
    \(\mealytostreami[(S, f, s)]\).
    For the initial outputs, we just need to show they are equal:
    \begin{align*}
        \initialoutput{\circuittostreami[\mealytocircuiti[(S, f, s)]]}{\listvar{a}}
         & =
        \proj{1}\mleft(\circuittofunci[g](\listvar{v}, \gamma_\leq(s), \listvar{a})\mright)
        \\
         & =
        \proj{1}\mleft(\circuittofunci[\mealytofunc[\gamma_\leq(f)]](\gamma_\leq(s), \listvar{a})\mright)
        \\
         & =
        \proj{1}\mleft(\gamma_\leq(f)(\gamma_\leq(s), \listvar(a))\mright)
        \\
         & =
        \proj{1}\left(
        \gamma_\leq(
            \mealyfunctiontransition{f}(s, \listvar{a})),
        \mealyfunctionoutput{f}(s, \listvar{a})\right)
        \\
         & =
        \mealyfunctionoutput{f}(s, \listvar{a})
    \end{align*}
    For the stream derivatives, we need to show that as states vary over
    \(s \in S\), the stream derivative of \(
    \circuittostreami[\mealytocircuiti[(S, f, s)]]
    \) is the \(\gamma_\leq\)-encoding of \(\mealytostreami[(S, f, s)]\).
    \begin{align*}
        \streamderivative{\left(\circuittostreami[\mealytocircuiti[(S, f, s)]]\right)}{\listvar{a}}
         & =
        \proj{0}\left(
        \circuittofunci[g](\listvar{v}, \gamma_\leq(s), \listvar{a})
        \right)
        \\
         & =
        \proj{0}\mleft(\circuittofunci[\mealytofunc[\gamma_\leq(f)]](\gamma_\leq(s), \listvar{a})\mright)
        \\
         & =
        \proj{0}\mleft(\gamma_\leq(f)(\gamma_\leq(s), \listvar(a))\mright)
        \\
         & =
        \proj{0}\left(
        \gamma_\leq(
            \mealyfunctiontransition{f}(s, \listvar{a})),
        \mealyfunctionoutput{f}(s, \listvar{a})\right)
        \\
         & =
        \gamma_\leq(
        \mealyfunctiontransition{f}(s, \listvar{a}))
    \end{align*}
    The initial outputs and stream derivatives agree, so
    \(\mealytostream = \circuittostreami \circ \mealytocircuiti\).
\end{proof}

On its own, this is a nice result to have; if we only know the specification of
a circuit in terms of a (monotone) Mealy machine, we can use the PROP morphism
\(\mealytocircuiti\) to generate a circuit in \(\scircsigma\) which has the
same behaviour as a stream function.
However, this is but one ingredient in our ultimate goal: the completeness of
the denotational semantics.

\subsection{Completeness of the denotational semantics}\label{sec:denotational-completeness}

We want \(\streami\) to be a \emph{complete} denotational semantics for digital
circuits.
This means that for every stream function \(f \in \streami\), there must be at
least one one circuit in \(\scircsigma\) such that its behaviour under
\(\interpretation\) is \(f\).

\begin{cor}\label{thm:circuit-stream-correspondence}
    \(
    \circuittostreami
    \circ
    \mealytocircuiti
    \circ
    \streamtomealyi
    =
    \id[\streami]
    \).
\end{cor}
\begin{proof}
    This follows immediately from \autoref{thm:mealy-to-circuit} and
    \autoref{prop:mealy-stream-id}, as we have that \(
    \circuittostreami
    \circ
    \mealytocircuiti
    \circ
    \streamtomealyi
    =
    \mealytostreami
    \circ
    \streamtomealyi
    =
    \id[\streami]
    \).
\end{proof}

There is no isomorphism between \(\scircsigma\) and \(\streami\)
as many circuits may have the same semantics but different syntax.
Instead, we can work in \emph{equivalence classes} of syntactic circuits based
on their behaviour in \(\streami\).

\begin{defi}[Denotational equivalence]
    Two sequential circuits are \emph{denotationally equivalent}
    under \(\interpretation\), written \(
    \adjustimage{valign=c,margin=0pt,page=89}{tikzfigures}
    \approx_{\interpretation}
    \adjustimage{valign=c,margin=0pt,page=108}{tikzfigures}
    \) if \(
    \circuittostream[
        \adjustimage{valign=c,margin=0pt,page=98}{tikzfigures}
    ]{\interpretation}
    =
    \circuittostream[
        \adjustimage{valign=c,margin=0pt,page=109}{tikzfigures}
    ]{\interpretation}
    \).
    Let \(\scircsigmai\) be the result of quotienting \(\scircsigma\) by \(
    \approx_{\interpretation}
    \).
\end{defi}

Every morphism in \(\scircsigmai\) is a class of circuits that share the
same behaviour under \(\interpretation\).
As we have a PROP morphism \(\mealytocircuiti \circ \streamtomealyi\), we know
that for every such behaviour there must be at least one such syntactic circuit,
and subsequently exactly one equivalence class \(\scircsigmai\).
Using \autoref{thm:circuit-stream-correspondence}, we know that all the circuits in
this equivalence class have the same behaviour as the original stream function,
so we can derive an isomorphism between \(\scircsigmai\) and
\(\streami\).

\begin{cor}
    \(\scircsigmai \cong \streami\).
\end{cor}

This gives us, for the first time, a fully compositional, sound and complete,
denotational semantics for sequential circuits with delay and (possibly
non-delay-guarded) feedback.
This formal model will serve as a backdrop against the operational and
algebraic semantics presented in the remainder of this paper.

\subsection{Denotational semantics for generalised circuits}

Although we have discussed the denotational semantics in terms of circuit
signatures with only a single width of wires, it is straightforward to extend
the results to categories
generated over \emph{generalised} circuit signatures

In the semantic categories \(\funci\), \(\streami\), and \(\mealyi\), the
morphisms are all variants on functions of the form
\(\valuetuple{m} \to \valuetuple{n}\) that operate on powers of elements in
\(\values\): one element for each (single-bit) input or output wire.
In the generalised setting, these input and output wires may not all be the
same width, so the input and output sets must be \emph{powers of powers} of
values.

\begin{nota}
    Given a set \(A\) and a word \(\listvar{v} \in \natplus^\star\) of
    length \(n\), we write \(
    A^{\listvar{v}}
    \coloneqq
    A^{\listvar{v}(0)}
    \times
    A^{\listvar{v}(1)}
    \times
    \dots
    \times
    A^{\listvar{v}(n-1)}
    \).
\end{nota}

Note that for a word \(\listvar{m} \coloneqq 11\dots1\) of length \(k\), the set
\(A^{\listvar{m}} = A^1 \times A^1 \times \dots A^1\) is isomorphic to \(A^k\),
much like how setting the set of colours in a coloured PROP to the singleton
recovers a PROP with a single colour.

The semantic categories can now be extended to these \emph{coloured} interfaces.

\begin{defi}
    Let \(\funcig\) be the \(\natplus\)-coloured PROP in which the morphisms
    \(\listvar{m} \to \listvar{n}\) are the monotone functions
    \(\valuetuple{\listvar{m}} \to \valuetuple{\listvar{n}}\).
    Let \(\streamig\) be the \(\natplus\)-coloured PROP in which the morphisms
    \(\listvar{m} \to \listvar{n}\) are the causal, monotone, and finitely
    specified stream functions \(
    \valuetuplestream{\listvar{m}} \to \valuetuplestream{\listvar{n}}
    \).
    Let \(\mealyig\) be the \(\natplus\)-coloured PROP in which the morphisms
    \(\listvar{m} \to \listvar{n}\) are the monotone
    \((\valuetuple{\listvar{m}}, \valuetuple{\listvar{n}})\)-Mealy machines.
\end{defi}

The various PROP morphisms between these categories are defined in a similar way
to the single-width versions, but now we have to account for the structural
generators for each \(n \in \natplus\), as well as the bundlers.

\begin{defi}
    Let \(\morph{\circuittofuncig}{\ccircsigmag}{\funcig}\) be the coloured PROP
    morphism with action defined as in \autoref{fig:circuittofuncig}.
    \begin{figure}
        \begin{align*}
            \circuittofuncig[
                \adjustimage{valign=c,margin=0pt,page=110}{tikzfigures}
            ]
             & \coloneqq
            \gateinterpretation[p]
            \\[0.75em]
            \circuittofuncig[
                \adjustimage{valign=c,margin=0pt,page=38}{tikzfigures}
            ]
             & \coloneqq
            (\listvar{v}) \mapsto (\listvar{v}, \listvar{v})
            \\[0.75em]
            \circuittofuncig[
                \adjustimage{valign=c,margin=0pt,page=40}{tikzfigures}
            ]
             & \coloneqq
            (\listvar{v}) \mapsto ()
            \\[0.75em]
            \circuittofuncig[
                \adjustimage{valign=c,margin=0pt,page=39}{tikzfigures}
            ]
             & \coloneqq
            (\listvar{v}, \listvar{w})
            \mapsto \listvar{v} \sqcup \listvar{w}
            \\[0.75em]
            \circuittofuncig[
                \adjustimage{valign=c,margin=0pt,page=37}{tikzfigures}
            ]
             & \coloneqq
            () \mapsto \bot^n
            \\[0.75em]
            \circuittofuncig[
                \adjustimage{valign=c,margin=0pt,page=92}{tikzfigures}
            ]
             & \coloneqq
            (\listvar{v}) \mapsto (\listvar{v}(0), \listvar{v}(1), \dots, \listvar{v}(n-1))
            \\[0.75em]
            \circuittofuncig[
                \adjustimage{valign=c,margin=0pt,page=93}{tikzfigures}
            ]
             & \coloneqq
            (v_0, v_1, \dots, v_{n-1}) \mapsto (v_0v_1\dots v_{n-1})
        \end{align*}
        \caption{Action of the PROP morphism \(\circuittofuncig\)}
        \label{fig:circuittofuncig}
    \end{figure}
\end{defi}

The map from coloured circuits to Mealy machines proceeds in a similar
manner.

\begin{defi}
    Let \(\morph{\circuittomealyig}{\scircsigmag}{\mealyig}\) be the traced PROP
    morphism defined on generators as in \autoref{fig:circuittomealyig}.
    \begin{figure}
        \begin{align*}
            \circuittomealyig[
                \adjustimage{valign=c,margin=0pt,page=97}{tikzfigures}
            ]
             & \coloneqq
            (\{s\},
             &             & (\listvar{v}_0,\dots,\listvar{v}_{m-1}) \mapsto
            \left\langle
            s,
            \gateinterpretation[g]\left(
            \listvar{v}_0,\listvar{v}_1,\dots,\listvar{v}_{m-1}
            \right)
            \right\rangle,
             &             & s)
            \\
            \circuittomealyig[
                \adjustimage{valign=c,margin=0pt,page=38}{tikzfigures}
            ]
             & \coloneqq (
            \{s\},
             &             & \listvar{v}
            \mapsto
            \left\langle s, (\listvar{v},\listvar{v})\right\rangle,
             &             & s
            )
            \\
            \circuittomealyig[
                \adjustimage{valign=c,margin=0pt,page=39}{tikzfigures}
            ]
             & \coloneqq (
            \{s\},
             &             & (\listvar{v}, \listvar{w}) \mapsto
            \left\langle s, \listvar{v} \ljoin \listvar{w}\right\rangle,
             &             & s
            )
            \\
            \circuittomealyig[
                \adjustimage{valign=c,margin=0pt,page=40}{tikzfigures}
            ]
             & \coloneqq
            (
            \{s\},
             &             & \listvar{v} \mapsto
            \left\langle s, s\right\rangle,
             &             & s
            )
            \\
            \circuittomealyig[
                \adjustimage{valign=c,margin=0pt,page=92}{tikzfigures}
            ]
             & \coloneqq
            (
            \{s\},
             &             & (v_0,v_1,\dots,v_{n-1}) \mapsto ((v_0), (v_1), \dots, (v_{n-1})),
             &             & s
            )
            \\
            \circuittomealyig[
                \adjustimage{valign=c,margin=0pt,page=93}{tikzfigures}
            ]
             & \coloneqq
            (
            \{s\},
             &             & ((v_0), (v_1), \dots, (v_{n-1})) \mapsto (v_0,v_1,\dots,v_{n-1}),
             &             & s
            )
            \\
            \circuittomealyig[
                \adjustimage{valign=c,margin=0pt,page=76}{tikzfigures}
            ]
             & \coloneqq
            (
            \{s_{\listvar{v}}, s_\bot\},
             &             & \{
            s_{\listvar{v}} \mapsto \left\langle{s_\bot,\listvar{v}}\right\rangle,
            s_\bot \mapsto \left\langle{s_\bot,\bot}\right\rangle
            \},
             &             & s_{\listvar{v}}
            )
            \\
            \circuittomealyig[
                \adjustimage{valign=c,margin=0pt,page=83}{tikzfigures}
            ]
             & \coloneqq
            (
            \{ s_{\listvar{v}} \,|\, \listvar{v} \in \valuetuple{n}\},
             &             & (s_{\listvar{v}}, \listvar{w}) \mapsto
            \left\langle{s_{\listvar{w}},\listvar{v}}\right\rangle,
             &             & s_{\bot^n}
            )
        \end{align*}
        \caption{Action of the PROP morphism \(\circuittomealyig\)}
        \label{fig:circuittomealyig}
    \end{figure}
\end{defi}

Although morphisms in \(\mealyig\) have different interfaces to those in
\(\streamig\), they are still monotone Mealy machines so it is simple to
translate them into stream functions or coloured circuits.

\begin{defi}
    Let the three coloured PROP morphisms
    \(\morph{\mealytostreamig}{\mealyig}{\streamig}\),
    \(\morph{\streamtomealyig}{\streamig}{\mealyig}\) and
    \(\morph{\mealytocircuitig}{\mealyig}{\scircsigmag}\) be defined as before,
    and let \(\morph{\circuittostreamig}{\scircsigmag}{\streamig}\)
    be defined as \(\mealytostreamig \circ \circuittomealyig\).
\end{defi}

By putting all these coloured PROP morphisms together, we can show the same
results as we did in the previous section.

\begin{thm}
    \(\mealytostreamig = \circuittostreamig \circ \mealytostreamig\) and
    \(\circuittostreamig \circ \mealytocircuitig \circ \streamtomealyig =
    \id[\streamig]\).
\end{thm}

As before, we derive a notion of denotational equivalence for generalised
circuits.

\begin{defi}
    Two generalised sequential circuits are \emph{denotationally equivalent}
    under \(\interpretation\), written \(
    \adjustimage{valign=c,margin=0pt,page=111}{tikzfigures}
    \approx_{\interpretation}^+
    \adjustimage{valign=c,margin=0pt,page=112}{tikzfigures}
    \) if \(
    \circuittostreamig[
        \adjustimage{valign=c,margin=0pt,page=98}{tikzfigures}
    ]
    =
    \circuittostreamig[
        \adjustimage{valign=c,margin=0pt,page=109}{tikzfigures}
    ]
    \).
    Let \(\scircsigmaig\) be the result of quotienting \(\scircsigma\) by \(
    \approx_{\interpretation}^+
    \).
\end{defi}

\begin{cor}
    \(\scircsigmaig \cong \streamig\).
\end{cor}

\section{Operational semantics}

The behaviour of two circuits in \(\scircsigma\) can be compared by examining
their corresponding stream functions.
However, translating a circuit into a stream function obscures the internal
structure of the circuit, much like representing a combinational circuit by its
truth table.
To better relate behaviour and structure in circuits we now define an
\emph{operational semantics} which evaluates circuits in a stepwise manner.
An informal operational semantics was presented in~\cite{ghica2017diagrammatic}
but only for the case of \emph{closed} circuits with \emph{delay-guarded}
feedback; in this section we drop these two requirements and present a sound
and complete operational semantics for \emph{all} sequential circuits.

An operational semantics is defined in terms of \emph{reductions}.
Here we are motivated by \emph{mechanising} circuit reduction;
for a set of reductions to be suitable then there should be a terminating
strategy for evaluating inputs to circuits.

\begin{nota}[Reduction]
    A \emph{reduction} from circuit \(
    \adjustimage{valign=c,margin=0pt,page=89}{tikzfigures}
    \) to \(
    \adjustimage{valign=c,margin=0pt,page=108}{tikzfigures}
    \), is denoted \(
    \adjustimage{valign=c,margin=0pt,page=89}{tikzfigures}
    \reduction
    \adjustimage{valign=c,margin=0pt,page=108}{tikzfigures}
    \).
    If there are reductions \(
    \adjustimage{valign=c,margin=0pt,page=89}{tikzfigures}
    \reduction
    \adjustimage{valign=c,margin=0pt,page=108}{tikzfigures}
    \reduction
    \cdots
    \reduction
    \adjustimage{valign=c,margin=0pt,page=113}{tikzfigures}
    \), we write \(
    \adjustimage{valign=c,margin=0pt,page=89}{tikzfigures}
    \reductions
    \adjustimage{valign=c,margin=0pt,page=113}{tikzfigures}
    \).
    A reduction \(
    \adjustimage{valign=c,margin=0pt,page=89}{tikzfigures}
    \reduction
    \adjustimage{valign=c,margin=0pt,page=108}{tikzfigures}
    \) is \emph{sound} if we have that \(
    \circuittostreami[
        \adjustimage{valign=c,margin=0pt,page=89}{tikzfigures}
    ]
    =
    \circuittostreami[
        \adjustimage{valign=c,margin=0pt,page=108}{tikzfigures}
    ].
    \)
\end{nota}

A reduction is effectively a directed equation that can be applied to circuits
in \(\scircsigma\), usually transforming a circuit into a simpler one, for
example by applying a gate to some input values.
This means that the only equations that hold `on the nose' are the axioms of
STMCs; diagrams can be deformed in order to expose redexes.
Any other reductions must be specified explicitly; in this section we will
present a set of reduction rules with which we can evaluate the streams of
inputs applied to circuits in a step-by-step, syntactic manner.

\subsection{Feedback}\label{sec:feedback}

One of the major issues that comes with trying to reduce circuits in
\(\scircsigma\) is the presence of feedback.
Without proper attention, one could end up infinitely unfolding, not producing
any output values.
The first portion of our operational semantics revolves around some
\emph{global transformations} to make a circuit suitable for reduction.

The first observation we make does not even need anything new to be defined as
it follows immediately from axioms of STMCs.

\begin{lem}[Global trace-delay form]\label{lem:trace-delay}
    For a sequential circuit \(
    \adjustimage{valign=c,margin=0pt,page=89}{tikzfigures}
    \) there exists a combinational circuit \(
    \adjustimage{valign=c,margin=0pt,page=114}{tikzfigures}
    \) and \(\overline{v} \in \valuetuple{z}\) such that \(
    \adjustimage{valign=c,margin=0pt,page=89}{tikzfigures}
    =
    \adjustimage{valign=c,margin=0pt,page=115}{tikzfigures}
    \) by axioms of STMCs.
\end{lem}
\begin{proof}
    By applying the axioms of traced categories; any trace can be `pulled'
    to the outside of a term by superposing and tightening.
    For the delays, a trace can be introduced using yanking and then the
    same procedure as above followed.
\end{proof}

\begin{exa}
    The SR NOR latch circuit from \autoref{ex:sr-latch} is assembled into global
    trace-delay form in \autoref{fig:sr-latch-global-trace-delay}.
\end{exa}

\begin{figure}
    \centering
    \adjustimage{valign=c,margin=0pt,page=116}{tikzfigures}
    \caption{
        The SR NOR latch from \autoref{ex:sr-latch} in global trace-delay form
    }
    \label{fig:sr-latch-global-trace-delay}
\end{figure}

This form is evocative of what we saw when mapping from Mealy machines to
circuits in the previous section, but rather than the state being determined by
one word, the instantaneous values and the delays are kept separate.

\begin{defi}[Pre-Mealy form]\label{def:pre-mealy}
    A sequential circuit is in \emph{pre-Mealy form} if it is in the form \(
    \adjustimage{valign=c,margin=0pt,page=117}{tikzfigures}
    \).
\end{defi}

Our first reduction transforms a circuit from global trace-delay form to
pre-Mealy form.

\begin{lem}\label{lem:mealy-rule}
    The following rule is sound: \[
        \adjustimage{valign=c,margin=0pt,page=118}{tikzfigures}
        \reduction[(\mealyeqn)]
        \adjustimage{valign=c,margin=0pt,page=119}{tikzfigures}
    \]
\end{lem}
\begin{proof}
    It is a simple exercise to check the corresponding stream functions.
\end{proof}

\begin{figure}
    \centering
    \adjustimage{valign=c,margin=0pt,page=120}{tikzfigures}
    \caption{
        Applying the \((\mealyeqn)\) rule to the circuit in
        \autoref{fig:sr-latch-global-trace-delay}
    }
    \label{fig:sr-latch-pre-mealy}
\end{figure}

By assembling a circuit into global trace-delay form and
applying the \((\mealyeqn)\) rule, we can construct a word \(\listvar{s}\)
from the juxtaposition of \(\bot\) elements for each register combined with
the instantaneous values \(\listvar{v}\) i.e.\ using the notation of the above
lemma \(\listvar{s} \coloneqq \bot^y\listvar{v}\).
This word represents the initial state of the circuit, but it is by no means
unique: it depends on how the circuit is put into global trace-delay form.
What matters most is that we \emph{can} do it.

\begin{cor}
    For any sequential circuit \(
    \adjustimage{valign=c,margin=0pt,page=89}{tikzfigures}
    \), there exists at least one valid application of the Mealy rule.
\end{cor}

\begin{exa}
    The result of applying the \((\mealyeqn)\) rule to the circuit in
    \autoref{fig:sr-latch-global-trace-delay} is shown in
    \autoref{fig:sr-latch-pre-mealy}.
    Here the initial state word is just \(\bot\).
\end{exa}

The result of applying the \((\mealyeqn)\) reduction still differs from the
image of \(\mealytocircuiti\) as it may have a trace with no delay on it: an
instance of \emph{non-delay-guarded feedback}.

The mere mention of non-delay-guarded feedback may trigger alarm bells in
the minds of those well-acquainted with circuit design.
It is often common in industry to enforce that circuits have no
non 5.13-delay-guarded feedback; one might ask if we should also enforce this
tenet in order to stick to `well-behaved' circuits.

\begin{rem}
    \emph{Categories with feedback}~\cite{katis2002feedback} are a weakening of
    traced categories that remove the yanking axiom: this effectively makes all
    traces delay-guarded.
    \emph{Categories with delayed trace}~\cite{sprunger2019differentiable}
    weaken this further by removing the sliding axiom, so no components can be
    `pushed round' into the next tick of execution.
    Neither of these are suitable for us as we actually \emph{want} to allow
    non-delay-guarded feedback.
\end{rem}

In fact, careful use of non-delay-guarded feedback can still result in useful
circuits as a clever way of sharing
resources~\cite{malik1994analysis,riedel2004cyclic,mendler2012constructive}.
The minimal circuit to implement a function often \emph{must} be
constructed using cycles~\cite{rivest1977necessity,riedel2003synthesis}.

\begin{exa}\label{ex:cyclic-combinational}
    A particularly famous circuit~\cite{malik1994analysis} which is useful
    despite the presence of non-delay-guarded feedback is shown in
    \autoref{fig:cyclic-combinational}, where \(
    \adjustimage{valign=c,margin=0pt,page=99}{tikzfigures}
    \) and \(
    \adjustimage{valign=c,margin=0pt,page=121}{tikzfigures}
    \) are arbitrary combinational circuits.
    The trapezoidal gate is a \emph{multiplexer}; it has a vertical
    \emph{control} input and two horizontal \emph{data} inputs.
    The multiplexer is defined as \(
    \adjustimage{valign=c,margin=0pt,page=122}{tikzfigures}
    \coloneqq
    \adjustimage{valign=c,margin=0pt,page=123}{tikzfigures}
    \).
    The multiplexer is effectively an if statement: when the control is
    \(\belnapfalse\) the output is the first data input and when it is
    \(\belnaptrue\) the output is the second data input.

    The circuit in \autoref{fig:cyclic-combinational} has no state and its trace is
    global so it is already in pre-Mealy form, and has
    non-delay-guarded feedback.
    Despite this, it produces useful output when the control signal is true or
    false:
    when the control signal is \(\belnapfalse\) then the behaviour of the
    circuit is \(
    \circuittofunc[
        \adjustimage{valign=c,margin=0pt,page=124}{tikzfigures}
    ]{\belnapinterpretation}
    \) and when the control is \(\belnaptrue\) then the behaviour is \(
    \circuittofunc[
        \adjustimage{valign=c,margin=0pt,page=125}{tikzfigures}
    ]{\belnapinterpretation}
    \).
    The feedback is just used as a clever way to share circuit components.
\end{exa}

\begin{figure}
    \centering
    \adjustimage{valign=c,margin=0pt,page=126}{tikzfigures}
    \quad
    \adjustimage{valign=c,margin=0pt,page=127}{tikzfigures}
    \caption{
        A useful cyclic combinational circuit
        \cite[Fig. 1]{mendler2012constructive}, and a possible interpretation in
        \(\scircsigma\).
    }
    \label{fig:cyclic-combinational}
\end{figure}

A combinational circuit surrounded by non-delay-guarded feedback still
implements a function, as there are no delay components.
Nevertheless, non-delay-guarded feedback does still block our path to future
transformations, so it must be eliminated.
Using a methodology also employed by~\cite{riedel2012cyclic}, we turn to the
Kleene fixed-point theorem.

\begin{lem}\label{lem:monotone-fixpoint}
    For a monotone function \(\morph{f}{\values^{n+m}}{\valuetuple{n}}\) and
    \(i \in \nat\), let \(\morph{f^i}{\valuetuple{m}}{\valuetuple{n}}\) be
    defined as \(f^0(x)  = f(\bot^n,x)\) and \(f^{k+1}(x) = f(f^k(x), x)\).
    Let \(c\) be the length of the longest chain in the lattice
    \(\valuetuple{n}\).
    Then, for \(j > c\), \(f^c(x) = f^{j}(x)\).
\end{lem}
\begin{proof}
    Since \(f\) is monotone and \(\values^{n}\) is finite, the function
    \(\listvar{s} \mapsto f(\listvar{s}, \listvar{v})\) for arbitrary
    \(\listvar{v} \in \values{m}\) has a least fixed point by the Kleene
    fixed-point theorem.
    This will either be some word \(\listvar{r} \in \valuetuple{m}\) or the
    maximal element \(\top^m\).
    The most iterations of \(f\) it would take to obtain this fixed point is
    \(c\), i.e.\ the function produces a value one step up the lattice each
    time.
\end{proof}

\begin{defi}[Iteration]\label{def:iteration}
    Given a combinational circuit \(
    \adjustimage{valign=c,margin=0pt,page=128}{tikzfigures},
    \)
    its \emph{\(i\)-th iteration} \(
    \adjustimage{valign=c,margin=0pt,page=129}{tikzfigures}
    \) is defined inductively over \(i\) in the following way: \[
        \adjustimage{valign=c,margin=0pt,page=130}{tikzfigures}
        \coloneqq
        \adjustimage{valign=c,margin=0pt,page=131}{tikzfigures}
        \qquad
        \adjustimage{valign=c,margin=0pt,page=132}{tikzfigures}
        \coloneqq
        \adjustimage{valign=c,margin=0pt,page=133}{tikzfigures}
    \]
\end{defi}

The trace in \(\streami\) is by the least fixed point, computed by repeatedly
applying \(f\) to itself starting from \(\bot\).
The above lemma gives a fixed upper bound for the number of times we need to
iterate \(f\) to reach this fixed point, based on the size of the lattice.
We can use this in the syntactic setting.

\begin{defi}[Unrolling]\label{def:unrolling}
    For an interpretation with values \(\values\), the \emph{unrolling}
    of a combinational circuit \(
    \adjustimage{valign=c,margin=0pt,page=128}{tikzfigures}
    \), written \(
    \adjustimage{valign=c,margin=0pt,page=134}{tikzfigures}
    \), is defined as \(
    \adjustimage{valign=c,margin=0pt,page=135}{tikzfigures}
    \) where \(c\) is the length of the longest chain in \(\valuetuple{x}\).
\end{defi}

Using these constructs we can eliminate non-delay-guarded feedback around a
combinational circuit.

\begin{prop}\label{prop:instant-feedback}
    The instant feedback rule \(
    \adjustimage{valign=c,margin=0pt,page=136}{tikzfigures}
    \reduction[(\instantfeedbackeqn)]
    \adjustimage{valign=c,margin=0pt,page=137}{tikzfigures}
    \) is sound.
\end{prop}
\begin{proof}
    By \autoref{lem:monotone-fixpoint}, applying the function \(
    (\listvar{x}) \mapsto \proj{0}\left(\circuittofunci[
        \adjustimage{valign=c,margin=0pt,page=105}{tikzfigures}
    ]\right)(\listvar{x}, \listvar{v})\) to itself \(c\) times reaches a
    fixed point.
    The circuit is combinational so each element of the output
    \(\circuittostreami[
        \adjustimage{valign=c,margin=0pt,page=105}{tikzfigures}
    ](\sigma)(i)\) is a function; this means that \autoref{lem:monotone-fixpoint}
    can be applied to each element.
\end{proof}

\begin{exa}\label{ex:sr-latch-unrolled}
    In \autoref{fig:sr-latch-unrolled}, the \(\instantfeedbackeqn\) is applied to
    the SR latch circuit in pre-Mealy form from
    \autoref{fig:sr-latch-global-trace-delay}.
\end{exa}

\begin{exa}
    In \autoref{fig:cyclic-combinational-unrolled}, the \((\instantfeedbackeqn)\)
    rule is applied to the cyclic combinational circuit from
    \autoref{fig:cyclic-combinational}.
\end{exa}

\begin{figure}
    \centering
    \scalebox{0.6}{\adjustimage{valign=c,margin=0pt,page=138}{tikzfigures}}
    \caption{
        Applying the \((\instantfeedbackeqn)\) rule to the circuit in
        \autoref{fig:sr-latch-pre-mealy}
    }
    \label{fig:sr-latch-unrolled}
\end{figure}
\begin{figure}
    \centering
    \scalebox{0.55}{
        \adjustimage{valign=c,margin=0pt,page=139}{tikzfigures}
    }
    \caption{
        Applying the \((\instantfeedbackeqn)\) rule to the circuit in
        \autoref{fig:cyclic-combinational}
    }
    \label{fig:cyclic-combinational-unrolled}
\end{figure}

If applied locally for every feedback loop, the \((\instantfeedbackeqn)\)
rule would cause an exponential blowup, but if a circuit is in global
trace-delay form, the rule need only be applied once to the global loop.
Although the value of \(c\) increases as the number of feedback wires increases,
it only does so linearly in the height of the lattice.

\begin{rem}
    \cite{mendler2012constructive} uses a ternary set of values and monotone
    functions to present \emph{constructive} circuits: the circuits
    that stabilise to unique Boolean values for all Boolean inputs.
    This definition excludes circuits that oscillate between two values, as
    these are not considered to be monotone circuits.
    Conversely, in our model such circuits \emph{can} be monotone.
    For example a gate-level circuit may alternate between \(\belnaptrue\) and
    \(\belnapfalse\) because these occupy the same level of the lattice.
\end{rem}

With a method to eliminate non-delay-guarded feedback, we can establish the
class of circuits which will act as the keystone of both the operational
semantics in this section and the algebraic semantics of the next.

\begin{defi}[Mealy form]\label{def:delay-guarded}
    A sequential circuit
    \adjustimage{valign=c,margin=0pt,page=89}{tikzfigures}
    is in \emph{Mealy form} if it is in the form \(
    \adjustimage{valign=c,margin=0pt,page=140}{tikzfigures}
    \).
\end{defi}

\begin{thm}\label{thm:all-mealy-form}
    For a sequential circuit
    \(\adjustimage{valign=c,margin=0pt,page=89}{tikzfigures}\), there
    exists at least one combinational circuit \(
    \adjustimage{valign=c,margin=0pt,page=141}{tikzfigures}
    \) and values \(\listvar{s} \in \valuetuple{x}\) such that \(
    \adjustimage{valign=c,margin=0pt,page=89}{tikzfigures}
    \reductions
    \adjustimage{valign=c,margin=0pt,page=142}{tikzfigures}
    \) by applying \((\mealyeqn)\) followed by \((\instantfeedbackeqn)\).
\end{thm}
\begin{proof}
    Any circuit can be assembled into global trace-delay form by
    \autoref{lem:trace-delay} and furthermore transformed into pre-Mealy form by
    using \((\mealyeqn)\).
    Since the core of a circuit in pre-Mealy form is combinational and has a
    non-delay-guarded trace, \((\instantfeedbackeqn)\) can be applied to it to
    produce a circuit with only delay-guarded feedback: a circuit in Mealy form.
\end{proof}

Non-delay-guarded feedback can be exhaustively unrolled because the circuit
essentially models a function despite the presence of the trace: this means that
we can transform the circuit without having to `look into the future'.
This is not the case for delay-guarded feedback as the internal state of the
circuit may depend on future inputs.
Indeed, a circuit with delay-guarded feedback may never `settle' on one
internal configuration but rather oscillate between multiple states.
This is simply a facet of sequential circuits and there is nothing we can do
about that.
What we \emph{can} do is show how to \emph{process} inputs at a given tick of
the clock.

\subsection{Productivity}\label{sec:productivity}

It is not possible to reduce an open circuit to some output values, as there will
be open wires awaiting the next inputs.
Nevertheless, if we precompose a circuit with some inputs we can provide some
rules for propagating them across the circuit.

Formally, for a sequential circuit \(
\adjustimage{valign=c,margin=0pt,page=89}{tikzfigures}
\) and values \(
\listvar{v} \in \valuetuplestream{m}
\), this corresponds to finding reductions such that \(
\adjustimage{valign=c,margin=0pt,page=143}{tikzfigures}
\reductions
\adjustimage{valign=c,margin=0pt,page=144}{tikzfigures}
\).
We first consider the combinational case, with our final global transformation.

\begin{lem}[Streaming]\label{lem:streaming}
    The following \emph{streaming rule} is sound: \[
        \adjustimage{valign=c,margin=0pt,page=145}{tikzfigures}
        \reduction[(\streamingeqn)]
        \adjustimage{valign=c,margin=0pt,page=146}{tikzfigures}
    \]
\end{lem}
\begin{proof}
    Once again this can be shown by considering the stream semantics.
    First note that by unfolding the notation, \(
    \adjustimage{valign=c,margin=0pt,page=145}{tikzfigures}
    \coloneqq
    \adjustimage{valign=c,margin=0pt,page=147}{tikzfigures}
    \).
    The streaming rule is then effectively `pushing' the combinational circuit
    \(\adjustimage{valign=c,margin=0pt,page=99}{tikzfigures}\) across the join.

    The join is \emph{not} a natural transformation so this does not hold in
    general, but because one argument is an instantaneous value and the other
    is a delay, at least one of the inputs to the join will be \(\bot\) for a
    given circuit.
    As the interpretations of combinational circuits must be
    \(\bot\)-preserving, the circuit can safely be pushed across the join and
    delay.
\end{proof}

The streaming rule shows that when a combinational circuit is applied to an
input with an instantaneous and a delayed component, the circuit can be copied
so that one copy handles what is happening `now' and the other handles what is
happening `later'.

\begin{exa}\label{ex:streaming}
    Pulsing the signals \(\belnapfalse\belnaptrue\) to the inputs of an SR NOR
    latch starts the procedure for `setting' the latch, causing it to output
    \(\belnaptrue\belnapfalse\).
    \autoref{fig:sr-latch-streamed} shows how the \(\streamingeqn\) rule is applied
    to the unrolled SR NOR circuit from
    \autoref{fig:sr-latch-unrolled} with these inputs to create a copy for what
    is happening `now' and another for what is happening `later'.
\end{exa}

\begin{figure*}
    \centering
    \scalebox{0.45}{\adjustimage{valign=c,margin=0pt,page=148}{tikzfigures}}
    \caption{
        Applying \(\streamingeqn\) with inputs \(\belnaptrue\belnapfalse\) to
        the circuit from \autoref{fig:sr-latch-unrolled}
    }
    \label{fig:sr-latch-streamed}
\end{figure*}

As there is a delay on the bottom argument of the join, the output of a streamed
circuit at the current tick is now contained entirely in the top argument of the
join.
The final rules we present will reduce this copy to values, as desired.

\begin{lem}[Value rules]
    The following \emph{value rules} are sound:
    \begin{gather*}
        \adjustimage{valign=c,margin=0pt,page=149}{tikzfigures}
        \reduction[(\forkeqn)]
        \adjustimage{valign=c,margin=0pt,page=150}{tikzfigures}
        \qquad
        \adjustimage{valign=c,margin=0pt,page=151}{tikzfigures}
        \reduction[(\joineqn)]
        \adjustimage{valign=c,margin=0pt,page=152}{tikzfigures}
        \\
        \adjustimage{valign=c,margin=0pt,page=153}{tikzfigures}
        \reduction[(\stubeqn)]
        \adjustimage{valign=c,margin=0pt,page=42}{tikzfigures}
        \qquad
        \adjustimage{valign=c,margin=0pt,page=154}{tikzfigures}
        \reduction[(\gateeqn)]
        \adjustimage{valign=c,margin=0pt,page=155}{tikzfigures}
    \end{gather*}
\end{lem}
\begin{proof}
    Straightforward by considering the interpretations of values as stream
    functions.
\end{proof}

Reducing the `now' core is the only time in which exhaustive application is
required, as more is involved than just copying circuit components.

\begin{lem}\label{lem:reduce-core-confluent}
    Applying the value rules is confluent.
\end{lem}
\begin{proof}
    There are no overlaps between the rules.
\end{proof}

\begin{lem}\label{lem:reduce-core-terminating}
    For a combinational circuit \(
    \adjustimage{valign=c,margin=0pt,page=36}{tikzfigures}
    \) and \(\listvar{v} \in \valuetuple{m}\), there exists
    \(\listvar{w} \in \valuetuple{n}\) such that applying the value
    rules to \(
    \adjustimage{valign=c,margin=0pt,page=156}{tikzfigures}
    \) terminates at \(
    \adjustimage{valign=c,margin=0pt,page=157}{tikzfigures}
    \).
\end{lem}
\begin{proof}
    By induction on the structure of \(
    \adjustimage{valign=c,margin=0pt,page=36}{tikzfigures}
    \).
\end{proof}

These rules are all we need to propagate input values across a circuit.

\begin{cor}\label{cor:mealy-form-productivity}
    For a circuit \(
    \adjustimage{valign=c,margin=0pt,page=158}{tikzfigures}
    \) there exist \(
    \listvar{t} \in \valuetuple{x}
    \) and \(
    \listvar{w} \in \valuetuple{n}
    \) such that \(
    \adjustimage{valign=c,margin=0pt,page=158}{tikzfigures}
    \reductions
    \adjustimage{valign=c,margin=0pt,page=159}{tikzfigures}
    \) by applying \(\streamingeqn\) once followed by the value rules
    exhaustively.
\end{cor}

\begin{exa}
    \autoref{fig:sr-latch-value-rules} shows how the value rules are applied to
    the streamed circuit from \autoref{fig:sr-latch-streamed}.
    After performing all the reductions exhaustively on the `now' circuit, the
    next state is \(\belnaptrue\), the first output is \(\bot\) and
    the second is \(\belnapfalse\).
    While the next state and second output make sense (if we apply Set, the
    state of the latch should turn true and the negated output false), the first
    output may raise eyebrows.
    This arises due to the presence of the delay; it will take another cycle to
    produce the expected output \(\belnaptrue\belnapfalse\).
\end{exa}

\begin{figure*}
    \centering
    \scalebox{0.45}{\adjustimage{valign=c,margin=0pt,page=148}{tikzfigures}}
    \(\reductions\)
    \\[1em]
    \scalebox{0.45}{\adjustimage{valign=c,margin=0pt,page=160}{tikzfigures}}
    \(\reductions\)
    \\[1em]
    \scalebox{0.45}{\adjustimage{valign=c,margin=0pt,page=161}{tikzfigures}}
    \caption{
        Using the value rules to reduce the streamed SR NOR latch circuit from
        \autoref{fig:sr-latch-streamed}.
    }
    \label{fig:sr-latch-value-rules}
\end{figure*}

By now putting together all the components in this section and the previous,
we have a productive strategy for processing inputs to \emph{any} sequential
circuit.

\begin{cor}[Productivity]\label{cor:productivity}
    For sequential circuit \(
    \adjustimage{valign=c,margin=0pt,page=89}{tikzfigures}
    \) and inputs \(\listvar{v} \in \valuetuple{m}\), there exists
    \(\listvar{w} \in \valuetuple{n}\) such that \(
    \adjustimage{valign=c,margin=0pt,page=143}{tikzfigures}
    \reductions
    \adjustimage{valign=c,margin=0pt,page=144}{tikzfigures}
    \) by applying \((\mealyeqn)\), \((\instantfeedbackeqn)\) and
    \((\streamingeqn)\) once successively followed by the value rules
    exhaustively.
\end{cor}

\begin{rem}
    As we saw in \autoref{cor:mealy-form-productivity}, applying
    \((\streamingeqn)\) followed by the value rules to a circuit in Mealy form
    produces another circuit in Mealy form.
    This means that for one circuit and a long waveform stream of inputs,
    \((\mealyeqn)\) and \((\instantfeedbackeqn)\) need only be applied
    \emph{once} at the very start before processing values.
\end{rem}

\begin{rem}
    This style of operational semantics differs from some other approaches in
    the area, such as the work on signal flow graphs~\cite{bonchi2021survey}.
    In these works, the operational semantics is specified in terms of
    the state transitions that take place in a circuit over time.
    For example, the rule that applies to the fork in signal flow graphs is
    \[
        t \triangleright \adjustimage{valign=c,margin=0pt,page=162}{tikzfigures}
        \xrightarrow[k\ k]{k}
        t+1 \triangleright \adjustimage{valign=c,margin=0pt,page=162}{tikzfigures}
    \]
    where \(t\) is the current timestep, \(k\) is the input signal and
    \(k\ k\) is the (forked) output signal.
    Note that the fork itself does not change; the `computation' occurring is
    contained entirely within the inputs and outputs.

    In our world of digital circuits we are more interested in propagating
    values to see how this affects the internal structure of a circuit; this is
    another instance of how we are working with a \emph{causal} rather than
    \emph{relational} semantics.
    This means we specify inputs as explicit components, and the reductions
    actually change the structure of the circuit.
    \[
        \adjustimage{valign=c,margin=0pt,page=163}{tikzfigures}
        \reduction
        \adjustimage{valign=c,margin=0pt,page=164}{tikzfigures}
    \]
\end{rem}

\subsection{Observational equivalence}\label{sec:observational}

In the denotational semantics, we defined the relation of
\emph{denotational equivalence}, in which circuits are related if their
denotations as streams are equal.
For operational semantics we have another notion of relation on circuits: that
of \emph{observational equivalence}.
This is due to Morris~\cite{morris1969lambdacalculus}, who named it
`extensional equivalence': essentially, two processes are observationally
equivalent if they cannot be distinguished by their input-output behaviour.

Testing for observational equivalence is traditionally performed by checking
that a program behaves the same in all \emph{contexts}.
For digital circuits, this means that for all possible streams of inputs, the
circuit produces the same outputs.
Of course, there are infinitely many streams of inputs, despite the set of
values being finite.
Fortunately, since circuits are constructed from a finite number of
\emph{components}, we need not check them all.

\begin{lem}\label{lem:number-of-states}
    Let \(
    \adjustimage{valign=c,margin=0pt,page=98}{tikzfigures}
    \) be a sequential circuit with \(c\) delays.
    Then applying \autoref{cor:productivity} successively to a Mealy form of
    this circuit will produce at most \(|\values|^c\) unique states.
\end{lem}
\begin{proof}
    The only varying elements of the state word are contributed by
    the \(c\) delay components, as the values transition to \(\bot\).
\end{proof}

To test all of the possible internal states of a circuit, we must use
sequences of inputs long enough in time to guarantee that every possible state
of a circuit is triggered.

\begin{nota}[Waveform]\label{def:waveform}
    The empty waveform is defined as \(
    \adjustimage{valign=c,margin=0pt,page=165}{tikzfigures}
    \coloneqq
    \adjustimage{valign=c,margin=0pt,page=166}{tikzfigures}
    \).
    Given values \(\overline{v} \in \valuetuple{n}\) and sequence \(
    \overline{\underline{w}} \in (\valuetuple{n})^\star
    \), the waveform for sequence \(
    \overline{v} \streamcons \overline{\underline{w}}
    \) is drawn as \(
    \adjustimage{valign=c,margin=0pt,page=167}{tikzfigures}
    \coloneqq
    \adjustimage{valign=c,margin=0pt,page=168}{tikzfigures}
    \).
\end{nota}

As a circuit with \(c\) delay components has at most \(|\values|^c\) states,
to fully evaluate the behaviour of a circuit it suffices to check every
waveform of length \(|\values|^c + 1\).
This is because even if such a waveform causes all \(|\values|^c\) states to
occur, the final element must produce a previously visited state, as there are
no other states that could arise.

\begin{cor}\label{cor:repeated-state}
    Given a circuit in Mealy form \(
    \adjustimage{valign=c,margin=0pt,page=169}{tikzfigures}
    \) and inputs \(\listlistvar{v} \in \valuetupleseq{m}\) of length
    \(|\values|^c + 1\), there exists a state \(
    \listvar{r} \in \valuetuple{x}
    \), an input sequence \(
    \listlistvar{u} \in \valuetupleseq{m}
    \) and output sequences \(
    \listlistvar{w},\listlistvar{z} \in \valuetupleseq{n}
    \) such that applying \autoref{cor:productivity} yields the
    following reduction pattern: \begin{gather*}
        \adjustimage{valign=c,margin=0pt,page=170}{tikzfigures}
        \reductions
        \adjustimage{valign=c,margin=0pt,page=171}{tikzfigures}
        \reductions
        \adjustimage{valign=c,margin=0pt,page=172}{tikzfigures}
    \end{gather*}
\end{cor}

This means that every possible behaviour of a circuit can be evaluated using
a finite number of sequences.
This can be used to define our notion of observational equivalence for digital
circuits.

\begin{defi}[Observational equivalence of circuits]
    We say that two sequential circuits \(
    \adjustimage{valign=c,margin=0pt,page=89}{tikzfigures}
    \) and \(
    \adjustimage{valign=c,margin=0pt,page=108}{tikzfigures}
    \) with no more than \(c\) delays are said to be
    \emph{observationally equivalent under} \(\interpretation\), written \(
    \adjustimage{valign=c,margin=0pt,page=98}{tikzfigures}
    \sim_{\interpretation}
    \adjustimage{valign=c,margin=0pt,page=109}{tikzfigures}
    \) if applying productivity produces the same output
    waveforms for all input waveforms \(
    \listlistvar{v} \in \valuetupleseq{m}\) of length
    \(|\values^c| + 1\).
\end{defi}

Observational equivalence is our semantic relation for operational semantics,
which relates two circuits based on their \emph{execution}.
To ensure it is suitable, it must be sound and complete with respect to the
denotational semantics.

\begin{thm}\label{thm:operational-denotational}
    Two sequential circuits \(
    \adjustimage{valign=c,margin=0pt,page=89}{tikzfigures}
    \) and \(
    \adjustimage{valign=c,margin=0pt,page=108}{tikzfigures}
    \) are observationally equivalent \(
    \adjustimage{valign=c,margin=0pt,page=89}{tikzfigures}
    \sim_\interpretation
    \adjustimage{valign=c,margin=0pt,page=108}{tikzfigures}
    \) if and only if \(
    \circuittostreami[
        \adjustimage{valign=c,margin=0pt,page=89}{tikzfigures}
    ]
    =
    \circuittostreami[
        \adjustimage{valign=c,margin=0pt,page=108}{tikzfigures}
    ]
    \).
\end{thm}
\begin{proof}
    The \(\onlyifdir\) direction follows by \autoref{cor:repeated-state}, as every
    possible internal configuration of the circuit will be tested.
    For \(\ifdir\), if \(
    \circuittostreami[
        \adjustimage{valign=c,margin=0pt,page=89}{tikzfigures}
    ]
    =
    \circuittostreami[
        \adjustimage{valign=c,margin=0pt,page=108}{tikzfigures}
    ]
    \), then this means \(
    \circuittostreami[
        \adjustimage{valign=c,margin=0pt,page=89}{tikzfigures}
    ](\listlistvar{v} \streamcons \sigma)
    =
    \circuittostreami[
        \adjustimage{valign=c,margin=0pt,page=108}{tikzfigures}
    ](\listlistvar{v} \streamcons \sigma)
    \) for any \(\sigma,\tau \in \valuetuplestream{m}\).
    By definition of \(\circuittostreami\), we then have that \(
    \circuittostreami[
        \adjustimage{valign=c,margin=0pt,page=173}{tikzfigures}
    ](\sigma)
    =
    \circuittostreami[
        \adjustimage{valign=c,margin=0pt,page=174}{tikzfigures}
    ](\sigma)
    \).
    Since this holds for \emph{all} sequences \(\listlistvar{v}\), it must hold
    for those of length \(|\values|^c + 1\), so the condition for observational
    equivalence is met.
\end{proof}

To verify that this is the `best' equivalence relation, we turn to a
definition of observational equivalence in terms of universal
properties~\cite{gordon1998operational}.
Gordon states that a relation is an \emph{adequate} observational semantics if
it only relates circuits that have the same denotational semantics;
observational equivalence is defined as the largest adequate congruence.

\begin{cor}
    \(\sim_\interpretation\) is the largest adequate congruence on
    \(\scircsigma\).
\end{cor}
\begin{proof}
    For \(\sim_\interpretation\) to be a congruence it must be preserved by
    composition, tensor and trace, and for it to be the largest there must be
    no denotationally equal circuit it does not relate.
    These, along with adequacy, all follow by
    \autoref{thm:operational-denotational}.
\end{proof}

This makes \(\sim_\interpretation\) a suitable notion of observational
equivalence for sequential circuits.

\begin{defi}
    Let \(\scircsigmaobs\) be defined as
    \(\scircsigma / \sim_{\interpretation}\).
\end{defi}

\begin{cor}
    There is an isomorphism \(\scircsigmai \cong \scircsigmaobs\).
\end{cor}

This means that reasoning with the operational semantics is equivalent to
reasoning denotationally with the stream function semantics; if two circuits
have the same stream functions then they can be shown to be observationally
equivalent, and vice versa.
As the results of the previous section give us an upper bound on the length of
waveforms required to establish observational equivalence, we have a
terminating strategy for comparing digital circuits syntactically.
Unfortunately, this is still an \emph{exponential} upper bound, so it is
infeasible to check for the equivalence of circuits with more than a few delay
components.
Nevertheless, the operational semantics gives us a straightforward way to
\emph{evaluate} circuits while respecting their internal structure, unlocking
more insight as to \emph{why} circuits are behaving the way they are.

Moreover, while it may be infeasible to check \emph{every single possible input}
to a circuit, it is often the case that one knows a particular input is fixed.
By precomposing the circuit with appropriate infinite waveforms to represent the
fixed inputs, insight and potential optimisations may be gleaned; this is known
as \emph{partial evaluation}, which will be examined more
in \autoref{sec:applications}.

\subsection{Operational semantics for generalised circuits}

When dealing with arbitrary-width wires, the only part of the operational
semantics that does not completely generalise in the obvious way are the value
rules.

\begin{lem}[Generalised value rules]
    The following \emph{generalised value rules} are sound:
    \begin{gather*}
        \adjustimage{valign=c,margin=0pt,page=175}{tikzfigures}
        \reduction[(\forkeqn)]
        \adjustimage{valign=c,margin=0pt,page=176}{tikzfigures}
        \qquad
        \adjustimage{valign=c,margin=0pt,page=177}{tikzfigures}
        \reduction[(\joineqn)]
        \adjustimage{valign=c,margin=0pt,page=178}{tikzfigures}
        \\[1em]
        \adjustimage{valign=c,margin=0pt,page=179}{tikzfigures}
        \reduction[(\stubeqn)]
        \adjustimage{valign=c,margin=0pt,page=42}{tikzfigures}
        \qquad
        \adjustimage{valign=c,margin=0pt,page=154}{tikzfigures}
        \reduction[(\gateeqn)]
        \adjustimage{valign=c,margin=0pt,page=155}{tikzfigures}
        \\[1em]
        \adjustimage{valign=c,margin=0pt,page=180}{tikzfigures}
        \reduction[(\spliteqn)]
        \adjustimage{valign=c,margin=0pt,page=181}{tikzfigures}
        \qquad
        \adjustimage{valign=c,margin=0pt,page=182}{tikzfigures}
        \reduction[(\combineeqn)]
        \adjustimage{valign=c,margin=0pt,page=183}{tikzfigures}
    \end{gather*}
\end{lem}

\begin{lem}
    Applying the generalised value rules is confluent.
\end{lem}
\begin{proof}
    There are no overlaps between the rules.
\end{proof}

\begin{lem}l
    For a generalised combinational circuit \(
    \adjustimage{valign=c,margin=0pt,page=184}{tikzfigures}
    \) and \(\listvar{v} \in \valuetuple{\listvar{m}}\), there exists a word
    \(\listvar{w} \in \valuetuple{\listvar{n}}\) such that applying the value
    rules exhaustively to \(
    \adjustimage{valign=c,margin=0pt,page=156}{tikzfigures}
    \) terminates at \(
    \adjustimage{valign=c,margin=0pt,page=157}{tikzfigures}
    \).
\end{lem}

With these rules, the inputs to a generalised circuit can be processed.

\begin{cor}
    For generalised circuit \(
    \adjustimage{valign=c,margin=0pt,page=185}{tikzfigures}
    \) there exist \(
    \listvar{t} \in \valuetuple{\listvar{x}}
    \) and \(
    \listvar{w} \in \valuetuple{\listvar{n}}
    \) such that \(
    \adjustimage{valign=c,margin=0pt,page=185}{tikzfigures}
    \reductions
    \adjustimage{valign=c,margin=0pt,page=186}{tikzfigures}
    \) by applying \((\streamingeqn)\) once followed by the generalised value
    rules exhaustively.
\end{cor}

\begin{cor}[Generalised productivity]
    For sequential circuit \(
    \adjustimage{valign=c,margin=0pt,page=111}{tikzfigures}
    \) and inputs \(\listvar{v} \in \valuetuple{\listvar{m}}\), there exists
    \(\listvar{w} \in \valuetuple{\listvar{n}}\) such that \(
    \adjustimage{valign=c,margin=0pt,page=187}{tikzfigures}
    \reductions
    \adjustimage{valign=c,margin=0pt,page=188}{tikzfigures}
    \) by applying \(\mealyeqn\), \(\instantfeedbackeqn\) and \(\streamingeqn\)
    once successively followed by the value rules exhaustively.
\end{cor}

Since register components can now hold words rather than just values, for
observational equivalence we must consider longer input waveforms.

\begin{defi}[Register width]
    Given a generalised sequential circuit \(
    \adjustimage{valign=c,margin=0pt,page=111}{tikzfigures}
    \), let \(c_n\) be the number of \(n\)-width delay components
    \(\adjustimage{valign=c,margin=0pt,page=83}{tikzfigures}\); the
    \emph{register width} of \(
    \adjustimage{valign=c,margin=0pt,page=111}{tikzfigures}
    \) is computed as \(\Sigma_{n \in \nat}\ c_n \cdot n\).
\end{defi}

\begin{defi}
    We say that two generalised sequential circuits \(
    \adjustimage{valign=c,margin=0pt,page=111}{tikzfigures}
    \) and \(
    \adjustimage{valign=c,margin=0pt,page=112}{tikzfigures}
    \) with register width at most \(c\) are said to be
    \emph{observationally equivalent under} \(\interpretation\), written \(
    \adjustimage{valign=c,margin=0pt,page=98}{tikzfigures}
    \sim^+_{\interpretation}
    \adjustimage{valign=c,margin=0pt,page=109}{tikzfigures}
    \) if applying productivity produces the same output
    waveforms for all input waveforms \(
    \listlistvar{v} \in \valuetupleseq{\listvar{m}}\) of length
    \(|\values^c| + 1\).
\end{defi}

The observational equivalence results from the previous section then generalise
nicely to the multicoloured case.

\begin{thm}
    Given two sequential circuits \(
    \adjustimage{valign=c,margin=0pt,page=111}{tikzfigures}
    \) and \(
    \adjustimage{valign=c,margin=0pt,page=112}{tikzfigures}
    \), we have that \(
    \adjustimage{valign=c,margin=0pt,page=111}{tikzfigures}
    \sim^+_{\interpretation}
    \adjustimage{valign=c,margin=0pt,page=112}{tikzfigures}
    \) if and only if \(
    \circuittostreami[
        \adjustimage{valign=c,margin=0pt,page=111}{tikzfigures}
    ]
    =
    \circuittostreami[
        \adjustimage{valign=c,margin=0pt,page=112}{tikzfigures}
    ]
    \).
\end{thm}

\begin{cor}
    \(\sim^+_\interpretation\) is the largest adequate congruence on
    \(\scircsigmag\).
\end{cor}

\begin{defi}
    Let \(\scircsigmagobs\) be defined as \(\scircsigmag / \sim^+_{\interpretation}\).
\end{defi}

\begin{cor}
    There is an isomorphism \(\scircsigmaig \cong \scircsigmagobs\).
\end{cor}

\subsection{Mechanising the operational semantics}

The operational semantics lends itself well to pen-and-paper reasoning: one can
draw circuits and apply the reduction rules in sequence to evaluate them.
However, for anything more complex than toy examples, this would become quite
unwieldy as the number of gates quickly increases.
Ideally one would \emph{mechanise} the evaluation process: we could provide
the circuit and inputs to a computer and let it perform the procedure
automatically.

While string diagrams as topological structures are not exactly the most
favourable structure for providing to a computer, there has recently been an
extensive line of work into string diagram rewriting using
\emph{hypergraphs}~\cite{bonchi2022string,bonchi2022stringa,bonchi2022stringb}.
Of particular relevance to us is string diagram rewriting for settings with a
traced comonoid structure~\cite{ghica2026rewriting}, which is exactly the
structure we work with in digital circuits.

In the string diagram hypergraph rewriting approach, string diagram terms are
interpreted as cospans of hypergraphs; in \emph{loc. cit.}\ it has been shown
precisely which classes of cospans correspond to different types of string
diagrams.
For example, any such cospan of hypergraphs corresponds to a string diagram
term in a symmetric monoidal setting equipped with a \emph{Frobenius} structure.

\begin{exa}
    Recall the SR NOR latch from \autoref{ex:sr-latch}.
    Its interpretation as a cospan of hypergraphs is illustrated in
    \autoref{fig:latch-graph}, in which the black dots are nodes and the
    white boxes are hyperedges.
    Nodes are annotated with numbers to indicate if they are in the image of
    the left leg of the cospan (inputs) or the image of the right leg of the
    cospan (outputs).
\end{exa}

\begin{figure}
    \centering
    \adjustimage{valign=c,margin=0pt,page=189}{tikzfigures}
    \caption{
        Interpretation of the SR NOR latch from \autoref{ex:sr-latch} as
        a cospan of hypergraphs
    }
    \label{fig:latch-graph}
\end{figure}

Reductions in the operational semantics can then be performed using
\emph{double pushout} (DPO) rewriting~\cite{ehrig1976parallelism}, in which
rewrites are identified by using a `reverse pushout' to remove the left side
of a reduction from a larger graph, followed by a regular pushout to paste in
the right side of the reduction.

\begin{exa}
    Consider the simple term rewrite \(
    \adjustimage{valign=c,margin=0pt,page=190}{tikzfigures}
    \grewrite
    \adjustimage{valign=c,margin=0pt,page=191}{tikzfigures}
    \).
    An example of how this could be applied as a double pushout rewrite is
    shown in \autoref{fig:dpo}.
    The rewrite rule itself occupies the top row of the diagram and the graph in
    which the rewrite is being applied in on the left of the middle row.
    The rewrite context (the target graph with the instance of the rewrite rule
    removed) sits in the middle of the diagram, and the rewritten graph on its
    right.
\end{exa}

\begin{figure}
    \center
    \begin{tikzcd}[column sep = small, row sep = small]
    \scalebox{0.8}{\adjustimage{valign=c,margin=0pt,page=390}{tikzfigures}}
    \arrow{d}
    &
    \arrow{l}
    \scalebox{0.8}{\adjustimage{valign=c,margin=0pt,page=391}{tikzfigures}}
    \arrow{d}
    \arrow{r}
    &
    \scalebox{0.8}{\adjustimage{valign=c,margin=0pt,page=392}{tikzfigures}}
    \arrow{d}
    \\
    \scalebox{0.8}{\adjustimage{valign=c,margin=0pt,page=393}{tikzfigures}}
    &
    \arrow{l}
    \scalebox{0.8}{\adjustimage{valign=c,margin=0pt,page=394}{tikzfigures}}
    \arrow{r}
    &
    \scalebox{0.8}{\adjustimage{valign=c,margin=0pt,page=395}{tikzfigures}}
    \\
    &
    \scalebox{0.8}{\adjustimage{valign=c,margin=0pt,page=396}{tikzfigures}}
    \arrow{ul}
    \arrow{u}
    \arrow{ur}
\end{tikzcd}
    \caption{
        Example of a DPO rewrite
    }
    \label{fig:dpo}
\end{figure}

While the hypergraph approach is a useful way to implement the operational
semantics, it is by no means an essential part of it: reductions on circuits may
be performed without ever appealing to the combinatorial graph perspective.
That being said, hypergraph string diagram rewriting is a powerful tool that
allows the theoretical basis to be applied to deeply complicated real-world
circuits.

\section{Algebraic semantics}

The previous section gives an upper bound on the length of waveforms required to
establish observational equivalence, so we have a terminating strategy
for comparing the behaviour of digital circuits using a
pseudo-normal form.
Unfortunately, this is still an \emph{exponential} upper bound, so it is
infeasible to check for equivalence of circuits with more than a few
delay components.

It is often the case that circuits differ by only a few components;
perhaps one is trying to show that two similar implementations are the same.
In this case, it is more feasible to find the parts of the circuit that differ
and then check if they have the same behaviour.
With the syntactic representation, we can reason about these subcircuits
\emph{algebraically} using equations.
The final contribution of this paper is to define a \emph{sound and complete}
algebraic semantics for sequential circuits.

\begin{rem}
    An `equational theory' was presented in \cite{ghica2016categorical},
    but the equations were used to quotient the syntax as an ad-hoc semantics,
    and soundness and completeness were not considered.
    Here we use the stream semantics to guide our choice in equations.
\end{rem}

When defining such an equational theory, there may be several different sound
and complete formulations.
Ideally, we want to stick to simple \emph{local} equations that concern the
interactions of concrete generators as much as possible, but as we will see
we will sometimes have no choice but to define \emph{families} of equations
parameterised over some arbitrary subcircuit.

\subsection{Normalising circuits}\label{sec:normalising}

How does one start when trying to define a complete set of equations for some
framework?
Usually the strategy is to define enough equations to bring any term to some
sort of (pseudo-)\emph{normal form}; the theory is then complete if terms with
the same semantics have the same normal form.

We have already seen something that looks a bit like a normal form: the
\emph{Mealy form} from the previous section.
This is by no means a true normal form, as there are many different Mealy forms
that represent the same behaviour.
Nevertheless, it is a useful starting point so we will need equations to bring
circuits to Mealy form in our theory.

Instead of just turning the Mealy reduction rules into equations, we will show
how Mealy form can be derived using smaller equations.

\begin{figure}
    \centering
    \(
    \equationdisplay{
        \adjustimage{valign=c,margin=0pt,page=192}{tikzfigures}
    }{
        \adjustimage{valign=c,margin=0pt,page=193}{tikzfigures}
    }{
        \joinunitleqn
    }
    \qquad
    \equationdisplay{
        \adjustimage{valign=c,margin=0pt,page=194}{tikzfigures}
    }{
        \adjustimage{valign=c,margin=0pt,page=195}{tikzfigures}
    }{
        \joinunitreqn
    }
    \)
    \\[0.25em]
    \rule{\textwidth}{0.1mm}
    \\[0.5em]
    \(
    \equationdisplay{
        \adjustimage{valign=c,margin=0pt,page=196}{tikzfigures}
    }{
        \adjustimage{valign=c,margin=0pt,page=197}{tikzfigures}
    }{
        \bottomdelayeqn
    }
    \qquad
    \equationdisplay{
        \adjustimage{valign=c,margin=0pt,page=136}{tikzfigures}
    }{
        \adjustimage{valign=c,margin=0pt,page=137}{tikzfigures}
    }{
        \instantfeedbackeqn
    }
    \)
    \\[0.25em]
    \rule{\textwidth}{0.1mm}
    \caption{
        Set of Mealy equations
        \(\mealyequations\).
    }
    \label{fig:mealy-equations}
\end{figure}

\begin{defi}
    The set \(\mealyequations\) of \emph{Mealy equations} in
    \autoref{fig:mealy-equations} are sound.
\end{defi}
\begin{proof}
    The first two rules hold as the join is a monoid in the stream semantics.
    The \((\bottomdelayeqn)\) holds because the semantics of the delay
    component are to output a \(\bot\) value first and then the (delayed)
    inputs: as the semantics of the \(
    \adjustimage{valign=c,margin=0pt,page=29}{tikzfigures}
    \) component are to \emph{always} produce \(\bot\), then it does not make a
    difference how delayed it is.
    The final equation is the instant feedback rule, which is sound by
    \autoref{prop:instant-feedback}.
\end{proof}

\begin{prop}\label{prop:mealy-equations}
    Given a sequential circuit \(
    \adjustimage{valign=c,margin=0pt,page=89}{tikzfigures}
    \), there exists a combinational circuit \(
    \adjustimage{valign=c,margin=0pt,page=141}{tikzfigures}
    \) and values \(\listvar{s} \in \valuetuple{x}\) such that \(
    \adjustimage{valign=c,margin=0pt,page=98}{tikzfigures}
    =
    \adjustimage{valign=c,margin=0pt,page=198}{tikzfigures}
    \) in \(\scircsigma / \mealyequations\).
\end{prop}
\begin{proof}
    Any circuit can be assembled into global trace-delay form solely using the
    axioms of STMCs.
    From this, a circuit in pre-Mealy form can be obtained by translating
    delays and values into registers using the following equations: \[
        \adjustimage{valign=c,margin=0pt,page=199}{tikzfigures}
        \eqaxioms[(\joinunitleqn)]
        \adjustimage{valign=c,margin=0pt,page=200}{tikzfigures}
        \qquad\quad
        \adjustimage{valign=c,margin=0pt,page=201}{tikzfigures}
        \eqaxioms[(\joinunitreqn)]
        \adjustimage{valign=c,margin=0pt,page=202}{tikzfigures}
        \eqaxioms[(\bottomdelayeqn)]
        \adjustimage{valign=c,margin=0pt,page=203}{tikzfigures}
    \]
    Subsequently a circuit in Mealy form can be obtained by applying the
    \((\instantfeedbackeqn)\) rule.
\end{proof}

\(\scircsigma / \mealyequations\) is a category in which all circuits are equal
to at least one circuit in Mealy form.
In general, there will be many Mealy forms depending on the ordering one picks
for the delays and value; our task is to provide equations to map any two
denotationally equivalent circuits to the \emph{same} Mealy form.

Even if the combinational cores of two Mealy forms have the same behaviour, they
may not have the same structure.
To reduce the number of cores we have to consider, we will first establish
equations for translating any combinational circuit into some canonical circuit.
We already met a method for determining what this canonical circuit is: the
functional completeness map \(\mealytofunc\) from \(\funci\) to \(\scircsigma\).

\begin{defi}[Normalised circuit]
    A sequential circuit \(
    \adjustimage{valign=c,margin=0pt,page=89}{tikzfigures}
    \) is \emph{normalised} if it is in the image of \(\mealytofunc\).
\end{defi}

As a shorthand, we will often abuse notation and write \(
\mealytofunc[f]
\coloneqq
\adjustimage{valign=c,margin=0pt,page=204}{tikzfigures}
\).
Recall that even though \(\mealytofunc\) maps into \(\scircsigma\), every
circuit in its image has combinational behaviour.
This is quite an important distinction to make, so we will give it a proper
name.

\begin{defi}[Essentially combinational]
    A sequential circuit is \emph{essentially combinational} if it is in the
    form \(
    \adjustimage{valign=c,margin=0pt,page=101}{tikzfigures}
    \).
\end{defi}

Such circuits are sequential circuits that exhibit combinational behaviour: any
value components are only used to introduce constants which do not alter over
time.

As the normalised version of a given circuit is interpretation-dependent, there
is no standard set of equations for normalising a circuit.
Instead, these must be specified on an interpretation-by-interpretation basis.

\begin{defi}[Normalising equations]
    For a complete interpretation \((\interpretation,\mealytofunc)\), a set of
    equations \(\normalisingequations\) is \emph{normalising} if any
    essentially combinational circuit \(
    \adjustimage{valign=c,margin=0pt,page=89}{tikzfigures}
    \) is equal to a circuit in the image of \(\mealytofunc\) by equations in
    \(\normalisingequations\).
\end{defi}

\begin{defi}[Normalisable interpretation]
    A complete interpretation \(\interpretation\) is called \emph{normalisable} if there
    exists a set of normalising equations \(\normalisingequations\).
\end{defi}

\begin{exa}
    The gate-level interpretation is normalisable; the set of normalising
    equations consists primarily of the standard completeness equations for
    Boolean logic extended to act on Belnap values.
    The complete set is extensive, so in the interests of brevity the equations
    and associated proofs can be found in \cite[Sec.\ 6.5]{kaye2024foundations}.
\end{exa}

The normalising equations for a given interpretation can be used to translate a
combinational core into a circuit from which it is easy to read off a truth
table.

\begin{thm}\label{thm:normalising}
    For every sequential circuit \(
    \adjustimage{valign=c,margin=0pt,page=89}{tikzfigures}
    \) in a normalisable complete interpretation
    \((\interpretation,\mealytofunc)\) over \(\Sigma\), there exists essentially
    combinational \(
    \adjustimage{valign=c,margin=0pt,page=205}{tikzfigures}
    \) and \(\listvar{s} \in \valuetuple{x}\) such that \(
    \adjustimage{valign=c,margin=0pt,page=89}{tikzfigures}
    =
    \adjustimage{valign=c,margin=0pt,page=206}{tikzfigures}
    \) in \(\scircsigma / \mealyequations + \normalisingequations\).
\end{thm}
\begin{proof}
    By \autoref{prop:mealy-equations}, \(
    \adjustimage{valign=c,margin=0pt,page=89}{tikzfigures}
    =
    \adjustimage{valign=c,margin=0pt,page=207}{tikzfigures}
    \) and by \(\normalisingequations\), \(
    \adjustimage{valign=c,margin=0pt,page=208}{tikzfigures}
    =
    \adjustimage{valign=c,margin=0pt,page=209}{tikzfigures}
    \).
\end{proof}

\subsection{Encoding equations}\label{sec:encoding}

A circuit in Mealy form is a syntactic representation of a Mealy machine: the
combinational core is the Mealy function, and the registers are the initial
state.
It is important to determine the states that the circuit might assume, as these
dictate whether or not an equation is valid.

\begin{defi}[States]
    Let \(\morph{f}{\valuetuple{x+m}}{\valuetuple{x+n}}\) be a
    monotone function and let \(\listvar{s} \in  \valuetuple{x}\) be a state.
    Then the \emph{states of \(f\) from \(\listvar{s}\)}, denoted
    \(S_{f,\listvar{s}}\), is the smallest set containing \(\listvar{s}\) and
    closed under \(
    \listvar{r}
    \mapsto
    \proj{0}\mleft(f(\listvar{r},\listvar{v})\mright)
    \) for any \(\listvar{v} \in \valuetuple{m}\).
\end{defi}

\begin{exa}\label{ex:circuit-states}
    Consider the circuit \(
    \adjustimage{valign=c,margin=0pt,page=210}{tikzfigures}
    \).
    The semantics of the combinational core are clearly
    \((s, r) \mapsto \left(s \lor r, s, s \lor r\right)\), where the first two characters are the
    next state and the third is the output.
    The initial state is \(\bot\belnaptrue\), so the subsequent states are
    \((\bot \lor \belnaptrue, \bot) = (\belnaptrue,\bot)\) and
    \((\belnaptrue \lor \bot, \belnaptrue) = (\belnaptrue, \belnaptrue)\).
    As \((\belnaptrue \lor \belnaptrue, \belnaptrue) = (\belnaptrue, \belnaptrue)\),
    there are no more circuit states and the complete set is
    \(\{(\bot, \belnaptrue),(\belnaptrue,\bot),(\belnaptrue,\belnaptrue)\}\).

    Note that as the output of the circuit is computed as \(s \lor r\), for each
    circuit state the output is \(\belnaptrue\).
    This means that the circuit is denotationally equivalent to \(
    \adjustimage{valign=c,margin=0pt,page=211}{tikzfigures}
    \), but this circuit only has a single state \(\belnaptrue\).
\end{exa}

As the above example dictates, there may be many denotationally equivalent
circuits with different state sets.
A major part of an equational theory for circuits will be to translate between
the states of different circuits in a behaviour-preserving manner.
To do this we turn back to the notion of Mealy homomorphisms:

Much like how we can view regular Mealy machines as \(\set\)-valued coalgebra,
we can view monotone Mealy machines as \(\pos_\bot\)-valued coalgebra.
In this setting, the Mealy homomorphisms are \(\bot\)-preserving monotone
functions and as such can be interpreted as circuits.
This means that for two circuits \(
\adjustimage{valign=c,margin=0pt,page=212}{tikzfigures}
\) and \(
\adjustimage{valign=c,margin=0pt,page=213}{tikzfigures}
\), a suitable encoder and decoder can be constructed from the the pair of
homomorphisms
\((S_{f,\listvar{s}},f) \to (S_{g,\listvar{t}},g)\)
and
\((S_{g,\listvar{t}},g) \to (S_{f,\listvar{s}},f)\).

These homomorphisms map between states that the circuits
can assume, \emph{not} the entire set of words that can fit into the state.
This means that encoding and decoding circuits cannot be inserted arbitrarily
but only in certain contexts.
For this reason any equation that involves using an encoder and decoder must
take into account the entire circuit.

\begin{prop}[Encoding equation]\label{prop:encoding-equation}
    For a normalised circuit \(
    \adjustimage{valign=c,margin=0pt,page=214}{tikzfigures}
    \) and \(\listvar{s} \in \valuetuple{x}\), let
    \(\morph{\mathsf{enc}}{S_{f,\listvar{s}}}{\valuetuple{y}}\) and
    \(\morph{\mathsf{dec}}{\valuetuple{y}}{S_{f, \listvar{s}}}\) be monotone
    functions such that the composite \(
    \morph{\mathsf{dec} \circ \mathsf{enc}}{S_{f, \listvar{s}}}{S_{f, \listvar{s}}}
    \) is a Mealy homomorphism.
    Then the \emph{encoding equation} \((\encodingequation)\) in
    \autoref{fig:encoding-equation} is sound, where
    \(\mathsf{enc}_\mathsf{m}\) and \(\mathsf{dec}_\mathsf{m}\) are monotone
    completions as in \autoref{def:monotone-completion}.
\end{prop}
\begin{proof}
    Let \(g\) be the map \(\listvar{r} \mapsto
    \circuittostreami[\adjustimage{valign=c,margin=0pt,page=215}{tikzfigures}]
    \); by \autoref{prop:mealy-form-image} we know that \(
    \mealyoutput{g(\listvar{t})}{\listvar{v}}
    =
    \proj{1}(f(\mathsf{dec}(\mathsf{enc}(\listvar{t})), \listvar{v}))
    \) and \(
    \mealytransition{g(\listvar{t})}{\listvar{v}}
    =
    g(\proj{0}(f(\mathsf{dec}(\mathsf{enc}(\listvar{t})), \listvar{v})))
    \).
    For a given state \(\listvar{t} \in S_{f, \listvar{s}}\) we have that \(
    \mealyoutput{g(\listvar{t})}{\listvar{v}}
    =
    \proj{1}(f(\listvar{t}), \listvar{v})
    \) and that \(
    \mealytransition{g(\listvar{t})}{\listvar{v}}
    \) shares outputs and transitions with \(
    g(\proj{0}(f(\listvar{t})), \listvar{v})
    \) as \(\mathsf{dec} \circ \mathsf{enc}\) is a Mealy homomorphism.
    As \(
    \adjustimage{valign=c,margin=0pt,page=216}{tikzfigures}
    \coloneqq
    g(\listvar{s})
    \) and \(\listvar{s} \in S_{f,\listvar{s}}\),
    every subsequent stream derivative will also be of the form
    \(g(\listvar{t})\) where \(\listvar{t} \in S_{f,\listvar{s}}\), so the
    equation is sound.
\end{proof}

\begin{rem}
    The encoding equation is an equation \emph{schema}: this is required because
    the width of a circuit state can be arbitrarily large, and each extra bit
    adds a whole new set of Mealy homomorphisms to consider.
\end{rem}

\begin{figure}
    \centering
    \scalebox{0.9}{\(
        \equationdisplay{
            \adjustimage{valign=c,margin=0pt,page=217}{tikzfigures}
        }{
            \adjustimage{valign=c,margin=0pt,page=218}{tikzfigures}
        }{
            \encodingequation
        }
        \)}
    \\[0.5em]
    \rule{\textwidth}{0.1mm}
    \\[0.75em]
    \scalebox{0.9}{\(
        \equationdisplay{
            \adjustimage{valign=c,margin=0pt,page=219}{tikzfigures}
        }{
            \adjustimage{valign=c,margin=0pt,page=220}{tikzfigures}
        }{
            \gateeqn
        }\)}
    \quad
    \scalebox{0.9}{\(\equationdisplay{
            \adjustimage{valign=c,margin=0pt,page=163}{tikzfigures}
        }{
            \adjustimage{valign=c,margin=0pt,page=164}{tikzfigures}
        }{
            \forkeqn
        }\)}
    \quad
    \scalebox{0.9}{\(\equationdisplay{
            \adjustimage{valign=c,margin=0pt,page=221}{tikzfigures}
        }{
            \adjustimage{valign=c,margin=0pt,page=222}{tikzfigures}
        }{
            \joineqn
        }
        \)}
    \\[0.5em]
    \rule{\textwidth}{0.1mm}
    \\[0.75em]
    \scalebox{0.9}{\(
        \equationdisplay{
            \adjustimage{valign=c,margin=0pt,page=223}{tikzfigures}
        }{
            \adjustimage{valign=c,margin=0pt,page=42}{tikzfigures}
        }{
            \stubeqn
        }\)}
    \quad
    \scalebox{0.9}{\(\equationdisplay{
            \adjustimage{valign=c,margin=0pt,page=224}{tikzfigures}
        }{
            \adjustimage{valign=c,margin=0pt,page=225}{tikzfigures}
        }{
            \delayforkeqn
        }\)}
    \quad
    \scalebox{0.9}{\(\equationdisplay{
            \adjustimage{valign=c,margin=0pt,page=196}{tikzfigures}
        }{
            \adjustimage{valign=c,margin=0pt,page=197}{tikzfigures}
        }{
            \bottomdelayeqn
        }
        \)}
    \\[0.5em]
    \rule{\textwidth}{0.1mm}
    \\[0.75em]
    \scalebox{0.9}{\(
        \equationdisplay{
            \adjustimage{valign=c,margin=0pt,page=226}{tikzfigures}
        }{
            \adjustimage{valign=c,margin=0pt,page=227}{tikzfigures}
        }{
            \forkuniteqn
        }\)}
    \quad
    \scalebox{0.9}{\(\equationdisplay{
            \adjustimage{valign=c,margin=0pt,page=228}{tikzfigures}
        }{
            \adjustimage{valign=c,margin=0pt,page=229}{tikzfigures}
        }{
            \joinassoceqn
        }
        \)}
    \quad
    \scalebox{0.9}{\(
        \equationdisplay{
            \adjustimage{valign=c,margin=0pt,page=230}{tikzfigures}
        }{
            \adjustimage{valign=c,margin=0pt,page=231}{tikzfigures}
        }{
            \streamingeqn
        }
        \)}
    \\[0.5em]
    \rule{\textwidth}{0.1mm}
    \\[0.75em]
    \scalebox{0.9}{\(
        \equationdisplay{
            \adjustimage{valign=c,margin=0pt,page=232}{tikzfigures}
        }{
            \adjustimage{valign=c,margin=0pt,page=233}{tikzfigures}
        }{
            \joincommeqn
        }\)}
    \quad
    \scalebox{0.9}{\(\equationdisplay{
            \adjustimage{valign=c,margin=0pt,page=234}{tikzfigures}
        }{
            \adjustimage{valign=c,margin=0pt,page=235}{tikzfigures}
        }{
            \joinforkeqn
        }
        \)}
    \\[0.25em]
    \rule{\textwidth}{0.1mm}
    \caption{
        Set \(\encodingequations\) of equations for encoding circuit states
    }
    \label{fig:encoding-equation}
\end{figure}

The encoding equation only inserts encoder circuits; to actually change the
state we need some more equations.

\begin{lem}
    The equations on the bottom four rows of \autoref{fig:encoding-equation} are
    sound.
\end{lem}
\begin{proof}
    It is a straightforward exercise to compare the stream functions.
\end{proof}

To show the final result we must prove some lemmas; first we show how we can
`pump' a value out of an infinite waveform.

\begin{lem}\label{lem:unroll-waveform}
    For \(v \in \values\), \(
    \adjustimage{valign=c,margin=0pt,page=236}{tikzfigures}
    =
    \adjustimage{valign=c,margin=0pt,page=237}{tikzfigures}
    \) using the encoding equations.
\end{lem}
\begin{proof}
    The proof is straightforward and is illustrated in
    \autoref{fig:unroll-waveform}.
\end{proof}
\begin{figure}
    \centering
    \begin{gather*}
        \scalebox{0.9}{\adjustimage{valign=c,margin=0pt,page=236}{tikzfigures}}
        \coloneqq
        \scalebox{0.9}{\adjustimage{valign=c,margin=0pt,page=238}{tikzfigures}}
        \eqaxioms[(\joinforkeqn)]
        \scalebox{0.9}{\adjustimage{valign=c,margin=0pt,page=239}{tikzfigures}}
        \eqaxioms[(\forkeqn)]
        \\[1em]
        \scalebox{0.9}{\adjustimage{valign=c,margin=0pt,page=240}{tikzfigures}}
        \eqaxioms[(\delayforkeqn)]
        \scalebox{0.9}{\adjustimage{valign=c,margin=0pt,page=241}{tikzfigures}}
        \coloneqq
        \scalebox{0.9}{\adjustimage{valign=c,margin=0pt,page=242}{tikzfigures}}
        =
        \scalebox{0.9}{\adjustimage{valign=c,margin=0pt,page=237}{tikzfigures}}
    \end{gather*}
    \caption{Proof of \autoref{lem:unroll-waveform}}
    \label{fig:unroll-waveform}
\end{figure}

The next lemma shows how the familiar `streaming' rule from the operational
semantics can be derived equationally.

\begin{lem}\label{lem:generalised-streaming}
    For a combinational circuit \(
    \adjustimage{valign=c,margin=0pt,page=99}{tikzfigures}
    \), \(
    \adjustimage{valign=c,margin=0pt,page=145}{tikzfigures}
    =
    \adjustimage{valign=c,margin=0pt,page=146}{tikzfigures}
    \) by the encoding equations.
\end{lem}
\begin{proof}
    This is by induction on the structure of \(
    \adjustimage{valign=c,margin=0pt,page=99}{tikzfigures}
    \).
    The case for the primitive is immediate by \((\streamingeqn)\).
    For \(\adjustimage{valign=c,margin=0pt,page=30}{tikzfigures}\) we have
    that \[
        \adjustimage{valign=c,margin=0pt,page=243}{tikzfigures}
        \eqaxioms[(\joinforkeqn)]
        \adjustimage{valign=c,margin=0pt,page=244}{tikzfigures}
        \eqaxioms[(\delayforkeqn)]
        \adjustimage{valign=c,margin=0pt,page=245}{tikzfigures}
    \]
    The proof for \(\adjustimage{valign=c,margin=0pt,page=31}{tikzfigures}\) is
    illustrated in \autoref{fig:generalised-streaming-join}.
    The case for \(\adjustimage{valign=c,margin=0pt,page=32}{tikzfigures}\) is
    trivial, and the case for \(\adjustimage{valign=c,margin=0pt,page=29}{tikzfigures}\)
    follows by \((\comonoiduniteqnletter)\) and \((\bottomdelayeqn)\).
    The cases for \(\adjustimage{valign=c,margin=0pt,page=246}{tikzfigures}\) and
    \(\adjustimage{valign=c,margin=0pt,page=247}{tikzfigures}\) follow by axioms of STMCs.
    Since the underlying circuit is combinational, for the inductive cases we just
    need to check composition and tensor, which are trivial.
\end{proof}
\begin{figure}
    \begin{gather*}
        \adjustimage{valign=c,margin=0pt,page=248}{tikzfigures}
        \eqaxioms[(\monoidassoceqnletter)]
        \adjustimage{valign=c,margin=0pt,page=249}{tikzfigures}
        \eqaxioms[(\monoidassoceqnletter)]
        \adjustimage{valign=c,margin=0pt,page=250}{tikzfigures}
        \eqaxioms[(\monoidcommeqnletter)]
        \\[1em]
        \adjustimage{valign=c,margin=0pt,page=251}{tikzfigures}
        =
        \adjustimage{valign=c,margin=0pt,page=252}{tikzfigures}
        \eqaxioms[(\monoidassoceqnletter)]
        \adjustimage{valign=c,margin=0pt,page=253}{tikzfigures}
        \eqaxioms[(\monoidassoceqnletter)]
        \\[1em]
        \adjustimage{valign=c,margin=0pt,page=254}{tikzfigures}
        \eqaxioms[(\monoidassoceqnletter)]
        \adjustimage{valign=c,margin=0pt,page=255}{tikzfigures}
    \end{gather*}
    \caption{Proof of \autoref{lem:generalised-streaming} for the join case}
    \label{fig:generalised-streaming-join}
\end{figure}

We next show how the encoding equations can be used to translate combinational
circuits with inputs into values.

\begin{lem}\label{lem:combinational-circuit-inputs}
    Let \(\adjustimage{valign=c,margin=0pt,page=36}{tikzfigures}\) be a
    combinational circuit such that \(
    \circuittofunci[\adjustimage{valign=c,margin=0pt,page=99}{tikzfigures}]
    =
    g
    \).
    Then \(
    \adjustimage{valign=c,margin=0pt,page=156}{tikzfigures}
    =
    \adjustimage{valign=c,margin=0pt,page=256}{tikzfigures}
    \) by the encoding equations.
\end{lem}
\begin{proof}
    For the same reasoning as \autoref{lem:reduce-core-terminating}, the
    \((\gateeqn)\), \((\forkeqn)\), \((\joineqn)\) and \((\stubeqn)\) equations
    can be used to show that there exists \(\listvar{w} \in \valuetuple{n}\)
    such that \(
    \adjustimage{valign=c,margin=0pt,page=36}{tikzfigures}
    =
    \adjustimage{valign=c,margin=0pt,page=157}{tikzfigures}
    \).

    Now we need to show that \(
    \circuittostreami[
        \adjustimage{valign=c,margin=0pt,page=156}{tikzfigures}
    ]
    =
    \circuittostreami[
        \adjustimage{valign=c,margin=0pt,page=256}{tikzfigures}
    ]
    \).
    By functoriality of \(\circuittostreami\), \(
    \circuittostreami[
        \adjustimage{valign=c,margin=0pt,page=156}{tikzfigures}
    ]
    =
    \circuittostreami[
        \adjustimage{valign=c,margin=0pt,page=257}{tikzfigures}
    ] \seq
    \circuittostreami[
        \adjustimage{valign=c,margin=0pt,page=99}{tikzfigures}
    ]
    \).
    By \autoref{lem:sequential-combinational-semantics} we know that \(
    \circuittostreami[
        \adjustimage{valign=c,margin=0pt,page=99}{tikzfigures}
    ](\sigma)(i) = \circuittofunci[
        \adjustimage{valign=c,margin=0pt,page=99}{tikzfigures}
    ] = g(\sigma)(i)\) for all \(\sigma \in \valuetuplestream{m}\) and
    \(i \in \nat\).
    Since \(\circuittostreami[
        \adjustimage{valign=c,margin=0pt,page=257}{tikzfigures}
    ] = \listvar{v} \streamcons \bot \streamcons \bot \streamcons \dots\), we
    have that \(
    \circuittostreami[
        \adjustimage{valign=c,margin=0pt,page=156}{tikzfigures}
    ]
    =
    g(\listvar{v}) \streamcons \bot \streamcons \bot \streamcons \dots
    \), which is the interpretation of \(
    \adjustimage{valign=c,margin=0pt,page=256}{tikzfigures}
    \).
    As the equations are sound they must preserve the stream semantics, so
    \(\listvar{w} = g(\listvar{v})\).
\end{proof}

Finally, we use the above lemma to show how values can be applied to
\emph{essentially} combinational circuits.

\begin{lem}\label{lem:essentially-combinational-applied}
    Let \(\morph{f}{\valuetuple{m}}{\valuetuple{n}}\) be a monotone function
    with \(
    \adjustimage{valign=c,margin=0pt,page=204}{tikzfigures}
    \coloneqq
    \adjustimage{valign=c,margin=0pt,page=258}{tikzfigures}
    \).
    Then \(
    \adjustimage{valign=c,margin=0pt,page=259}{tikzfigures}
    =
    \adjustimage{valign=c,margin=0pt,page=260}{tikzfigures}
    \).
\end{lem}
\begin{proof}
    Let \(h \coloneqq \circuittofunci[
        \adjustimage{valign=c,margin=0pt,page=261}{tikzfigures}
    ]\); using \autoref{lem:combinational-circuit-inputs}, we have that \(
    \adjustimage{valign=c,margin=0pt,page=259}{tikzfigures}
    =
    \adjustimage{valign=c,margin=0pt,page=262}{tikzfigures}
    \).
    So we must show that \(f(\listvar{w}) = h(\listvar{v}, \listvar{w})\).
    \begin{align*}
        f(\listvar{w})
         & =
        \circuittostreami[
            \adjustimage{valign=c,margin=0pt,page=204}{tikzfigures}
        ](\listvar{w} \streamcons \bot^\omega)(0)
         &
        \text{\autoref{def:functional-completeness}}
        \\
         & \coloneqq
        \circuittostreami[
            \adjustimage{valign=c,margin=0pt,page=258}{tikzfigures}
        ](\listvar{w} \streamcons \bot^\omega)(0)
        \\
         & =
        \circuittostreami[
            \adjustimage{valign=c,margin=0pt,page=261}{tikzfigures}
        ](\listvar{v}^\omega, \listvar{w} \streamcons \bot^\omega)(0)
        \\
         & =
        \circuittofunci[
            \adjustimage{valign=c,margin=0pt,page=261}{tikzfigures}
        ](\listvar{v}, \listvar{w})
         &
        \text{\autoref{lem:sequential-combinational-semantics}}
        \\
         & =
        h(\listvar{v}, \listvar{w})
    \end{align*}
    This completes the proof.
\end{proof}

With these lemmas in our toolkit, we can now show that the encoding equations
allow us to translate a circuit into one with an encoded state, and therefore
translate between the state sets of any two denotationally equivalent circuits.

\begin{thm}\label{thm:encoding}
    For a circuit \(
    \adjustimage{valign=c,margin=0pt,page=214}{tikzfigures}
    \) and \(\listvar{s} \in \valuetuple{x}\), the
    equation \(
    \adjustimage{valign=c,margin=0pt,page=263}{tikzfigures}
    =
    \adjustimage{valign=c,margin=0pt,page=264}{tikzfigures}
    \) is derivable by the equations in \(\encodingequations\).
\end{thm}
\begin{proof}
    We have that \(
    \adjustimage{valign=c,margin=0pt,page=217}{tikzfigures}
    =
    \adjustimage{valign=c,margin=0pt,page=216}{tikzfigures}
    \) by the \((\encodingequation)\) equation; we need to `push' the encoder \(
    \adjustimage{valign=c,margin=0pt,page=265}{tikzfigures}
    \) through the state.
    Although the encoder is sequential, by the definition of \(\mealytofunc\),
    it must be of the form \(
    \adjustimage{valign=c,margin=0pt,page=258}{tikzfigures}
    \) by definition of complete interpretations.
    This means we have:
    \begin{align*}
        \adjustimage{valign=c,margin=0pt,page=266}{tikzfigures}
         & \coloneqq
        \adjustimage{valign=c,margin=0pt,page=267}{tikzfigures}
        \\[1em]
         & =
        \adjustimage{valign=c,margin=0pt,page=268}{tikzfigures}
         &
        \text{\autoref{lem:unroll-waveform}}
        \\[1em]
         & =
        \adjustimage{valign=c,margin=0pt,page=269}{tikzfigures}
         &
        \text{\autoref{lem:generalised-streaming}}
        \\[1em]
         & =
        \adjustimage{valign=c,margin=0pt,page=270}{tikzfigures}
         &
        \text{\autoref{lem:essentially-combinational-applied}}
        \\[1em]&\coloneqq
        \adjustimage{valign=c,margin=0pt,page=271}{tikzfigures}
        \\[1em]
         & \coloneqq\adjustimage{valign=c,margin=0pt,page=272}{tikzfigures}
    \end{align*}
    The proof is completed by sliding the encoder around the trace.
\end{proof}

With the right encoders, the initial state of a circuit can be translated into
a different word, giving us a new circuit in Mealy form.
As all the involved components are essentially combinational, the circuit can
be normalised to produce a circuit in normalised Mealy form.

\subsection{Restriction equations}\label{sec:restriction}

We can now map the state set of one circuit to another using encodings.
Does this mean that the two circuits will now be structurally equal?
Unfortunately not: all it means is that the circuits agree on the set of
circuit states.

\begin{exa}\label{ex:restriction-example}
    Consider the following two circuits in \(\scirc{\belnapsignature}\): \[
        \adjustimage{valign=c,margin=0pt,page=273}{tikzfigures}
        \quad
        \adjustimage{valign=c,margin=0pt,page=274}{tikzfigures}
    \]
    Both circuits have circuit states \(\{\belnaptrue\belnapfalse\}\), but their
    combinational cores do \emph{not} have the same semantics.
    They only act the same because they receive certain inputs.
\end{exa}

The final family of equations required is one for mapping between combinational
circuits that agree on the subset of possible inputs they actually receive.

\begin{nota}
    Given sets \(A\), \(B\) and \(C\) where \(C \subseteq A\) and a function
    \(\morph{f}{A}{B}\), the \emph{restriction of \(f\) to \(C\)} is a function
    \(\morph{f|C}{C}{B}\), defined as \(f|C(c) \coloneqq f(c)\).
\end{nota}

\begin{defi}[Restriction equations]
    Let the schema of \emph{restriction equations} be defined as in
    \autoref{fig:restriction-equation}.
\end{defi}

\begin{exa}
    By a restriction equation, the circuits in \autoref{ex:restriction-example} are
    now equal, as the cores produce equal outputs for inputs where the state is
    \(\belnaptrue\belnapfalse\).
\end{exa}

\begin{figure}
    \centering
    \(\equationdisplay{
        \adjustimage{valign=c,margin=0pt,page=275}{tikzfigures}
    }{
        \adjustimage{valign=c,margin=0pt,page=276}{tikzfigures}
    }{\restrictionequation}\)
    \,\,
    where \(
    f|(S_{f,\listvar{s}} \times \valuetuple{\listvar{m}})
    =
    g|(S_{g,\listvar{s}} \times \valuetuple{\listvar{m}})
    \)
    \caption{The schema of \emph{restriction} equations}
    \label{fig:restriction-equation}
\end{figure}

\subsection{Completeness of the algebraic semantics}\label{sec:algebraic-completeness}

It is now possible to collect all the equations together and define a sound and
complete algebraic theory of sequential digital circuits.

\begin{defi}
    For a complete interpretation \(\interpretation\), let
    \(\mce_{\interpretation}\) be \(
    \mealyequations +
    \normalisingequations +
    \encodingequations +
    (\restrictionequation)
    \), and let \(\scircsigmae\) be defined as
    \(\scircsigma / \mce_{\interpretation}\).
\end{defi}

For this to be a \emph{complete} set, we must be able to translate
a circuit \(
\adjustimage{valign=c,margin=0pt,page=89}{tikzfigures}
\) into another circuit \(
\adjustimage{valign=c,margin=0pt,page=108}{tikzfigures}
\) with the same behaviour by only using these equations.

\begin{thm}
    For a complete interpretation \(\interpretation\), \(
    \adjustimage{valign=c,margin=0pt,page=89}{tikzfigures}
    =
    \adjustimage{valign=c,margin=0pt,page=108}{tikzfigures}
    \) in \(\scircsigmae\) if and only if \(
    \circuittostreami[
        \adjustimage{valign=c,margin=0pt,page=98}{tikzfigures}
    ]
    =
    \circuittostreami[
        \adjustimage{valign=c,margin=0pt,page=109}{tikzfigures}
    ]
    \).
\end{thm}
\begin{proof}
    All the equations are sound, so we only need to consider the \(\ifdir\)
    direction.
    Using \autoref{thm:normalising}, the circuits \(
    \adjustimage{valign=c,margin=0pt,page=98}{tikzfigures}
    \) and \(
    \adjustimage{valign=c,margin=0pt,page=109}{tikzfigures}
    \) can be brought to Mealy form, so we have that \(
    \circuittostreami[
        \adjustimage{valign=c,margin=0pt,page=277}{tikzfigures}
    ]
    =
    \circuittostreami[
        \adjustimage{valign=c,margin=0pt,page=278}{tikzfigures}
    ]
    \); this induces monotone Mealy machines
    \((S_{\hat{f}, \listvar{s}}, \hat{f})\)
    and \((S_{\hat{g}, \listvar{t}}, \hat{g})\).
    These Mealy machines have equal stream functions, so there are Mealy
    homomorphisms
    \(\morph{\phi}{S_{\hat{f}, \listvar{s}}}{S_{\hat{g}, \listvar{t}}}\) and
    \(\morph{\psi}{S_{\hat{g}, \listvar{t}}}{S_{\hat{f}, \listvar{s}}}\);
    as we are operating over Mealy coalgebra in \(\pos_\bot\) these are
    \(\bot\)-preserving monotone maps.
    The composite of \(\phi\) and \(\psi\) is also a monotone Mealy
    homomorphism so by \autoref{thm:encoding} we have that \[
        \adjustimage{valign=c,margin=0pt,page=277}{tikzfigures}
        =
        \adjustimage{valign=c,margin=0pt,page=279}{tikzfigures}.
    \]
    The circuit \(
    \adjustimage{valign=c,margin=0pt,page=280}{tikzfigures}
    \) is a composition of normalised circuits, so it is essentially
    combinational; when restricted to the set \(S_{\hat{g}, \listvar{t}}\) its
    truth table is the same as that of \(
    \adjustimage{valign=c,margin=0pt,page=281}{tikzfigures}
    \), as the encoder-decoder pair were defined precisely as the Mealy
    homomorphisms that translate between the two Mealy machines.
    Using the normalisation equations again, the encoded circuit can be
    brought into normalised Mealy form.
    Finally, the restriction equations can be used to translate from \(
    \adjustimage{valign=c,margin=0pt,page=279}{tikzfigures}
    \) into \(
    \adjustimage{valign=c,margin=0pt,page=278}{tikzfigures}
    \).
\end{proof}

As always, the soundness and completeness of the algebraic semantics means we
can establish another isomorphism of PROPs.

\begin{cor}
    \(\scircsigmai \cong \scircsigmae\).
\end{cor}

One might wonder how this improves on the operational approach, as the
normal form is quite complicated.
The beauty of the \emph{algebraic} semantics is that equations can be proven
as lemmas and used in the future as shortcuts; in time, the algebraicist will
build up a repertoire of equations and use them to bend circuits to
their will.

\subsection{Algebraic semantics for generalised circuits}

When it comes to lifting the algebraic semantics to the generalised case, all we
need to do is to extend some of the equations to arbitrary width wires, and to
add equations to handle the bundlers.

\begin{defi}
    Let the set \(\mealyequations^+\) be defined as \(\mealyequations\) but
    with equations \((\monoidunitleqn)\), \((\monoidunitreqn)\), and
    \((\bottomdelayeqn)\) for wires of each width \(n \in \natplus\).
\end{defi}

Since the set of normalising equations \(\normalisingequations\) is determined
by the interpretation, we do not need to do anything there.
The encoding equations need to be extended to act on all wire widths, and we
need to be able to handle encoders that contain bundlers.

\begin{defi}
    Let the set \(\encodingequations^+\) be defined as \(\encodingequations\)
    but with all equations adjusted to operate on wires of all widths
    \(n \in \natplus\), and with the addition of equations \[
        \equationdisplay{
            \adjustimage{valign=c,margin=0pt,page=180}{tikzfigures}
        }{
            \adjustimage{valign=c,margin=0pt,page=181}{tikzfigures}
        }{
            \spliteqn
        }
        \quad
        \equationdisplay{
            \adjustimage{valign=c,margin=0pt,page=182}{tikzfigures}
        }{
            \adjustimage{valign=c,margin=0pt,page=183}{tikzfigures}
        }{
            \combineeqn
        }
    \] on bundlers.
\end{defi}

Finally the restriction schema also needs to operate on arbitrary-width wires.

\begin{defi}
    Let \((\mathsf{Res}^+)\) be defined as \((\mathsf{Res})\) but extended to
    operate on wires of widths \(n \in \natplus\).
\end{defi}

Putting this all together gives us a set of equations for
generalised circuits.

\begin{defi}
    For a generalised interpretation \(\interpretation^+\), let \(
    \mathcal{E}^+_{\mathcal{I}}
    \coloneqq
    \mealyequations^+ +
    \mathcal{N}^+_{\mathcal{I}} +
    \encodingequations^+ +
    (\mathsf{Res^+})
    \), and let \(\scircsigmage\) be defined as
    \(\scircsigmag / \mathcal{E}^+_{\mathcal{I}}\).
\end{defi}

This set of equations is sound and complete.

\begin{thm}
    For a functionally complete generalised interpretation \(\interpretation\),
    \(
    \adjustimage{valign=c,margin=0pt,page=111}{tikzfigures}
    =
    \adjustimage{valign=c,margin=0pt,page=112}{tikzfigures}
    \) in \(\scircsigmage\) if and only if \(
    \circuittostreamig[
        \adjustimage{valign=c,margin=0pt,page=98}{tikzfigures}
    ]
    =
    \circuittostreamig[
        \adjustimage{valign=c,margin=0pt,page=109}{tikzfigures}
    ]
    \).
\end{thm}

Subsequently we obtain another isomorphism of categories.

\begin{cor}
    \(\scircsigmaig \cong \scircsigmage\).
\end{cor}

\section{Potential applications}\label{sec:applications}

So far, we have been concerned largely with \emph{theoretical} concepts; we have
shown how the categorical framework of sequential digital circuits is rigorous
enough to handle composing circuits in sequence, parallel or with the trace
without causing any of the three semantic models to become degenerate.

As existing circuit design technologies are incredibly successful,
our framework is not meant to \emph{replace} but to \emph{complement} them by
highlighting different perspectives on reasoning with sequential digital
circuits.
While the ideas provided in this section are certainly not industry-grade
applications, they are intended to demonstrate the potential of what
the compositional theory can bring to the table.

\subsection{Refactoring}\label{sec:tidy}

When building a circuit, it is desirable to reduce the number of wires
and components used; this reduces both the physical size of the circuit and its
power consumption.
We can use partial evaluation to transform a circuit into a more minimal form.

\begin{defi}[Refactoring rules]
    Let the \emph{refactoring rules} be defined as in \autoref{fig:tidy-rules}.
\end{defi}

Most of the refactoring rules are self-explanatory; the final rule is necessary in
order to deal with traced circuits with no outputs.
Since all circuits with no outputs have the same behaviour, we are permitted to
cut the trace to obtain a circuit we can apply more refactoring rules to.
As non-delay-guarded feedback is already handled by the
\((\instantfeedbackeqn)\) rule, we only need to consider the delay-guarded case.

\begin{figure}
    \centering
    \(
    \adjustimage{valign=c,margin=0pt,page=282}{tikzfigures}
    \reduction
    \adjustimage{valign=c,margin=0pt,page=283}{tikzfigures}
    \)
    \quad
    \(
    \adjustimage{valign=c,margin=0pt,page=223}{tikzfigures}
    \reduction
    \adjustimage{valign=c,margin=0pt,page=42}{tikzfigures}
    \)
    \quad
    \(
    \adjustimage{valign=c,margin=0pt,page=284}{tikzfigures}
    \reduction
    \adjustimage{valign=c,margin=0pt,page=285}{tikzfigures}
    \)
    \\[0.25em]
    \rule{\textwidth}{0.1mm}
    \\[0.5em]
    \(
    \adjustimage{valign=c,margin=0pt,page=192}{tikzfigures}
    \reduction
    \adjustimage{valign=c,margin=0pt,page=193}{tikzfigures}
    \)
    \quad
    \(
    \adjustimage{valign=c,margin=0pt,page=194}{tikzfigures}
    \reduction
    \adjustimage{valign=c,margin=0pt,page=195}{tikzfigures}
    \)
    \quad
    \(
    \adjustimage{valign=c,margin=0pt,page=226}{tikzfigures}
    \reduction
    \adjustimage{valign=c,margin=0pt,page=227}{tikzfigures}
    \)
    \\[0.25em]
    \rule{\textwidth}{0.1mm}
    \\[0.5em]
    \(
    \adjustimage{valign=c,margin=0pt,page=286}{tikzfigures}
    \reduction
    \adjustimage{valign=c,margin=0pt,page=287}{tikzfigures}
    \)
    \quad
    \(
    \adjustimage{valign=c,margin=0pt,page=288}{tikzfigures}
    \reduction
    \adjustimage{valign=c,margin=0pt,page=289}{tikzfigures}
    \)
    \quad
    \(
    \adjustimage{valign=c,margin=0pt,page=290}{tikzfigures}
    \reduction
    \adjustimage{valign=c,margin=0pt,page=291}{tikzfigures}
    \)
    \\[0.25em]
    \rule{\textwidth}{0.1mm}
    \\[0.5em]
    \caption{Rules for refactoring up circuits in Mealy form}
    \label{fig:tidy-rules}
\end{figure}

\begin{prop}
    Applying the refactoring rules to a Mealy form is a confluent and
    terminating procedure.
\end{prop}
\begin{proof}
    The refactoring rules always decrease the size of the circuit.
    The only choice is raised when there is a trace around a combinational
    circuit, but this does not change the internal structure of the subcircuit,
    so rule applications are prevented.
    Moreover, since all this rule does is cut a trace, it does not matter if
    this is performed all in one go, or each feedback loop is cut one by one.
\end{proof}

\subsection{Partial evaluation}\label{sec:partial}

Partial evaluation~\cite{jones1996introduction} is a paradigm used in software
optimisation in which programs are `evaluated as much as possible' while only
some of the inputs are specified.
For example, it may be the case that a particular input to a program is fixed
for long periods of time; using partial evaluation, we can define a program
specialised for this input.
This program might run significantly faster than the original.

There has been work into partial evaluation for hardware, such as constant
propagation~\cite{singh1996expressing,singh1999partial} and
unfolding~\cite{thompson2006bitlevel}.
However, this has been relatively informal, and can be made rigorous using the
categorical framework.
In this section we will focus on how we could extend the reduction-based
operational semantics to define automatic procedures for applying partial
evaluation to circuits.

\subsection{Shortcut rules}\label{sec:shortcut}

It is often the case that we know that some of the inputs to a circuit are
fixed.
This can be modelled by precomposing the relevant input with an
\emph{infinite waveform} \(
\adjustimage{valign=c,margin=0pt,page=292}{tikzfigures}
\).
We can propagate these waveforms across a circuit to see if we can reduce it to
a circuit \emph{specialised} for these inputs.

To propagate waveforms across circuits we need to derive a version of the
\((\gateeqn)\) rule for applying waveforms to primitives.
These rules are illustrated in \autoref{fig:waveform-rules}.

\begin{figure}
    \centering
    \(
    \adjustimage{valign=c,margin=0pt,page=293}{tikzfigures}
    \reduction
    \adjustimage{valign=c,margin=0pt,page=294}{tikzfigures}
    \)
    \quad
    \(
    \adjustimage{valign=c,margin=0pt,page=295}{tikzfigures}
    \reduction
    \adjustimage{valign=c,margin=0pt,page=296}{tikzfigures}
    \)
    \\[0.25em]
    \rule{\textwidth}{0.1mm}
    \\[0.5em]
    \(
    \adjustimage{valign=c,margin=0pt,page=297}{tikzfigures}
    \reduction
    \adjustimage{valign=c,margin=0pt,page=298}{tikzfigures}
    \)
    \\[0.25em]
    \rule{\textwidth}{0.1mm}
    \\[0.5em]
    \caption{Rules for infinite waveforms}
    \label{fig:waveform-rules}
\end{figure}

This is not the only way we can partially evaluate with some inputs.
In some interpretations, it may be that we learn something about the output of
a primitive with only some of the inputs specified.

\begin{exa}[Gate-level shortcuts]
    In the gate-level interpretation \(\belnapinterpretation\), if one applies a
    false value to an \(\andgate\) gate then it will output false regardless of
    the other input.
    Similarly, if one applies a true value to an \(\orgate\) gate it will output
    true.
    Conversely, if one applies a true value to an \(\andgate\) gate or a false
    value to an \(\orgate\) gate, it will act as the identity on the other
    input.
\end{exa}

These `shortcuts' can also be implemented as rules, as illustrated in
\autoref{fig:shortcut-waveform-rules}.
Note that here the value that `triggers' the shortcut must be contained within
an infinite waveform; if we applied the rule with just an instantaneous value,
this value would produce \(\bot\) on ticks after the first and the rule would
be unsound.

\begin{figure}
    \centering
    \(
    \adjustimage{valign=c,margin=0pt,page=299}{tikzfigures}
    \reduction
    \adjustimage{valign=c,margin=0pt,page=300}{tikzfigures}
    \)
    \quad
    \(
    \adjustimage{valign=c,margin=0pt,page=301}{tikzfigures}
    \reduction
    \adjustimage{valign=c,margin=0pt,page=302}{tikzfigures}
    \)
    \\[0.3em]
    \rule{\textwidth}{0.1mm}
    \\[0.5em]
    \(
    \adjustimage{valign=c,margin=0pt,page=303}{tikzfigures}
    \reduction
    \adjustimage{valign=c,margin=0pt,page=300}{tikzfigures}
    \)
    \quad
    \(
    \adjustimage{valign=c,margin=0pt,page=304}{tikzfigures}
    \reduction
    \adjustimage{valign=c,margin=0pt,page=302}{tikzfigures}
    \)
    \\[0.3em]
    \rule{\textwidth}{0.1mm}
    \\[0.5em]
    \(
    \adjustimage{valign=c,margin=0pt,page=305}{tikzfigures}
    \reduction
    \adjustimage{valign=c,margin=0pt,page=306}{tikzfigures}
    \)
    \quad
    \(
    \adjustimage{valign=c,margin=0pt,page=307}{tikzfigures}
    \reduction
    \adjustimage{valign=c,margin=0pt,page=308}{tikzfigures}
    \)
    \\[0.3em]
    \rule{\textwidth}{0.1mm}
    \\[0.5em]
    \(
    \adjustimage{valign=c,margin=0pt,page=309}{tikzfigures}
    \reduction
    \adjustimage{valign=c,margin=0pt,page=306}{tikzfigures}
    \)
    \quad
    \(
    \adjustimage{valign=c,margin=0pt,page=310}{tikzfigures}
    \reduction
    \adjustimage{valign=c,margin=0pt,page=308}{tikzfigures}
    \)
    \\[0.3em]
    \rule{\textwidth}{0.1mm}
    \\[0.5em]
    \caption{Gate-level shortcut rules for waveforms}
    \label{fig:shortcut-waveform-rules}
\end{figure}

\begin{exa}[Control switches]\label{ex:shortcut-switches}
    Recall that a \emph{multiplexer} is a circuit component constructed as \(
    \adjustimage{valign=c,margin=0pt,page=122}{tikzfigures}
    \coloneqq
    \adjustimage{valign=c,margin=0pt,page=123}{tikzfigures}
    \).
    The first input is a \emph{control} which specifies which of the two other
    input signals is produced as the output signal.
    It is often the case that these control signals will be fixed for long
    periods of time; perhaps they specify some sort of global circuit
    configuration.

    Consider the circuit \(
    \adjustimage{valign=c,margin=0pt,page=311}{tikzfigures}
    \), in which the control signal to the multiplexer determines which of two
    subcircuits will become the output.
    We will assume that the control signal is held at false, and reduce
    accordingly by instantiating the rule in \autoref{fig:waveform-rules} detailing
    the interaction of gates and waveforms to the \(\notgate\) case; the steps
    of reasoning are illustrated in \autoref{fig:shortcut-control}.
    \begin{figure}
        \begin{gather*}
            \adjustimage{valign=c,margin=0pt,page=312}{tikzfigures}
            \coloneqq
            \adjustimage{valign=c,margin=0pt,page=313}{tikzfigures}
            \reduction
            \\[1em]
            \adjustimage{valign=c,margin=0pt,page=314}{tikzfigures}
            \reduction
            \adjustimage{valign=c,margin=0pt,page=315}{tikzfigures}
            \reduction
            \\[1em]
            \adjustimage{valign=c,margin=0pt,page=316}{tikzfigures}
            \reduction
            \adjustimage{valign=c,margin=0pt,page=317}{tikzfigures}
            \reduction
            \adjustimage{valign=c,margin=0pt,page=318}{tikzfigures}
        \end{gather*}
        \caption{
            Reducing the circuit in \autoref{ex:shortcut-switches} by using the
            shortcut rules in \autoref{fig:shortcut-waveform-rules}}
        \label{fig:shortcut-control}
    \end{figure}
\end{exa}

\subsection{Shortcuts after streaming}

The rules in the previous sections are intended for use on circuits before we
even apply values to them.
However, there is still potential for partial evaluation when we consider the
outputs of a circuit one step at a time.
To do this, we can apply variants of the shortcut rules \emph{after} performing
streaming for some inputs.
These variants are illustrated in \autoref{fig:shortcuts}.

\begin{figure}
    \centering
    \(
    \adjustimage{valign=c,margin=0pt,page=319}{tikzfigures}
    \reduction
    \adjustimage{valign=c,margin=0pt,page=320}{tikzfigures}
    \)
    \quad
    \(
    \adjustimage{valign=c,margin=0pt,page=321}{tikzfigures}
    \reduction
    \adjustimage{valign=c,margin=0pt,page=322}{tikzfigures}
    \)
    \quad
    \(
    \adjustimage{valign=c,margin=0pt,page=323}{tikzfigures}
    \reduction
    \adjustimage{valign=c,margin=0pt,page=324}{tikzfigures}
    \)
    \\[0.4em]
    \rule{\textwidth}{0.1mm}
    \\[0.5em]
    \(
    \adjustimage{valign=c,margin=0pt,page=325}{tikzfigures}
    \reduction
    \adjustimage{valign=c,margin=0pt,page=326}{tikzfigures}
    \)
    \quad
    \(
    \adjustimage{valign=c,margin=0pt,page=327}{tikzfigures}
    \reduction
    \adjustimage{valign=c,margin=0pt,page=320}{tikzfigures}
    \)
    \quad
    \(
    \adjustimage{valign=c,margin=0pt,page=328}{tikzfigures}
    \reduction
    \adjustimage{valign=c,margin=0pt,page=322}{tikzfigures}
    \)
    \\[0.4em]
    \rule{\textwidth}{0.1mm}
    \\[0.5em]
    \(
    \adjustimage{valign=c,margin=0pt,page=329}{tikzfigures}
    \reduction
    \adjustimage{valign=c,margin=0pt,page=324}{tikzfigures}
    \)
    \quad
    \(
    \adjustimage{valign=c,margin=0pt,page=330}{tikzfigures}
    \reduction
    \adjustimage{valign=c,margin=0pt,page=326}{tikzfigures}
    \)
    \\[0.4em]
    \rule{\textwidth}{0.1mm}
    \caption{Examples of `instantaneous' shortcut rules}
    \label{fig:shortcuts}
\end{figure}

\begin{exa}[Blocking boxes]\label{ex:blocking-boxes}
    Consider the circuit \(
    \adjustimage{valign=c,margin=0pt,page=331}{tikzfigures}
    \), which contains a `blackbox' combinational circuit \(
    \adjustimage{valign=c,margin=0pt,page=99}{tikzfigures}
    \) with unknown behaviour.

    Even though we cannot directly reduce the blackbox, if we set the first
    input to false and use the shortcut rule above, we can still produce an
    output value.
    \[
        \adjustimage{valign=c,margin=0pt,page=332}{tikzfigures}
        \reduction
        \adjustimage{valign=c,margin=0pt,page=333}{tikzfigures}
        \reduction
        \adjustimage{valign=c,margin=0pt,page=334}{tikzfigures}
    \]
\end{exa}

As well as removing redundant blackboxes, judicious use of shortcut
reductions can dramatically reduce the reductions needed to get the outputs of a
circuit.

\subsection{Protocols}\label{sec:uncertain}

Sometimes we may not know the exact inputs to a circuit, but know that they make
up a fixed subset of all possible inputs, or they follow some sort of protocol.
We can implement this in our reduction framework with \emph{uncertain values}
which we either know nothing about or know can only take some specified values.

\begin{defi}
    Let \(\scircsigmap\) be the result of extending \(\scircsigma\) with value
    generators for each word \(\listvar{v?} \in \freemon{\values}\).
\end{defi}

The additional value generators indicate that there are multiple
possible values.
When a circuit contains uncertain values \(
\adjustimage{valign=c,margin=0pt,page=335}{tikzfigures},
\adjustimage{valign=c,margin=0pt,page=336}{tikzfigures},
\dots
\adjustimage{valign=c,margin=0pt,page=337}{tikzfigures},
\) where the maximum length of a given \(v_i\) is \(k\), there are \(k\)
possible value assignments.
For a given assignment \(i < k\), each value will produce a concrete value
defined as \(v_i(j)\) if \(|v_i| > j\) or \(\bot\) otherwise.

To avoid confusion with our syntax sugar for arbitrary-width values, we will
always end uncertain value lists with \(?\).
When writing out specific uncertain words, we delimit the elements with vertical
bars like \(\mathsf{f}|\mathsf{t}\) to allude to the fact that this value is
either the first \emph{or} the second element.

\begin{exa}
    If a circuit contains uncertain values \(
    \adjustimage{valign=c,margin=0pt,page=338}{tikzfigures}
    \) and \(
    \adjustimage{valign=c,margin=0pt,page=339}{tikzfigures}
    \) in a circuit, then there are two universes to consider, one where the
    values output \(\belnapfalse\belnaptrue\) and one where they output
    \(\belnaptrue\belnapfalse\).
    If we add in another uncertain value with three possible values, \(
    \adjustimage{valign=c,margin=0pt,page=340}{tikzfigures}
    \), we now have three possible universes, in which the values output
    \(\belnapfalse\belnaptrue\belnaptrue\),
    \(\belnaptrue\belnapfalse\belnapfalse\), and
    \(\bot\bot\top\) respectively.
\end{exa}

To reason with uncertain values in the reductional framework we need to add
rules for processing them.
Once again it is useful to have versions for both waveforms and values, for
reasoning before and during execution.

\begin{defi}[Uncertain rules]
    The \emph{uncertain rules} are listed in \autoref{fig:uncertain-rules}.
\end{defi}

After applying uncertain values to a primitive, it may turn out that all the
possibilities are in fact the same.
This removes any uncertainty, and means the value can be treated as an
ordinary value in future reductions and outputs.

\begin{figure}
    \centering
    \(
    \adjustimage{valign=c,margin=0pt,page=341}{tikzfigures}
    \reduction
    \adjustimage{valign=c,margin=0pt,page=342}{tikzfigures}
    \)
    \\[0.4em]
    \rule{\textwidth}{0.1mm}
    \\[0.5em]
    \(
    \adjustimage{valign=c,margin=0pt,page=343}{tikzfigures}
    \reduction
    \adjustimage{valign=c,margin=0pt,page=344}{tikzfigures}
    \)
    \quad
    \(
    \adjustimage{valign=c,margin=0pt,page=345}{tikzfigures}
    \reduction
    \adjustimage{valign=c,margin=0pt,page=346}{tikzfigures}
    \)
    \\[0.4em]
    \rule{\textwidth}{0.1mm}
    \\[0.5em]
    \(
    \adjustimage{valign=c,margin=0pt,page=347}{tikzfigures}
    \reduction
    \adjustimage{valign=c,margin=0pt,page=348}{tikzfigures}
    \)
    \quad
    \(
    \adjustimage{valign=c,margin=0pt,page=349}{tikzfigures}
    \reduction
    \adjustimage{valign=c,margin=0pt,page=350}{tikzfigures}
    \)
    \quad
    \(
    \adjustimage{valign=c,margin=0pt,page=351}{tikzfigures}
    \reduction
    \adjustimage{valign=c,margin=0pt,page=352}{tikzfigures}
    \)
    \\[0.4em]
    \rule{\textwidth}{0.1mm}
    \\[0.5em]
    \(
    \adjustimage{valign=c,margin=0pt,page=353}{tikzfigures}
    \reduction
    \adjustimage{valign=c,margin=0pt,page=74}{tikzfigures}
    \)
    \,\,
    if \(\forall i,j < |\listvar{v?}|\), \(\listvar{v?}(i) = \listvar{v?}(j) = v\)
    \\[0.4em]
    \rule{\textwidth}{0.1mm}
    \\[0.5em]
    \(
    \adjustimage{valign=c,margin=0pt,page=354}{tikzfigures}
    \reduction
    \adjustimage{valign=c,margin=0pt,page=292}{tikzfigures}
    \)
    \,\,
    if \(\forall i,j < |\listvar{v?}|\), \(\listvar{v?}(i) = \listvar{v?}(j) = v\)
    \\[0.4em]
    \rule{\textwidth}{0.1mm}
    \caption{Rules for uncertain values}
    \label{fig:uncertain-rules}
\end{figure}

\begin{exa}[Protocols]\label{ex:protocols}
    One sticking point that arises when using the categorical framework is the
    presence of the \(\bot\) and \(\top\) values, which would not normally
    be explicitly provided to a circuit.
    These values mean that some well-known Boolean identities do not always hold.
    By using uncertain values, we can specify the values that \emph{will} be
    applied to a circuit and apply reductions that are not valid in general but
    are in this context.

    In the example in \autoref{fig:protocols-example}, setting the first two
    inputs to true/false inverses reduces the circuit to one with combinational
    behaviour.
    \begin{figure*}
        \begin{gather*}
            \adjustimage{valign=c,margin=0pt,page=355}{tikzfigures}
            \reduction
            \\[0.25em]
            \adjustimage{valign=c,margin=0pt,page=356}{tikzfigures}
            \reduction
            \\[0.25em]
            \adjustimage{valign=c,margin=0pt,page=357}{tikzfigures}
            \reduction
            \\[0.25em]
            \adjustimage{valign=c,margin=0pt,page=358}{tikzfigures}
            \reduction
            \\[0.25em]
            \adjustimage{valign=c,margin=0pt,page=359}{tikzfigures}
            \reduction
            \adjustimage{valign=c,margin=0pt,page=360}{tikzfigures}
            \reduction
            \\[0.25em]
            \adjustimage{valign=c,margin=0pt,page=361}{tikzfigures}
            \reduction
            \adjustimage{valign=c,margin=0pt,page=362}{tikzfigures}
            \reduction
            \adjustimage{valign=c,margin=0pt,page=363}{tikzfigures}
        \end{gather*}
        \caption{Reducing a circuit using protocols}
        \label{fig:protocols-example}
    \end{figure*}
\end{exa}

\subsection{Layers of abstraction}\label{sec:abstraction}

Circuits can be viewed at multiple levels of abstraction.
One could drop down to the level of transistors, as illustrated in
\cite[Sec.\ 4.1]{ghica2017diagrammatic}, or become more abstract, setting the
generators to be \emph{subcircuits} such as arithmetic operations.

The levels of abstraction need not remain isolated.
Using \emph{layered explanations}~\cite{lobski2022string}, multiple signatures
can be mixed in one diagram, with the subcircuits acting as `windows' into
different levels of abstraction, and drawn using `functorial boxes'
\cite{mellies2006functorial}.

\begin{exa}[Implementation]
    Suppose one is working in a high-level signature \(\Sigma_+\) containing a
    generator \(
    \adjustimage{valign=c,margin=0pt,page=364}{tikzfigures}
    \), representing an \emph{IP core}: a circuit that has a known behaviour but
    with an unknown implementation.
    This component can be left as a blackbox and evaluated as
    demonstrated above.

    The designer then attempts to design their own implementation using the
    gate-level signature \(\belnapsignature\).
    To synthesise the final circuit, a map is defined from generators in \(
    \Sigma_+
    \) to morphisms in \(\scirc{\belnapsignature}\), which induces a functor
    \(\scirc{\Sigma_+} \to \scirc{\belnapsignature}\).
    Different implementations can be defined as different maps, and hence
    different functors, as illustrated in \autoref{fig:implementations}.
    These circuits can then be tested to see if they act as intended.
    \begin{figure*}
        \begin{gather*}
            \adjustimage{valign=c,margin=0pt,page=365}{tikzfigures}
            \xLeftarrow{\text{Imp.\ 1}}
            \adjustimage{valign=c,margin=0pt,page=366}{tikzfigures}
            \xRightarrow{\text{Imp.\ 2}}
            \adjustimage{valign=c,margin=0pt,page=367}{tikzfigures}
        \end{gather*}
        \caption{Using different implementations of a circuit}
        \label{fig:implementations}
    \end{figure*}
\end{exa}

\subsection{Refining circuits}\label{sec:refining}

A key part of circuit design comes in \emph{optimising circuits}: making them
run as fast as possible and reduce the \emph{clock cycle}.

\begin{exa}[Retiming]
    The clock cycle of a circuit is determined by the longest paths between
    registers. Altering the paths between registers can be achieved using
    \emph{retiming}~\cite{leiserson1991retiming}: moving registers across gates.
    This is modelled by the streaming rule (\autoref{lem:streaming});
    forward retiming (streaming left to right) is always possible
    but for backward retiming (streaming right to left), the value
    in the register must be in the image of the gates.
\end{exa}

The streaming rule permits retiming using the composite register construct, but
can also be used to retime raw delay components.

\begin{lem}[Timelessness]
    For any primitive \(\adjustimage{valign=c,margin=0pt,page=368}{tikzfigures}\),
    \(
    \adjustimage{valign=c,margin=0pt,page=369}{tikzfigures} =
    \adjustimage{valign=c,margin=0pt,page=370}{tikzfigures}
    \).
\end{lem}
\begin{proof}
    \(
    \adjustimage{valign=c,margin=0pt,page=369}{tikzfigures} =
    \adjustimage{valign=c,margin=0pt,page=371}{tikzfigures} =
    \adjustimage{valign=c,margin=0pt,page=372}{tikzfigures} =
    \adjustimage{valign=c,margin=0pt,page=373}{tikzfigures} =
    \adjustimage{valign=c,margin=0pt,page=370}{tikzfigures}
    \)
\end{proof}

When reasoning equationally, the behaviour of the circuits on either side of the
equation must have exactly the same behaviour.
However, when reasoning with circuits it is sometimes the case that this is too
strict an assertion; we are looking for circuits that output the same outputs
but over a shorter period of time.
This means we may wish to use transformations that only `morally' preserve the
behaviour of a circuit.

\begin{defi}
    For two finite sequences \(
    \listlistvar{v},\listlistvar{w} \in (\valuetuple{m})^k
    \), we say that \(\listlistvar{w}\) is a \emph{stretching} of
    \(\listlistvar{v}\), written \(\listlistvar{v} \ll \listlistvar{w}\), if
    \(\listlistvar{w}\) contains the characters of \(\listlistvar{v}\) but
    possibly repeated or with additional \(\bot\) characters e.g.\ \(
    \belnaptrue\belnapfalse
    \ll
    \bot\bot\belnaptrue\belnaptrue\bot\belnapfalse
    \).
\end{defi}

\begin{defi}
    For two sequential circuits \(
    \adjustimage{valign=c,margin=0pt,page=89}{tikzfigures}
    \) and \(
    \adjustimage{valign=c,margin=0pt,page=108}{tikzfigures}
    \) with \(c\) and \(c^\prime\) delay components respectively, we say that \(
    \adjustimage{valign=c,margin=0pt,page=89}{tikzfigures}
    \) is \emph{more efficient} than \(
    \adjustimage{valign=c,margin=0pt,page=108}{tikzfigures}
    \), written \(
    \adjustimage{valign=c,margin=0pt,page=89}{tikzfigures}
    \ll
    \adjustimage{valign=c,margin=0pt,page=108}{tikzfigures}
    \), if for all sequences \(\listlistvar{v},\listlistvar{w}\) produced by the
    productive operational semantics for inputs of length
    \(\mathsf{max}(c,c^\prime)\), \(\listlistvar{v} \ll \listlistvar{w}\).
\end{defi}

Including this notion of efficiency in algebraic reasoning allows us to reason
with \emph{inequalities} as well as equalities, so more efficient circuits can
be identified.
The simplest form of reasoning in this manner is where we have the
same circuit but guarded by different numbers of delays.

\begin{nota}
    We write \(
    \adjustimage{valign=c,margin=0pt,page=374}{tikzfigures}
    \) for the composition of \(p\) delay components, i.e.\ \(
    \adjustimage{valign=c,margin=0pt,page=375}{tikzfigures}
    \coloneqq
    \adjustimage{valign=c,margin=0pt,page=376}{tikzfigures}
    \) and \(
    \adjustimage{valign=c,margin=0pt,page=377}{tikzfigures}
    \coloneqq
    \adjustimage{valign=c,margin=0pt,page=378}{tikzfigures}
    \).
\end{nota}

\begin{lem}
    For a combinational circuit \(
    \adjustimage{valign=c,margin=0pt,page=36}{tikzfigures}
    \) and \(p,q \in \nat\) such that \(p < q\), then \(
    \adjustimage{valign=c,margin=0pt,page=379}{tikzfigures}
    \ll
    \adjustimage{valign=c,margin=0pt,page=380}{tikzfigures}
    \).
\end{lem}

By combining this with streaming and timelessness, we gain a useful tool for
translating between circuits with the same behaviour but different timings.

\begin{exa}
    One source of delay in circuits is the time gates take to process input
    signals.
    We can model this by inserting delay components after each gate, such as in
    the following circuit:
    \[
        \adjustimage{valign=c,margin=0pt,page=381}{tikzfigures}
    \]
    During reasoning we can permit these delays to be moved around, so long as
    when we finish any gates are still guarded by delays.
    \begin{gather*}
        \adjustimage{valign=c,margin=0pt,page=381}{tikzfigures}
        =
        \adjustimage{valign=c,margin=0pt,page=382}{tikzfigures}
        \gg
        \\[1em]
        \adjustimage{valign=c,margin=0pt,page=383}{tikzfigures}
        =
        \adjustimage{valign=c,margin=0pt,page=384}{tikzfigures}
        =
        \adjustimage{valign=c,margin=0pt,page=385}{tikzfigures}
    \end{gather*}
\end{exa}

While this is a somewhat contrived toy example, it is possible that this
technique could be applied to actual circuit optimisation procedures.

\begin{exa}[Pipelining]
    \emph{Pipelining}~\cite{parhi1999vlsi} is a technique in which more
    registers are inserted into a circuit to increase throughput.
    This can be emulated in the compositional framework by applying
    transformations locally to registers.
    Ordinarily, such transformations can obfuscate a circuit's behaviour since
    the state space dramatically changes.
    In the compositional model, the structure of the circuit is left relatively
    untouched so this is less of an issue.
\end{exa}

Not all circuit transformations are for the purpose of improving performance.
Sometimes additional components must be bolted onto a circuit for \emph{testing}
purposes.

\begin{exa}[Scan chains]
    A common way of testing circuits is by using a
    \emph{scan chain}~\cite{mourad2000principles}, a way of forcing the
    inputs to flipflops to test how specific states affect the outputs of the
    circuit.
    Adding a flipflop to a scan chain requires some extra inputs: the
    \(\mathsf{scan}_\mathsf{en}\) wire toggles if the flipflop operates in
    normal mode or if it takes \(\mathsf{scan}_\mathsf{in}\) as its value.
    \begin{gather*}
        \adjustimage{valign=c,margin=0pt,page=386}{tikzfigures}
        \xRightarrow{\text{scan}}
        \\[1em]
        \adjustimage{valign=c,margin=0pt,page=387}{tikzfigures}
    \end{gather*}
\end{exa}

One could factor in these transformations when designing the circuit, but this
can obfuscate the design of the actual logic.
Additionally, applying these transformations where the remaining part of the
circuit is \emph{not} combinational can be quite complex.
With the compositional approach the two tasks can be kept isolated by using
blackboxes, layered explanations, and graphical reasoning.

\section{Conclusion}\label{sec:conclusion}

\begin{figure}
    \centering
    \begin{tikzcd}[column sep=large]
    \streami
    \arrow[swap]{dd}{\streamtomealyi}
    \arrow[<->, bend left]{drr}{\cong}
    &
    \funci
    \arrow{l}{\functostreami}
    &
    \\
    &
    \ccircsigma
    \arrow[hookrightarrow]{d}
    \arrow{u}
    &
    \scircsigmai
    \\
    \mealyi
    \arrow[bend left=60]{uu}{\mealytostream}
    \arrow[bend right=60, swap]{r}{\mealytocircuiti}
    &
    \scircsigma
    \arrow{uul}{\circuittostream{\interpretation}}
    \arrow{l}{\circuittomealy{\interpretation}}
    \arrow{ur}{/ \approx}
    \arrow{r}{/ \mce}
    \arrow[swap]{dr}{/ \sim}
    &
    \scircsigmae
    \arrow[<->]{u}{\cong}
    \\
    &
    &
    \scircsigmaobs
    \arrow[<->]{u}{\cong}
\end{tikzcd}
    \caption{Relating categorical models of circuits}
    \label{fig:map}
\end{figure}

We have presented three ways of showing equivalence of
circuits: \emph{denotationally} by using the stream semantics,
\emph{operationally} by using the productive reduction strategy, and
\emph{algebraically} by using the sound and complete equational theory.
This completes the research programme initiated
by~\cite{ghica2016categorical,ghica2017diagrammatic}; a summary of the various
categories and morphisms between them that make up the categorical framework of
circuits is shown in \autoref{fig:map}.

String diagrams as a graphical syntax for monoidal categories were introduced a
few decades ago~\cite{joyal1991geometry,joyal1996traced}, and there has since
been an explosion in their use for various applications, such as
cyclic lambda calculi~\cite{hasegawa1997recursion,ghica2023string}, fixpoint
operators~\cite{hasegawa2003uniformity},
quantum protocols~\cite{abramsky2004categorical}, signal flow
diagrams~\cite{bonchi2014categorical,bonchi2015full}, linear
algebra~\cite{bonchi2017interacting,zanasi2015interacting,bonchi2019graphical,boisseau2022graphical},
dynamical systems~\cite{baez2015categories,fong2016categorical}, electrical
circuits~\cite{boisseau2022string} and automatic
differentiation~\cite{alvarez-picallo2023functorial}.
While all these frameworks use compositionality in some way, the nature of
digital circuits mean they differ to ours.
In many of the above applications, the join and the fork form a
\emph{Frobenius structure}, making the wires bidirectional.
This means the trace is constructed as
\(\adjustimage{valign=c,margin=0pt,page=388}{tikzfigures}\),
which degenerates to
\(\adjustimage{valign=c,margin=0pt,page=389}{tikzfigures}\) in the presence of
\emph{Cartesian} structure.
Indeed, in any compact closed category the product would automatically be a
coproduct (biproduct), which is a degeneracy incompatible with models of
digital circuits.

There are other settings that permit loops but retain unidirectionality of
wires.
\emph{Categories with feedback} were introduced in~\cite{katis2002feedback} as a
weakening of STMCs that removes the yanking axiom, enforcing that \emph{all}
traces are delay-guarded.
In~\cite{dilavore2021canonical} Mealy machines are characterised as a category
with feedback: this is compatible with our framework since all `instant
feedback' is expressed as fixpoints and only delay-guarded feedback remains.
\emph{Categories with delayed trace}~\cite{sprunger2019differentiable} weaken
the notion further by removing the sliding axiom; this prohibits the unfolding
rule so would be unsuitable.

Axiomatising fixpoint operators has been studied
extensively~\cite{bloom1993iteration,stefanescu2000network,simpson2000complete}.
Since any Cartesian traced category admits a fixpoint (or \emph{Conway})
operator~\cite{hasegawa1997recursion}, these equations can be expressed using
the Cartesian equations and axioms of STMCs.
Since our work takes place in a \emph{finite} lattice, we are able express a
fixpoint by iterating the circuit a finite number of times.
While this result is well-known from the denotational
perspective~\cite{stoltenberg-hansen1994mathematical}, it has not been used
before, perhaps surprisingly, to solve the problem of combinational feedback.
The interplay of causal streams and dataflow categories has also been studied
elsewhere: recently, a generalisation of causal streams known as
\emph{monoidal streams}~\cite{dilavore2022monoidala} has been developed to
provide semantics to dataflow programming.
Although this generalises some aspects of this paper, our approach differs in
the use of the finite lattice and monotone functions.

The correspondence between Mealy machines and digital circuits is a fundamental
result in automata theory~\cite{mealy1955method} applied extensively in circuit
design~\cite{kohavi2009switching}.
The links between Mealy machines and causal stream functions using coalgebras is
a more recent
development~\cite{rutten2005coinductive,rutten2005algebra,rutten2006algebraic}.
Mealy machines over meet-semilattices are introduced
in~\cite{bonsangue2008coalgebraic} to model a logical framework which includes
a fixpoint.
We also employ this technique in order to handle fixpoints, but also assemble
(monotone) Mealy machines into a PROP in order to bridge between
between stream functions and sequential circuits.

String diagrams are not efficient to work with computationally.
Instead they must be translated into combinatorial graphs; this was touched on
in \cite{ghica2017diagrammatic} using
\emph{framed point graphs}~\cite{kissinger2012pictures}.
Recent work in string diagram
rewriting~\cite{bonchi2022string,bonchi2022stringa,bonchi2022stringb} has used
\emph{hypergraphs} for rewriting modulo Frobenius structure.
This framework has been adapted for categories with a (co)monoid
structure~\cite{fritz2023free,milosavljevic2023string}, and \emph{traced}
comonoid structure~\cite{ghica2026rewriting}: the setting in which we model
digital circuits.
We have already begun to develop an automated framework for rewriting
circuits based on this work~\cite{kaye2026circuitcj}.

\paragraph*{Acknowledgments.}
Thanks to Miriam Backens and Chris Barrett for comments on earlier versions of
this paper, and to the anonymous reviewers for their invaluable feedback.

\bibliographystyle{alphaurl}
\bibliography{refs/refs}

\end{document}